\documentclass[sigconf,nonacm]{acmart}

\usepackage{latexsym}
\usepackage{amsfonts}
\usepackage{amsmath}
\usepackage{nccmath}
\usepackage{amssymb}
\usepackage{color}
\usepackage{epsfig}
\usepackage{xspace}
\usepackage{graphicx}
\usepackage{balance}  
\usepackage{subfigure}
\usepackage{balance}
\usepackage{epsfig}
\usepackage{multirow}
\usepackage{url}
\usepackage{makecell}
\usepackage{colortbl}
\usepackage{mathrsfs}
\usepackage{xcolor,listings}
\usepackage{stmaryrd}
\usepackage{textcomp}
\usepackage{mathtools}
\usepackage{float}
\usepackage{multicol}
\usepackage{algpseudocode}
\usepackage[linesnumbered, ruled, vlined]{algorithm2e}
\usepackage{newfloat}
\renewcommand\footnotetextcopyrightpermission[1]{} 
\allowdisplaybreaks

\usepackage[normalem]{ulem}
\usepackage{enumitem}
\setlist{topsep=0pt,noitemsep} \setitemize[1]{label=$\circ$}

\usepackage{soul}

\newcommand{\at}[1]{\protect\ensuremath{\mathsf{#1}}\xspace}

\newcommand{\eat}[1]{}

\newcommand{\sstab}{\rule{0pt}{8pt}\\[-1.8ex]}

\newcommand{\ra}{\rightarrow}

\newcommand{\bi}{\begin{itemize}}
\newcommand{\ei}{\end{itemize}}
        {\end{itemize}} 

\newcommand{\mat}[2]{{\begin{tabbing}\hspace{#1}\=\+\kill #2\end{tabbing}}}

\newcommand{\be}{\begin{enumerate}}
\newcommand{\ee}{\end{enumerate}}
\newcommand{\beqnn}{\begin{eqnarray*}}
\newcommand{\eeqnn}{\end{eqnarray*}}
\newcommand{\beqn}{\begin{eqnarray}}
\newcommand{\eeqn}{\end{eqnarray}}

\newcommand{\stitle}[1]{\vspace{1.6ex}\noindent{\bf #1}}
\newcommand{\etitle}[1]{\vspace{1ex}\noindent{\underline{\em #1}}}

\newcommand{\ie}{\emph{i.e.,}\xspace}
\newcommand{\eg}{\emph{e.g.,}\xspace}
\newcommand{\wrt}{\emph{w.r.t.}\xspace}

\renewcommand{\st}{\emph{s.t.}\xspace}

\newcommand{\SELECT}{\mbox{{\bf select}}\ }
\newcommand{\FROM}{\mbox{{\bf from}\ }}
\newcommand{\JOIN}{\mbox{{\bf join}\ }}
\newcommand{\AS}{\mbox{{\bf as}\ }}
\newcommand{\ON}{\mbox{{\bf on}\ }}
\newcommand{\MATCH}{\mbox{{\bf match}\ }}
\newcommand{\MAP}{\mbox{{\bf map}\ }}
\newcommand{\WHERE}{\mbox{\bf where}\ }

\newcommand{\Or}{\mbox{\bf or}\ }
\renewcommand{\And}{\mbox{\bf and}\ }

\newcommand{\Continue}{\mbox{\bf continue}\ }

\newcommand{\False}{\mbox{\em false}\xspace}
\newcommand{\True}{\mbox{\em true}\xspace}

\newcommand{\kw}[1]{{\ensuremath {\mathsf{#1}}}\xspace}

\newcounter{ccc}

\newcommand{\SQL}{\kw{SQL}}

\newcommand{\R}{{\mathcal R}}

\newcommand{\eop}{\hspace*{\fill}\mbox{$\Box$}}     
\newcounter{example}
\renewcommand{\theexample}{\arabic{example}}
\newenvironment{example}{
        \vspace{1.5ex}
        \refstepcounter{example}
        {\noindent\bf Example \theexample:}}{
        \eop\vspace{1.5ex}}

\newcommand{\nthesection}{\arabic{section}}
 \newcounter{theorem}
\newcounter{prop}
\renewcommand{\theprop}{\arabic{theorem}}
\newcounter{lemma}
\renewcommand{\thelemma}{\arabic{theorem}}
\newcounter{cor}
\renewcommand{\thecor}{\arabic{theorem}}

\newcounter{definition}[section]
\renewcommand{\thedefinition}{\nthesection.\arabic{definition}}

\newcounter{alg}[section]
\renewcommand{\thealg}{\nthesection.\arabic{alg}}

\newcounter{arule}
\renewcommand{\thearule}{\arabic{arule}}

\newcounter{claim}
\renewcommand{\theclaim}{\arabic{claim}}

\newcommand{\tbf}{\textbf{\textcolor{red}{X}}\xspace}

\renewcommand{\texttt}[1]{{\small\textsf{#1}}}

\definecolor{gray}{rgb}{0.5,0.5,0.5}

\newcommand{\F}{{\mathcal F}}

\renewcommand{\P}{{\mathcal P}}

\newcommand{\revise}[1]{{#1}}

\newcommand{\warn}[1]{{#1}}

\newcommand{\dbpedia}{\kw{DBpedia}}
\newcommand{\linkedmdb}{\kw{LinkedMDB}}
\newcommand{\yago}{\kw{YAGO}}

\newcommand{\movie}{\kw{movieLens}}

\newlist{myitemize}{itemize}{3}
\setlist[myitemize,1]{label=$\circ$,leftmargin=2.8ex}
\setlist[myitemize,2]{label=$\bullet$,leftmargin=2.8ex}
\setlist[myitemize,3]{label=$\diamond$, leftmargin=2.2ex}
\newenvironment{mbi}{\begin{myitemize}
        \setlength{\topsep}{0.6ex}\setlength{\itemsep}{0.16ex}\vspace{0.36ex}}
        {\end{myitemize}\vspace{0.16ex}}

        \newcommand{\mei}{\end{myitemize}\vspace{0.6ex}}

\newcommand{\D}{\mathcal{D}}

\newcommand{\kid}{\kw{id}}

\newcommand{\imdb}{\kw{IMDB}}
\newcommand{\dblp}{\kw{DBLP}}

\newcommand{\ResolveOp}{\kw{ResolveOp}}
\newcommand{\ResolveExp}{\kw{ResolveExp}}
\newcommand{\ResolveCond}{\kw{ResolveCond}}

\newcommand{\SystemName}{\kw{WhiteDB}} 
\newcommand{\SQLd}{$\kw{SQL_{\delta}}$\xspace}
\newcommand{\RA}{\kw{RA}}

\newcommand{\RAd}{$\kw{RA_{\delta}}$\xspace}
\newcommand{\RDBMS}{\kw{RDBMS}}

\newcommand{\join}{\bowtie}

\newcommand{\ModelName}{\kw{RG}} 
\newcommand{\explore}[1]{{\ensuremath{\vec\bowtie^{#1}}}}
\newcommand{\cexplore}[1]{\ensuremath{\vec{\bowtie}^{#1}_\theta}}
\newcommand{\iexplore}{\ensuremath{\:\vec{\bowtie}^A_{\theta^I}}}
\newcommand{\ijoin}{\ensuremath{\:{\bowtie}_{\theta^I}}}

\DeclareFloatingEnvironment[fileext=frm,placement={!ht},name=Frame]{myfloat}
\makeatletter
\newcommand{\removelatexerror}{\let\@latex@error\@gobble}
\makeatother

\newcommand{\nextnr}{\stepcounter{AlgoLine}\ShowLn}
\title{Joining Entities Across Relation and Graph with a Unified Model}

\author{Wenzhi Fu}

\affiliation{University of Edinburgh}

\email{wenzhi.fu@ed.ac.uk}

\date{}

\begin{document}
\fancyhead{}

\begin{abstract}
		This paper introduces \warn{\ModelName (\warn{\underline{R}elational \underline{G}enetic}) model,
		a  revised relational model to represent graph-structured data in \RDBMS while preserving its topology,
		for efficiently and effectively extracting data in different formats from disparate sources.}
		Along with:
		(a) \SQLd, an \SQL dialect augmented with graph pattern queries and
		tuple-vertex joins, such that one can extract graph properties via graph pattern matching, 
		and ``semantically'' match entities across relations
		and graphs\eat{,
		to enrich the semantics of the data};
		(b) a logical representation of graphs in \RDBMS,
		\warn{which introduces an exploration operator for efficient pattern querying,
			supports also browsing and updating graph-structured data;}\eat{
		\warn{that supports browsing, updating and induces more efficient operators to support the declartive query}
		as an interface \warn{that provides more efficient operators} to support declarative query, browsing and updating on
		graph-structured data;} 
		and
		(c) a strategy to uniformly evaluate \SQL,  pattern and
		hybrid queries that join tuples and vertices\eat{ if they
		refer to the same entity}, all inside an \RDBMS by 
		leveraging its optimizer \warn{without performance degradation on switching different execution engines}.
		\warn{A lightweight system, \SystemName, is developed as an implementation
		to evaluate the benefits it can actually bring on real-life data.}
		We empirically \warn{verified} that the \ModelName model enables the
		\warn{graph pattern queries to be answered} as \warn{efficiently} as
		\warn{in} native graph engines;
		\warn{can consider the access on graph and relation in any order for optimal plan};
		\warn{and supports effective data enrichment.}
		\eat{\revise{and effectively enrich the data.}}
\end{abstract}

\maketitle
\setcounter{page}{1}
\pagestyle{plain}

\vspace{-0.7ex}
\section{Introduction}
\label{sec-intro}
\vspace{-0.4ex}

\eat{
Joining the aligned entities
across data in different formats 
from disparate sources
that referring to the same real-world entity
is in an evident need to 
add values from the Variety of big data.
It is in practice to enrich data and incorporate additional useful information 
to make insightful and informed decisions,
which is highly valued
for its increasing practicality in a variety of applications~\cite{DE2,DE3, DE4,DE5, DE6, DE7, DatabaseBasedIntegrationTools}.
In particular, practitioners often want to recognize
entities in graph and relational data, then join them together 
according to their equivalent value or semantic connection.}

\warn{Data enrichment aims at extracting properties and attributes 
in different formats
from disparate sources, and incorporating additional
useful information for users to make insightful
and informed decisions,
which is highly valued
for its increasing practicality in a variety of applications~\eat{DE2,DE5, DatabaseBasedIntegrationTools}\cite{DE2,DE5,DE4, DE6, DE7, DE1, DE3, DatabaseBasedIntegrationTools}.\eat{
The data enrichment market ``was valued
at US\$1.7 Bn in 2021, and is expected to grow at US\$3.4 Bn by 2030'';
the market size is predicted to ``grow at a CAGR of 8.5\%''
from 2022-2030~\cite{DE3}. Data enrichment has been increasingly
practiced in a variety of applications, such as fraud prevention,
insurance and retail~\cite{DE2}. }
In particular, to enrich the semantics of relational data with graph properties and enrich their graph queries with
relational predicates, 
\warn{{\em unified data analytics} is required for 
the practitioners to identify the aligned entities across graphs and relations dynamically
and join them together}.}
\looseness=-1

\begin{figure}[t]
	\centerline{\includegraphics[scale=1]{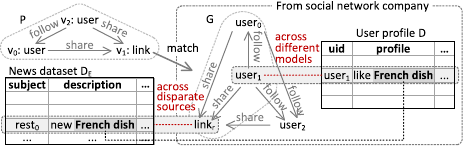}}
	\centering
	\vspace{-2.4ex}
	\caption{\revise{An example hybrid query}}
	\label{fig-example}
	\vspace{-4.4ex}
\end{figure}

\begin{example}
\label{exa-running}
\warn{
Below is a recommendation task from a social media company,
which maintains: 
(1) a relational database $\D$ of detailed user portrait and profile, and 
(2) a graph $G$ of the ``follow'' among users and the links the users share.
The connection between $G$ and $\D$ can be maintained for the vertices and the tuples
to be matched together with aligned user id.\eat{
The company can maintain the connections between 
$\D$ and $G$ with the aligned user id.
Meanwhile, the company can also access an external
database $\D_E$ of the news data extracted from different sources.}
Meanwhile, the company would also monitor an external
database $\D_E$ of on-going events from disparate sources,
and decide whether to push those news to its users according to their profile
and the users they are following.
\looseness=-1

The company then needs to:
(1) match a pattern $P$ in $G$ to find the links shared by both a user and its follower,
(2) extract the events from $\D_E$ relevant to the shared links, and
(3) access the profile of the follower from $\D$ to find whether they would be interested 
	in those events.
As demonstrated in Figure~\ref{fig-example},
$rest_0$ can be recommended to $user_1$ to try on their new French dishes.}
\end{example}

\eat{The need for dynamically enriching data is evident in practice. It is
to extract and merge with only semantically relevant information
without introducing irrelevant ``noise''. It is important
for unified data analytics and for adding values from the Variety of big data.}

\warn{The need for unified data analytics is evident in practice to extract value out of the Variety of big data~\cite{russom2011big}.
More specifically, to effectively and efficiently join the aligned entities across graph and relation
according to equivalent value or semantic connection, several questions need to be answered.\eat{
To efficiently and effectively add values from the Variety of big data,
a unified data analytics is in an evident need
to dynamically enrich data in practice.
Several questions then need to be answered:}} 
Is it possible to dynamically enrich relational entities
identified by \SQL queries with additional properties of
the same entities from graphs? How can we enrich
graph queries with relational properties? Can we support
these ``hybrid'' queries in a relational database management system (\RDBMS)
without switching between relation and graph engines? 
Is it possible to store graphs in the \RDBMS such that
the system is able to conduct graph pattern matching as efficient as
a native graph engines~\cite{GSQL}?
How can we efficiently evaluate the hybrid queries \warn{with an enlarged unified plan space} by leveraging
the optimizer of the \RDBMS? 
\looseness=-1

\eat{
It is important for unified data analytics and for adding values from the Variety of big data,
to meet the evident need for dynamically enriching data in practice.
However, to effectively and efficiently achieve this, several questions need to be answered: 
Is it possible to dynamically enrich relational entities
identified by \SQL queries with additional properties of
the same entities from graphs? How can we enrich
graph queries with relational properties? Can we support
these ``hybrid'' queries in a relational database management system (\RDBMS)
without switching between relation and graph engines? 
Is it possible to store graphs in the \RDBMS such that
the system is able to conduct graph pattern matching as efficient as
a native graph engines~\cite{GSQL}?
How can we efficiently evaluate the hybrid queries \warn{with an enlarged unified plan space} by leveraging
the optimizer of the \RDBMS? 
}

\vspace{0.36ex}
Several systems have been built atop an \RDBMS by storing 
graphs as relations~\cite{Jeffrey1,Jeffrey2,Cytosm,Verticagraph,
Vertexica,DaveJLXGZ16,NScaleSpark,Kaskade,GraphGen}.
These systems either translate specific
types of graph queries into \SQL and execute the queries
via the \RDBMS, or extract graph views of relations and
conducts graph analytics using another query engine.
These methods either incur heavy cost when answering
common graph queries, \eg graph pattern matching since
it involves multi-way join, as observed in~\cite{GSQL};
or do not allow users to write declarative queries
in their familiar language such as \SQL and do not
make use of the sophisticated optimizer of \RDBMS, as pointed
out by~\cite{GRFusion}. While some systems, \eg~\cite{GRFusion},
combine the two approaches and uniformly support both, 
none of the 
systems support hybrid queries specified in
Example~\ref{exa-running} to enrich data as demanded by
practitioners. 
In particular, no prior systems are able to
dynamically identify an entity $e$ by a tuple returned by an \SQL query,
extract the properties of the same entity encoded in a graph,
and enrich $e$ with the properties as additional attributes,
all inside the \RDBMS. 

\stitle{Contributions \& organization}.
In light of these, 
\warn{this paper introduces \ModelName (\warn{\underline{R}elational \underline{G}enetic}) model
	 to uniformly represent graphs in \RDBMS}, along with an implementation of it, \SystemName.
The novelty of it consists of the following.
(a) It supports a dialect of \SQL to dynamically enrich
relational entities with additional graph properties of
the same entities, and enrich graph pattern matching
with relational predicates. This departs from previous
works~\cite{Jeffrey1,Jeffrey2,Cytosm,Kaskade,Verticagraph,Vertexica,DaveJLXGZ16,NScaleSpark,GraphGen,GRFusion}.
(b) It provides a logical representation of graph-structured data in \RDBMS, 
which preserves its topology and provides more efficient operators 
to conduct graph pattern matching as efficient
as a native graph store.
(c) The \ModelName model enables an enlarged unified plan space to 
uniformly optimize native \SQL queries, graph pattern 
queries and hybrid queries \warn{by fully leveraging}
the sophisticated query optimizer of the \RDBMS,
and allow them to be executed all inside \RDBMS without performance degradation. 
We show the above advantages by empirically evaluating
the performance of \SystemName with real-life data.
\looseness=-1

\eat{
In light of these, we introduce \SystemName, a lightweight system for
dynamic data enrichment across relations and graphs. The novelty
of \SystemName consists of the following.
(a) It  supports hybrid queries to dynamically enrich
relational entities with additional graph properties of
the same entities, and enrich graph pattern matching
with relational predicates. This departs from previous
systems~\cite{Jeffrey1,Jeffrey2,Cytosm,Kaskade,Verticagraph,
Vertexica,DaveJLXGZ16,NScaleSpark,GraphGen,GRFusion}.
(b) \SystemName uniformly supports native \SQL queries, graph pattern 
queries and hybrid queries, all inside \RDBMS, and makes full
use of the sophisticated query optimizer of the \RDBMS. 
(c) It proposes a model to store graphs in the \RDBMS that
preserves the topology and locality of subgraphs, such that
\SystemName is able to conduct graph pattern matching as efficient
as a native graph store, and incurs no 
performance
degradation when evaluating hybrid queries.}

\vspace{0.36ex}
\warn{Pattern matching is focused as graph queries in this work, to prove the concept of unified evaluation for hybrid queries (see~\cite{full}).}
As surveyed
in~\cite{SahuMSLO2020ubiquity}, 
pattern matching dominates the topics of  \revise{studies}
on graph computations,
and \warn{also}\eat{pattern queries} have been widely used in social network analysis~\cite{fan2015association},
security~\cite{jiang2010identifying}, software
engineering~\cite{surian2010mining}, 
biology~\cite{royer2008unraveling} 
and chemistry~\cite{wegner2006data}, among other things.
Graph pattern matching has also been considered as one of the core functionalities for querying graphs
\cite{angles2018GCORE, alin2022Graph},
and is enabled as a feature in a number of
graph query languages,~\eg~\cite{francis2018cypher,
TinkerPop} and graph databases~\cite{tian2023world}.
For data enrichment in particular, pattern queries suffice to identify
graph entities and properties by specifying their topological structures.
We defer a full treatment of generic graph queries, especially considering the recursive feature~\cite{perez2009semantics},
to a later paper.
\looseness=-1

\vspace{0.36ex}
More specifically, we highlight the following.

\etitle{(1) \SQLd: A dialect of \SQL} (Section~\ref{sec-SQL}).
We propose \SQLd to support (a) \SQL queries, (b) graph pattern queries,
and (c) hybrid of the two by joining entities across relations and graphs.
\SQLd can dynamically join entities: it may identify an entity $e$
by an \SQL query (encoded as a tuple $t$), locate vertex $v$ that
represents $e$ in a graph $G$, extract properties of $v$ from $G$
via pattern queries, and enrich $e$ by augmenting $t$ with the graph
properties of $v$. 
That is, it ``joins'' tuple $t$ and vertex $v$ if they
refer to the same entity $e$.\eat{ Along the same lines, it may enrich
graph pattern queries pivoted at vertex $v$ 
by imposing predicates on the attributes of tuple $t$. }
Each \SQLd query returns a relation with \warn{deduced schema},
and is composable with other \SQLd queries, 
just like \SQL queries.

\etitle{(2) Minor extension of \RDBMS} 
(Section~\ref{sec-overview}).
We present the query evaluation workflow
enabled by the \ModelName model, inherited from the counterpart for relational query~\cite{stonebraker1990implementation,stonebraker1976design}
\warn{with few operators incorporated into the logical and physical query \revise{plans} 
and corresponding revises on the optimizer.}\eat{with only minor revises incorporated into the logical and physical query \revise{plans} and the optimizer.}
\warn{The advantages offered by the \ModelName model can thus be readily obtained by
existing \RDBMS as an extension.}

\etitle{(3) Relational model of graphs} (Section~\ref{sec-model}).
The \ModelName model introduces logical pointers to represent graph-structured data in \RDBMS
while preserving its topology, to uniformly manage graph and relational data without cross-model access. 
It also allows users to 
(a) query graph-structured data with the same efficiency as the native graph stores,
(b) browse the graphs stored in it to formulate queries, and 
(c) update the graphs, to manage the graph-structured data,
without breaking the physical data 
independence~\cite{stonebraker1975implementation}.

\eat{Extending the conventional relational model,
\ModelName makes use of 
logical
pointers to represent the \revise{topology} 
of graph data.
Based on \ModelName, graph data and relational data can be 
uniformly managed without cross-model access. 
\SystemName also allows users to (a) browse the graphs stored in it,
to help users formulate  queries, and (b) update the
graphs, to manage the graph-structured data. Capitalizing on
its 
\ModelName model of graphs, it provides efficient support
for these without breaking the physical data 
independence~\cite{stonebraker1975implementation}.}


\etitle{(4) Uniform query evaluation in \RDBMS} (Section~\ref{sec-eval}).
The \ModelName model enables an enlarged unified plan space to optimize \SQL queries, graph pattern queries
and hybrid queries, 
\warn{which allows the access on graph and relational data to be considered in any order
	for a more efficient query plan.}
The generated query plan can then be executed seamlessly 
all inside the \RDBMS without performance degradation.\looseness=-1
\eat{
Introducing the exploration operator enabled by the \ModelName model into the query plans
also supports a seamless querying of graphs and 
relational data all inside the \RDBMS without performance degradation.
Moreover, it introduces an
exploration operator in the query plans, which enables
native graph pattern matching algorithms to be executed in
\SystemName equivalently\eat{without changes}; }

\eat{The newly introduced exploration operator in the query plan allows 
native graph 
pattern matching algorithms to be 
performed equivalently in \SystemName, which  
further achieves seamless querying of graphs and 
relational data while providing a larger unified space for
optimizing query plans.}
 

\eat{
\etitle{(5) Browsing and updating graphs} (Section~\ref{sec-update}).
\SystemName also allows users to (a) browse the graphs stored in it,
to help users formulate  queries, and (b) update the
graphs, to manage the graph-structured data. Capitalizing on
its 
\ModelName model of graphs, it provides efficient support
for these without breaking the physical data 
independence~\cite{stonebraker1975implementation}.
}



\etitle{(5) Experimental study} (Section~\ref{sec-expt}).
We provide an implementation of the proposed \ModelName model
to evaluate the advantages it can actually bring \warn{on} real-life datasets.
We empirically verify the following.
On average,\warn{
(a) Integrating the exploration operator induced by the \ModelName model 
	into the query evaluation workflow enables \SystemName 
	to bring orders-of-magnitude performance improvement
	to other relational systems, 32500x to PostgreSQL and 112x to DuckDB, 
	and also to achieve a comparable performance, 1.01x, 
	to the most efficient native graph-based algorithm we tested.
(b) \SystemName can optimize the hybrid query in an enlarged unified plan space
	to pull the low selectivity join conditions down to 
	the pattern matching process for early pruning
	and evaluate it seamlessly without involving cross model access.
	Together, \warn{a}
	13.5x\eat{3.62 + 12.9 + 23.9} speedup is achieved.
(c) \SystemName still performs well as the pattern scales, 
	which can achieve 
    0.91x
	compared to the native graph algorithms as the patterns reach 7 edges.}
\looseness=-1


\vspace{0.8ex}
We discuss related work in Section~\ref{sec-related}
and future work in Section~\ref{sec-conclude}.

\looseness=-1

\vspace{-1.0ex}

\section{Expressing Hybrid Queries in SQL$_\delta$}
\label{sec-SQL}
\vspace{-0.4ex}

In this section, we introduce \SQLd, a dialect of \SQL to express
native \SQL queries, graph pattern queries and hybrid queries. 

\vspace{-0.5ex}
\stitle{Preliminaries}. 
\revise{Assume a countably infinite set $\mathbb{V}$ as the
underlying domain~\cite{AbHuVi1995}}.
A database schema is specified
as $\R = (R_1,\allowbreak \ldots,\allowbreak R_m)$,
where each $R_i$ is a relation schema with a fixed set of attributes.
A tuple $t$ of $R_i$ consists of a value \warn{in $\mathbb{V}$}
for each attribute of $R_i$. 
A relation of $R_i$ is a set of tuples of $R_i$. 
A database $\D$ of $\R$ is $(D_1, \ldots, D_m)$, where $D_i$ is
a relation of $R_i$ for $i \in [1, m]$~\cite{AbHuVi1995}.
Following Codd~\cite{Codd},
we assume  that each tuple $t$ represents an entity.
\looseness=-1

\eat{
\stitle{Horizontal partitioning}. 
(such a definition is copy from \cite{sacca1985partitioning})
A horizontal partitioning of a relation Ri in R is a set 
of relation fragments fh(Ri) = (Ri, . . . , Ri,j such that all tuples are complete and 
represented only once: 
(i) for all Riq in fh(Ri), Aiq = Ai; 
(ii) for all tuples t in Ri there is exactly one fragment in fh(Ri) that contains 
t; 
(iii) for all Riq in fh(Ri) and for all tuples t of R,, Ri contains t.
}

\eat{\begin{figure}[t]
	\vspace{-1.4ex}
	\centerline{\includegraphics[scale=1.4]{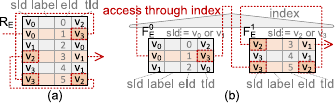}}
	\centering
	\vspace{-2.4ex}
	\caption{
		\revise{(a) Graph data in relations.
		(b) Table Partitions.}}
	\label{fig-relation}
	\vspace{-3ex}
\end{figure}

\begin{example}
\label{exa-relation}
Figure~\ref{fig-relation} (a) demonstrates a typical relational representation
of the data graph $G$ shown in Figure~\ref{fig-example} where all edges are 
stored in a single edge relation $R_E$.
\end{example} }

\eat{A {\em graph} is specified as $G = (V, E, L, F_A)$, where
(a) $V$ is a finite set of vertices;
(b) $E$ is a finite set of edges,
in which $(v, v')$ denotes an edge
from vertex $v$ to $v'$;
and
(c) each vertex $v \in V$  has label $L(v)$.
Here an edge label typifies predicates while vertex labels carry values.
(c) for $\forall v \in V$ (resp.~$\forall e \in E$),
$v$ (resp.~$e$) carries a tuple $F_A(v) = (A_1 = a_1,...,A_n = a_n)$ 
(resp.~$F_A(e) = (A_1 = a_1, \ldots , A_n = a_n)$)
of attributes.

The \emph{``label''} of vertex (resp.~edge) would 
therefore be regarded as an attribute with a special name.
}

\vspace{0.36ex}
A {\em graph} is specified as $G = (V, E,
L, F_A)$, where
(a) $V$ is a finite set of vertices;
(b) $E \subseteq V \times V$ is a set of edges,
in which $(v, v')$ denotes an edge
from vertex $v$ to $v'$;
(c) function $L$ assigns a label $L(v)$ (resp. $L(e)$) to each $v \in V$ (resp. $e \in E$);
and \warn{(d)} \warn{each vertex $v {\in} V$ (resp.~ edge $e {\in} E$)
carries a tuple $F_A(v) = (A_1 = a_1,...,A_n = a_n)$ 
(resp.~$F_A(e) = (B_1 = b_1, \ldots , B_m = b_m)$)
of {\em attributes} of a finite arity}, written as $v.A_i = a_i$ 
(resp. $e.B_i = b_i$), and $A_i \neq A_j$ (resp. $B_i \neq B_j$)
if $i \neq j$.
\looseness=-1

We assume the existence of a special attribute \kid at each vertex
$v$, denoting its vertex identity, 
such that for any two vertices $v$ and
$v'$ in $G$, if $v.\kid = v'.\kid$, 
then $v$ and $v'$ refer to the same entity.
Similarly, each edge $e$ is also associated with
an \kid to indicate edge identity.

\eat{
\vspace{0.36ex}
A {\em graph} is specified as $G = (V, E, L, F_A)$, where
(a) $V$ is a finite set of vertices;
(b) $E$ is a finite set of edges,
in which $(v, l, v')$ denotes an edge
from vertex $v$ to $v'$ that is labeled with  $l$;
(c) each vertex $v \in V$ has label $L(v)$;
and (d) each vertex $v\in V$ carries a tuple $F_A(v) = (A_1 = a_1, \ldots, A_n = a_n)$ of {\em attributes} of a finite arity, where $a_i$ is a constant,
written as $v.A_i = a_i$, and $A_i \neq A_j$ if $i\neq j$,
representing properties.
Different vertices may carry different
attributes, which are not constrained by a schema like relational
databases.

A {\em path $\rho$ from a vertex $v_0$} in  $G$ is a list
$\rho = (v_0, v_1, \ldots, v_n)$ such that $(v_{i-1}, l_{i-1},
v_{i})$ is an edge in $G$ for $i \in [1, n]$. The path is
{\em undirected} if for $i \in [1, n]$,
either $(v_{i-1}, l_{i-1}, v_{i})$
or  $(v_{i}, l_{i-1}, v_{i-1})$ is an edge in $E$, 
regardless of the orientation of the edge. 
}

\etitle{Patterns}. A {\em graph pattern} is defined as
$P = (V_P, E_P, L_P)$, where
$V_P$ (resp.~$E_P$) is a set
of pattern vertices (resp.~pattern edges) and
$L_P$ is a labeling function for $V_P$ and 
$E_P$ as above.
\looseness=-1

\etitle{Matches}. A {\em match of a  pattern $P $ in
a graph $G$} is a homomorphic mapping
$h$ 
from $V_P$ to $V$ such that (a) for each pattern 
vertex $u \in V_P$, $L_P(u) = L(h(u))$; and
(b) for each pattern edge $e = (u, u') \in E_P$, 
there exists an edge $(h(u), h(u'))$ in graph $G$ labeled $L_P(e)$. 
By slight abuse of notation,
we also denote \revise{$(h(u), L_P(e), h(u'))$} as $h(e)$.
\looseness=-1

\etitle{Pattern queries}. A {\em graph pattern query} is 
$P[\bar A]$, where (a) $P$ is a graph pattern, and (b)
$\bar A$ is a list of attributes such that each $A_i$ in $\bar A$
is attached to a pattern vertex or edge in $P$.\eat{;
in particular, $A_0$ is an \kid attribute attached to
a designated vertex in $P$, \ie entity of interest.  }

We represent the {\em result} $P[\bar A](G)$
of pattern 
$P[\bar A]$
over graph $G$ as a relation $S$ of schema $R_P$ with
attributes $\bar A$. More specifically,
each match $h$ of $P$ in $G$ is encoded
as a tuple $s$ in relation $S$,
such that 
\revise{each attribute $A_i$ in $\bar A$  
is associated} with
pattern vertex $u$ (resp. pattern edge $e$), $s[A_i] = h(u).A_i$
(resp. $s[A_i] = h(e).A_i$). 

\vspace{0.36ex}
Intuitively, besides topological matching, pattern queries 
extract important properties for entities of interests
from the graphs. These are organized in the form of relations
with \revise{a} deducible schema.

\eat{
\etitle{Patterns}. A {\em graph pattern} is defined as
$P[x_0, \bar x]$ = ($V_P$, $E_P$, $L_P$, $\mu$), where
(1) $V_P$ (resp.~$E_P$) is a set
of pattern vertices (resp.~edges) as  above;
(2) $L_P$ assigns a label to each vertex in $V_P$;
(3) $\bar x$ is a list of distinct variables, and
$\mu$ is a bijective mapping from
$\bar x$ to $V_P$; and
(4) $x_0$ is a designated variable in $\bar x$,
referred to as the {\em center} of $P$.

Intuitively,  $x_0$  denotes an entity of interest;
its properties are denoted by the vertices in 
$\bar x \setminus \{x_0\}$ and are specified by the
topological structure of pattern $P$. We consider connected
patterns $P$, such that $x_0$ is linked to each vertex
in $\bar x$ via an (undirected) path. \looseness=-1

For $x \in \bar x$, we use $\mu(x)$ and $x$ interchangeably.

\etitle{Matches}. A {\em match of a  pattern $P$ in
a graph $G$} is a homomorphic mapping
$h$ from the pattern vertices in $P$ to $G$ such that
(a) for each vertex $u \in V_P$,
$L_P(u) = L(h(u))$, and (b) for each pattern edge $(u, l, u')$ in
$P$, $(h(u), l, h(u'))$ is an edge in graph $G$.

\vspace{0.36ex}
We represent the match as a tuple $s$ of schema $R_P$,
which carries attributes $\bar B$, where each $B_i$ in
$\bar B$ denotes $h(x_i)$ for $x_i \in \bar x$.
In particular, $B_0 \in \bar B$ denotes the entity $h(x_0)$,
and $s[B_0]$ is the vertex $h(x_0)$. For $i \not\eq 0$,
$s[B_i] = L(h(x_i))$, \ie the label (value) of vertex $h(x_i)$.

We denote by $P[x_0, \bar x](G)$ the set of all matches of
pattern $P$ in graph $G$, which is a relation of the schema
$R_P$ of $P$.

\stitle{RA$_\delta$}. We now present a dialect of relational algebra
(\RA), denoted by \RAd. A query of \RAd has one of the following forms:
\vspace{-0.7ex}
\beqnn
Q &::=& R\ |\ \pi_{\bar A} Q\ |\ \sigma_C Q\ |\ Q_1 \join Q_2\ |\
Q_1 \cup Q_2\ |\ Q_1 \setminus Q_2 \ |\ \\
  &&  P[x_0, \bar x](G) \ |\ Q \join_\delta P[x_0, \bar x](G).
\eeqnn

  \vspace{-0.7ex}

\noindent
Here  (a) $R, \pi, \sigma, \join, \cup $ and $\setminus$ denote
relation atom, projection, selection, (natural) join, union
and set difference, respectively, just like their counterpart
in the classic \RA (see~\cite{AbHuVi1995}); 
(b) $P[x_0, \bar x](G)$ is a {\em graph pattern query} as defined above;
it returns a relation of schema $R_P$, where $G$ is (the name of)
a graph; 
and (c) $Q \join_\delta P[x_0, \bar x](G)$ ``joins''
a relational query and a graph pattern query, referred to as a 
{\em $\delta$-join}.\looseness=-1
}

\vspace{-0.7ex}
\stitle{RA$_\delta$}.
We present a dialect of 
relational algebra (\RA), 
denoted by 
\RAd. 
A query of \RAd 
has one of the following forms:
\vspace{-0.7ex}
\beqnn
Q &::=& R\ |\ \pi_{\bar A} Q\ |\ \sigma_C Q\ |\ Q_1 \join Q_2\ |\
Q_1 \cup Q_2\ |\ Q_1 \setminus Q_2 \ |\ \\
&&  
P[\bar A](G) \ |\ Q_1 \join_\delta Q_2.
\eeqnn

\vspace{-0.7ex}

\noindent
Here  (a) $R, \pi, \sigma$, $\join$ , $\cup $ and $\setminus$ 
denote
relation atom, projection, selection and (natural) join, 
set union and difference, 
respectively, as 
in the classic \RA 
(see~\cite{AbHuVi1995}); 
(b) $P[\bar A](G)$ is a {\em graph pattern query} as defined above;
it returns a relation of schema $R_P$, where $G$ is (the name of)
a graph; 
and (c) 
$Q_1 \join_\delta Q_2$ ``joins''
two queries \warn{according to their semantic connection} \revise{(see below), 
\eat{referred to as an {\em H-join} ({\em heterogeneous join}) and }denoted by
{\em $\delta$-join}.}

\etitle{The semantics of $\delta$-join}.
Suppose that $Q_1$ is
a relational query $Q$ and 
$Q_2$ is a graph pattern query $P[\bar A]$; 
the cases for joining queries of the same type are similar.
Assume that over a database $\D$ of schema
$\R$, sub-query $Q(\D)$ returns a relation $D$ of schema
$R_Q$ with attributes $\bar B$; and over graph $G$,
$P[\bar A](G)$ returns a relation $S$ of schema $R_P$.

The $\delta$-join $Q \join_\delta P[\bar A](G)$ works as
follows: 
for each result
tuple \warn{$t \in Q(D)$} and each tuple \warn{$s \in S = P[\bar A](G)$},
if $t$ and $s$ 
refer to the same real-world
entity (with $s[A_0]$ as the \kid), then it returns tuple $t$  
augmented with all attributes in $s$ \revise{except 
$A_0$}.
The schema of the result of the $\delta$-join can be readily deduced from $R_Q$ and $R_P$, 
\ie combining
attributes $\bar A$ and $\bar B$ excluding 
$A_0$.\eat{

\vspace{0.3ex}
Intuitively, for each tuple  $t$ dynamically computed
by query $Q(\D)$, 
this $\delta$-join locates entities $s$ that are
identified by $P[\bar A](G)$ 
and semantically refer to the same entity
as $t$; it extracts properties of $s$ and 
incorporates them into $t$
as additional attributes.}
\warn{This said}, graph pattern 
matching~\cite{angles2018GCORE,
alin2022Graph}
and ER (entity resolution)~\cite{altwaijry2015query, 
sismanis2009resolution} are combined here to 
query across relational and graph data\eat{, or graph and graph~\cite{francis2018cypher}}.

\eat{
\etitle{The semantics of $\delta$-join}.
Assume that over a database $\D$ of schema
$\R$, sub-query $Q(\D)$ returns a relation $D$ of schema
$R_Q$ with attributes $\bar A$; and over graph $G$,
$P[x_0, \bar x](G)$ returns a relation $S$ of schema $R_P$
with attributes $\bar B$; in particular, for each tuple
$s \in S$, $s[B_0]$ is the vertex (id) that maps to the
center $x_0$ of pattern $P$. 

The $\delta$-join $Q \join_\delta P[x_0, \bar x](G)$ works as
follows: for each tuple $t \in D$ and each tuple $s \in S$,
if tuple $t$ and vertex $s[B_0]$ refer to the same real-world
entity, then it returns tuple $t$ augmented with attributes
$s[\bar B']$, where $\bar B' = \bar B \setminus \{B_0\}$.
The schema of the result of the $\delta$ join can be readily
deduced from $R_Q$ and $R_P$.

\vspace{0.36ex}
Intuitively, for each tuple  $t$ dynamically computed
by query $Q(\D)$, the $\delta$-join locates vertices $v$ that are
identified by $P[x_0, \bar x](G)$ and semantically refer to the same entity
as $t$; it extracts the properties of $v$ and incorporates them into $t$
as additional attributes.
}

\begin{example}
\label{exa-join}
\eat{Following Example~\ref{exa-running}, give queries for
  (a) enriching \SQL   with graph properties (hybrid queries),
  and (b) pattern queries with relational predicates (\eg
  $\sigma_C (Q \join P[x_0, \bar x](G))$; push conditions $C$ down to
  attributes in $G$ and hence constrain matches of pattern $P$. 
  Justify why the results are more powerful than \SQL alone}
  \warn{
    The query in Example~\ref{exa-running} can be described with 
    \RAd as:
    $\D_E \join_\delta (P[v_2.vid](G) \join_{v_2.vid = uid} \D)$.
    It consists a pattern query, a join across graph and relational data
    and a $\delta$-join. 
    Other attributes extracted from pattern query are omitted as a simplification.}
\end{example}\looseness=-1

\eat{
Several methods are already in place to determine whether a
tuple in a relation and a vertex in a graph refer to the same
entity, \eg rule-based
JedAI~\cite{papadakis2020three}, parametric simulation \cite{HER},
and ML models Silk \cite{Silk},  MAGNN~\cite{fu2020magnn} and
EMBLOOKUP~\cite{lookup} \warn{add more}. \SystemName implements
\tbf inside \RDBMS to support $\delta$-join (Section~\ref{sec-eval}).
}

Several methods  are 
in place 
to embed 
ER in query processing, 
\ie to decide 
whether two tuples refer to the same 
entity with additional operator at query time, \eg 
ETJ/Lineage~\cite{agrawal2009trio},
polymorphic operators~\cite{altwaijry2015query},
hyper-edge~\cite{bhattacharya2006query},
\revise{parametric simulation \cite{HER}},
entity aggregation with mapping
\cite{sismanis2009resolution}. 
\warn{Such a procedure will be embedded as a black box 
to return the matched entity pairs from two input relations,
which allows all kinds of detailed methods to be adopted here.}\looseness=-1

\stitle{SQL$_\delta$}. We next present \SQLd. Over database
schema $\R = (R_1, \ldots, R_n)$ and graph $G$, an \SQLd query
has the form:

\vspace{-0.3ex}
\mat{1.0ex}{
	\SELECT\ \ \= $A_{1}$, \ldots, $A_{h}$\\
	\FROM\> $R_{1}$,  \ldots,  $R_{n}$,\\
	\> $P_1[\bar A_1](G)$,\ \ldots,\ $P_k[\bar A_k](G)$, \\
	\> $S_{1}\ \JOIN \ P_1'[\bar A_1'](G)$,\ \ldots,\
	   $S_{m}\ \JOIN  \ P_m'[\bar A_m'](G)$, \\
  \> $S_{1}'\ \MAP \ S_{1}''$,\ \ldots,\
     $S_{q}'\ \MAP \ S_{q}''$\\
	\WHERE\> {\sc Condition-1} \{\And|\Or\} 
	\ldots \{\And|\Or\}  {\sc  Condition-p}
}

\noindent
Here $S_{1}$, \ldots, $S_{m}$ are either relations 
of $\R$ or \SQLd
sub-queries over $\R$ and $G$.
An \SQLd query may embed a normal \SQL query,
a pattern query $P_i[\bar A_i](G)$,
a value-based hybrid query $S_{j}\ \JOIN \ P_j'[\bar A_j'](G)$,
\warn{and a $\delta$-join $S_{k}'\ \MAP\ S_{k}''$}. 
Each {\sc Condition} in the \WHERE clause is
an \SQL condition over relations $R_{1}$, \ldots, $R_{n}$,
and the result relations of pattern queries and $\delta$-joins.
Following~\cite{angles2018GCORE, alin2022Graph, francis2018cypher, francis2023researcher},
\revise{users may specify patterns with
``ASCII-art'' style paths}.

\eat{
\vspace{-0.3ex}
\mat{1.0ex}{
  \SELECT\ \ \= $A_{1}$, \ldots, $A_{h}$\\
  \FROM\> $R_{1}$,  \ldots,  $R_{n}$,\\
  \> $P_1[x_0, \bar x_1](G)$,\ \ldots,\ $P_k[x_0, \bar x_k](G)$, \\
  \> $S_{1}\ \kw{join}\ P_1'[x_0, \bar x_1'](G)$,\ \ldots,\
  $S_{m}\ \kw{join}\  P_m'[x_0, \bar x_m'](G)$\\
  \WHERE\> {\sc Condition-1} \{\And|\Or\} 
  \ldots \{\And|\Or\}  {\sc  Condition-p}
}
\vspace{-0.4ex}

\noindent
Here $S_{1}$, \ldots, $S_{m}$ are either relations in $\R$ or \SQLd
sub-queries over $\R$ and $G$.
An \SQLd query may embed a normal \SQL query,
a pattern query $P_i[x_0, \bar x_i](G)$,
and a $\delta$-join $S_{j}\ \kw{join}\ P_j'[x_0, \bar x_j'](G)$. 
Each {\sc Condition} in the \WHERE clause is
an \SQL condition over relations $R_{1}$, \ldots, $R_{n}$,
and the result relations of pattern queries and $\delta$ joins.
}

\begin{example}
\label{exa-SQL}
The \RAd query in Example~\ref{exa-join} can be written in \SQLd:
{\footnotesize
\mat{1.0ex}{
	\SELECT\ \ \= $*$ \ \ \FROM \ (\\
	\ \ \ \ \SELECT \ $v0.id$ \AS $vid_0$,  \ $v2.id$ \AS $vid_2$,  \ $D.profile$ \AS $user\_profile$\\
	\ \ \ \ \FROM \ (\\
	\ \ \ \ \ \ \SELECT \> \ \ \ \ \ \ \ \ \  $v0.id$ \AS $vid_0$,  \ $v2.id$ \AS $vid_2$  \ $v1.attr$ \AS $link\_attr$\\
	\ \ \ \ \ \ \MATCH \ \ \ $(v0: User) \text{-}[e0: Share] \text{->} (v1: Link)$\\
	\ \ \> \ \ \ \ \ \ \ \  $(v2: User)  \text{-}[e1: Share] \text{->} (v1: Link)$\\
	\ \ \> \ \ \ \ \ \ \ \  $(v2: User)  \text{-}[e2: Follow] \text{->} (v0: User)$ \\
  \ \ \ \ )\ \AS $P$ \ \JOIN \ $D$ \ON $P.vid_0 = D.uid$ \\
  )\ \AS $P\_R$ \ \MAP \ $D_E$} 
}
Where pattern $P$ 
is described with ``ASCII-art'' style paths.
\end{example}

\begin{figure}[t]
\vspace{-0.4ex}
\centerline{\includegraphics[scale=1.15]{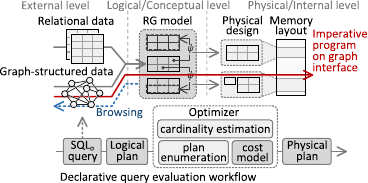}}
\centering
\vspace{-2.4ex}
\caption{Data representation and query evaluation overview}
\label{fig-overview}
\vspace{-4ex}
\end{figure}

\warn{
\stitle{Usability}.
\SQLd retains the capability of query composition~\cite{angles2018GCORE} in \SQL,
while the extensions it introduces are also all easy for the practitioners to accept:
The ``ASCII-art'' style description of pattern query is widely-recognized in the community of graph query language~\cite{alin2022Graph};
The ``map'' clause for ER can be simply regarded as a special kind of {\em join} without specifying conditions
(see~\cite{full} for more).}
\looseness=-1

As will be seen in Section~\ref{sec-eval}, all \SQLd queries can be
 evaluated inside an \RDBMS along the same lines as \SQL
queries without switching to another system\eat{
or invoking expensive graph operators}.

\looseness=-1

\vspace{-0.3ex}
\section{Hybrid Query Evaluation Overview}
\label{sec-overview}
\vspace{-0.2ex}

\eat{
In this section, we present an overview of \SystemName.
\revise{\SystemName aims to extend existing \RDBMS{s} with minimum changes
to uniformly support \SQLd, \ie native
\SQL, graph patterns and hybrid queries.}}

\vspace{0.36ex}
\warn{The \ModelName model provides a unified logical level representation of both the relations and graph-structured data from the {\em external}~\cite{date1999introduction, steel1975ansi} view,
with a physical data independence
to support the declarative hybrid query as shown in Figure~\ref{fig-overview}.
As will be seen in Section~\ref{sec-eval},
the \ModelName model enables an enlarged plan space to 
uniformly optimize the queries that operate on both graph and relational data,
and reduce the overhead incurred by switching between different
execution engines with heterogeneous data stores~\cite{tan2017enabling, gupta2016cross}.
Moreover, its memory layout
can also be efficiently implemented at the physical level.
\looseness=-1

\stitle{Workflow}. As shown in 
Figure~\ref{fig-overview}, \SQLd queries will be processed
along exactly the same lines as \SQL queries in
\RDBMS{s}~\cite{stonebraker1990implementation, stonebraker1976design,joseph2007architecture}.
An \SQLd query $Q_\delta$ 
will be first parsed into a logical plan,
then optimized to generate the corresponding physical plan and 
be executed to get the evaluation result.

We remark the following.

\sstab
(1) The query evaluation workflow enabled by the \ModelName model differs from the counterpart in \RDBMS{s}~\cite{graefe1993query}
only in a few simple operators included in the
logical and physical plans, and a slight revision of
the cost model accordingly.
These revises can be incorporated into the relational query evaluation,
along with the well-developed techniques (\eg plan enumeration)~\cite{ramakrishnan2003database},
to support hybrid queries.

\eat{These revises are all minor that allow the well-developed techniques for relational query evaluation (\eg plan enumeration)~\cite{ramakrishnan2003database}
to be inherited to support hybrid query.}

\sstab
(2) Along with the model, we provide a lightweight system, \SystemName, 
as an implementation to evaluate the actual benefits that
the proposed hybrid query evaluation workflow can bring on real-life data,
and prove the concept of dynamic data enrichment.
A native interface is also supported for procedural graph data access,
which allows it to support generic graph computation, not limited to 
the declarative graph pattern queries.
\revise{We defer imperative programming for graph computation to fu ture work.}}

\eat{Our ultimate goal is to extend existing \RDBMS with
minimum modifications while achieving 
seamless evaluation for hybrid queries, \ie~\SQLd queries.

\vspace{0.36ex}
The key module in \SystemName is the \ModelName model,
which will be introduced in Section~\ref{sec-model}. 
It acts as a unified representation of 
relational data and graph data at logical level 
(Figure~\ref{fig-overview}).
As will also be seen in Section~\ref{sec-model},
there exists efficient implementation
for our proposed \ModelName model, \eg memory layout 
at physical level. This model indeed 
leads to a larger room for optimizing
the query plans regarding 
seamless evaluation of hybrid queries.
Hence the overhead associated with accessing across 
heterogeneous data stores and execution engines~\cite{tan2017enabling, gupta2016cross} 
is eliminated at the very beginning.

\begin{figure}[t]
	\vspace{-1.4ex}
	\centerline{\includegraphics[scale=1.4]{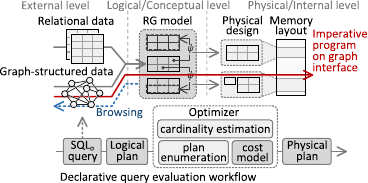}}
	\centering
	\vspace{-2.4ex}
	\caption{Data representation and query 
	processing}
	\label{fig-overview}
	\vspace{-3ex}
\end{figure}

\stitle{Workflow}. As shown in 
Figure~\ref{fig-overview}, the query processing
workflow in \SystemName coincides with 
the ones for relational
query processing~\cite{stonebraker1990implementation, stonebraker1976design}.
Given an \SQLd query, \SystemName 
first translates it into 
a corresponding logical plan. Then
various optimizations are applied before generating
the physical plan for actual execution to
get the query result.

\vspace{0.36ex}
Compared to conventional \RDBMS, 
the only difference is that a few new simple operators
are included in the logical and physical
plans to support the semantics induced by 
\ModelName model, \ie exploration, and the cost model
is slightly revised by taking these
operators into consideration (see Section~\ref{sec-eval}).
All the other well-developed techniques
for physical design and query optimization (\eg
cardinality estimation)~\eg~\cite{lightstone2010physical,ramakrishnan2003database}
in \RDBMS can be directly inherited here. 
In practice, any \ModelName model-based
system can be implemented as a minor extension 
of existing \RDBMS or a lightweight
stand-alone relational engine.
\looseness=-1

\etitle{Remark}. We remark that \SystemName 
also provides a native graph interface for
conducting procedural graph data access. 
Therefore, in principle, \SystemName can support all kinds of generic graph computation, not limited to 
the declarative graph pattern matching.
Such an imperative programming manner 
mainly relies on engineering efforts, 
which will not be discussed in this work.
}

\vspace{-0.3ex}
\section{Modelling Graphs in Relations}
\label{sec-model}
\vspace{-0.4ex}

\eat{
\warn{To do: make it concise; don't inflate your claim:
\mbi
\item the architecture of \SystemName; put it in a separate section if needed;
spell out the key modules and workflow
 \item relational model of graphs: logical design; don't use
 ``first-order''
 \item example: illustrate the extension
 \item justify: preserves topology/locality; speed up query processing;
 advantages over previous systems
  \item physical implementation
  \mei
  }
}

We \warn{revise the relational model to store graphs} (Section~\ref{subsec-define-model}),
present its physical implementation (Section~\ref{subsec-physical-implementation}),
and show how to
update and browse the graphs
maintained in relations (Section~\ref{sec-update}).

\vspace{-0.3ex}
\subsection{The RG Model}
\label{subsec-define-model}
\vspace{-0.2ex}

To encode 
graphs in relations, 
we propose a simple  \ModelName (\eat{\underline{R}elation-\revise{\underline{G}raph}}\underline{R}elational \underline{G}enetic)
model.
It \revise{extends conventional relations with logical pointers,
to preserve the topological structures of graphs when stored in relations,
and support} additional operators 
for uniform
query evaluation across graphs
and relations (Section~\ref{sec-eval}).
\eat{
which 
enable us to organize 
graph data
 in relations \wrt their
topological structures and
introduce additional operators 
for uniform
query evaluation across graphs
and relations (Section~\ref{sec-eval}).}

\eat{
\warn{
\etitle{Definition}.
In addition to the set of plain values,
$\mathbb{P}$ is another infinite countable set of symbol that distinct from $\mathbb{V}$, 
\eg $\mathbb{V} \cap \mathbb{P} = \emptyset$, that define the value set of the pointers.
}
}

Assume two disjoint \revise{countably infinite sets}
$\mathbb{V}$ and $\mathbb{P}$, denoting
the domains of \revise{data values} 
and \revise{logical} pointers 
(symbols), respectively.

\stitle{RG model}. Consider a database schema
$\R = (R_1, \ldots, R_m)$. The {\em \ModelName schema}
\wrt $\R$ is $\R^E = (R_1^E, \ldots, R_m^E)$, in which 
each $R_i^E$ extends the conventional 
schema $R_i$ such that the domain of 
attributes in $R_i^E$
is \revise{$\mathbb{V} \cup \mathbb{P}$ instead of $\mathbb{V}$},
\ie attributes can also be pointers \warn{in addition to the plain values}, which are  
references that ``virtually link'' to 
a subset of an 
extended relation.

\vspace{0.36ex}
An {\em extended database} 
$\D^E$
of $\R^E$ is $(D_1^E, \ldots, D_m^E, \phi)$, where
(1) $D_i^E$ is an extended relation, \ie a set of 
tuples of $R_i^E$ for $i \in [1, m]$; and 
(2) an injective \warn{partial} reference function 
$\phi: \bigcup_{i \in [1, m]} \mathcal{P}(D_i^E) \to \mathbb{P}$ 
maps  
\revise{subsets of an extended relation in $\D^E$ to references}
from  the domain $\P$.
Here $\mathcal{P}(D_i^E)$ is 
the power set of $D_i^E$.
%
We also denote by $\psi = \phi^{-1}$
the {\em dereference}
of a pointer to the specific subset 
$S$ in $\D^E$. This is well-defined since $\phi$ is injective; \revise{thus
$\psi(\phi(S)) = S$}.


\etitle{Regular form}. An extended database 
$\D^E$ is in the {\em 
regular form} if:
\eat{either $\psi(p_1) \cap \psi(p_2) = \emptyset$
or $\psi(p_1) = \psi(p_2)$ 
for any two  pointers $p_1$
and $p_2$ 
in $\D^E$.}
\warn{(1) For any two pointers $p_1$ and $p_2$ in $\D^E$,
either $\psi(p_1) \cap \psi(p_2) = \emptyset$
or $\psi(p_1) = \psi(p_2)$;
(2) For each extended relation $D_i^E$ in $\D^E$, 
for each of its attribute $A_{ij}$, there exists an extended relation $D_{ij}^E$ in $\D^E$,
s.t. $\bigcup\{\psi(t[A_{ij}]) | t\in D_i^E, t[A_{ij}] \in \mathbb{P}\}  \subseteq D_{ij}^E$,
\ie all pointers in a same column will not point to different relations.} 
In the sequel, we consider only extended
\revise{databases 
in the regular form. This suffices to 
model graph data}.
\eat{and can result
in an efficient physical implementation 
(see below). 
\looseness=-1
}
\eat{
\etitle{Fragmentation}. Consider a fragmentation
of extended relation $D_i^E$ that
divides the tuples of $D_i^E$ into
disjoint sets. We refer to each such set
as a {\em fragment} of $D_i^E$.
\warn{how to implement? 
can we have a definition without partition?}
}

\eat{
\etitle{Pointer binding}. Observe that the 
connection between pointers 
and tuples is implied by the 
dereference function $\psi$. 
\warn{Given an extended 
database instance $\D^E = (D_1^E, \ldots, D_m^E, \psi)$.
For $\forall T \subseteq D_i^E, i \in [1, m]$ 
and $T \neq \emptyset$,
we also denote by $(T)$ the unique reference of tuple set $T$.
It can be easily checked that $\warn{\phi}$ is well defined for $T \neq \emptyset$,
and there are $\psi(\warn{\phi}(T)) = T$ follows.}
}

\begin{figure}[t]
\centerline{\includegraphics[scale=1.05]{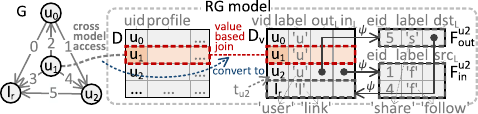}}
\centering
\vspace{-2.4ex}
\caption{Relations and graphs in the \ModelName model}
\label{fig-RX-model}
\vspace{-3.4ex}
\end{figure}

\eat{
\begin{example}
\label{exa-pointer}
\warn{
$D_V$ and $F_{in}^{v_0}$ in Figure~\ref{fig-RX-model} 
denote an extended relation and a fragment respectively.
The type of $D_V[out_L]$ (resp. $F_{in}^{v_0}[src_L]$) is 
the pointer toward fragment (resp. tuple),
where $\psi(t_{v_0}[out_L]) = F_{out}^{v_0}$ 
(resp. $\psi(t_{out}^{e_1}[dst_L]) = t_{v_3}$).}
\end{example}}

\etitle{Remark}.
\revise{The \ModelName model 
enhances the classical relational model on its 
``structural part''~\cite{codd1980data, codd1981relational}
with logical pointers.
The pointers} 
are at \revise{the} logical level, for referencing tuples.
\revise{Such pointers are along the same lines
as foreign keys, similarly for tuple identifiers and 
the ``database pointers'' in~\cite{nystrom2005snapshots},
for preserving the topology of graph data and
for optimizing hybrid/pattern queries.}

\warn{As will be seen in Section~\ref{sec-eval}, 
these pointers \warn{induce an ``exploration'' operator}
into logical query plan and facilitate the evaluation of
graph pattern queries, instead of merely supporting semantic optimization~\cite{ramakrishnan1994semantics}}\eat{; \eg 
we will introduce ``exploration'' operators in query plans
based on such pointers}. This allows us to uniformly manage relations
and graph data, without cross-model access
\cite{vyawahare2018hybrid, hassan2018grfusion, gimadiev2021combined}. Moreover,
it provides  a larger space for optimizing hybrid queries, \eg
pushing relational join conditions down into 
the pattern matching.
In contrast, 
pointers
in \RDBMS~\cite{garcia1992main,zhang2015memory,wu2017empirical}
are \revise{at the physical level and are}
mainly adopted inside a single tuple, pointing
to segments for storing one of its variable-length fields,
\eg VARCHAR in Postgres, 
or \warn{different versions for multi-version concurrency control}. Such physical pointers are
transparent to 
logical query \revise{plans}.
\eat{The logical pointers in \ModelName are along the same lines
as foreign keys, similarly for tuple identifies and 
the ``database pointers'' in~\cite{nystrom2005snapshots},
for preserving the topology of graph data and
for optimizing hybrid/pattern queries.}
\eat{Our pointers are also similar to foreign keys,
but the latter are constraints that help
semantic query optimization~\cite{ramakrishnan1994semantics}
instead of 
dynamic referencing,
similarly for tuple identifies and 
the ``database pointers'' in~\cite{nystrom2005snapshots}.}

\eat{
\warn{
	\etitle{Disjoint pointed relation (another name?)}. 
	Given an {\em extended database instance} $\D^E = (D_1^E, \ldots, D_m^E, \psi)$.
	Considering $D_i^E, i \in [1, m]$, 
	for $\forall p_0, p_1 \in \mathbb{P}$, 
	if $\psi(p_0) \subseteq D_i^E$ and $\psi(p_1) \subseteq D_i^E$,
	then there are $\psi(p_0) = \psi(p_1)$ or $\psi(p_0) \cap \psi(p_1) = \emptyset$,
	we say $D_i^E$ is a disjoint pointed relation in $\D^E$
}

\warn{
\etitle{Remark}.
We consider only the extended database instance when it only consists
of disjoint pointed relations, which is general enough for modeling 
graph data (see below). 
As would be shown in Section~\ref{subsec-physical-implementation},
a more efficient physical implementation can be supported in this case.
}
}

\stitle{Representing graphs}. To encode graph
$G = (V, E, L, F_A)$ \revise{in the 
\ModelName model, we  adopt} 
extended relations with the following
schemas.

\mbi
\item $\at{V(vid, label, out_{L}, in_{L})}$, where
each tuple $t$ of $\at{V}$ specifies 
information related to vertex $v$ with $v.\kid = 
t[\at{vid}$]. Here 
attributes $t[\at{out_{L}}]$ and $t[\at{in_{L}}]$
are pointers 
to those 
incoming
and outgoing edges incident to $v$, respectively \revise{(see below)}.
\item $\at{V^l_A(vid, A_1, \ldots, A_n)}$,
for encoding attributes \revise{of} 
vertices of 
a specific type (label) $l$~\cite{PG-Schema}.
\item $\at{E_{out}(eid, label, dst_{L})}$, where
each tuple $t$ of $\at{E_{in}}$ encodes edge $e$
such that $t.[\at{eid}] {=} e.\kid$, $t[\at{label}] {=} 
L(e)$ and $t[\at{dst_{L}}]$ is a pointer referencing the
destination vertex of $e$  in $\at{V}$-relation.
\item $\at{E_{in}(eid, label, src_{L})}$, which is
similar to $\at{E_{out}}$ but $\at{src_{L}}$ indicates 
a pointer referencing the source vertex instead.
\item $\at{E_A^l(eid, B_1, \ldots, B_m)}$,
for \revise{edge 
attributes 
with} 
label $l$.
\mei

\vspace{0.36ex}
\revise{A graph can be converted to such extended relations as follows.}
\eat{Then the conversion from a data 
graph $G$ to extended relations
\wrt~\ModelName model
can be readily achieved according to
the above semantics, with the following steps.}

\vspace{-0.3ex}
\sstab
(1)  All vertices in $G$ are maintained 
in a single extended relation $D_\at{V}$ of schema 
$R_\at{V}$,
by allocating vertex \kid's and labels appropriately.

\sstab
(2) Vertex attributes (resp. edge attributes)
are stored in  
multiple relations $D^{l_i}_\at{V_A}$ (resp. 
$D^{l_j}_\at{E_A}$) of 
schema \at{V^{l_i}_A} (resp. \at{E^{l_j}_A}),
\revise{one for each $l_i$ and $l_j$}, 
where the vertex labels $l_i$ and edge labels
$l_j$ are taken from $G$. This is in 
accordance with the recent attempts on 
defining schemas for 
property graphs~\cite{PG-Schema} 
and SQL/PGQ standards~\cite{GQL-standard,SQL-standard},
where the labels (types) of vertices or edges
also indicate allowed attributes.
In practice, the number of 
entity and relationship types in real-life 
knowledge graphs 
is relatively small~\cite{dbpedia,yago3link}. 
\revise{When attributes do not exist at some vertices/edges,
the corresponding fields are \kw{null}.}
\looseness=-1

\eat{
alternatively, we can split the attributes 
into classes based on the types (labels) of
vertices (resp. edges) as in~\cite{PG-Schema} and
use separate relations to store them.
}
\looseness=-1

\sstab
(3) 
\revise{Two relations $D_\at{out}$ and $D_\at{in}$
of schemas $\at{E_{out}}$ and $\at{E_{in}}$, 
respectively, 
maintain} the outgoing and
incoming edges of vertices. The outgoing
(resp. incoming) edges of each vertex 
$v$ correspond to a subset $F_\at{out}^v$
(resp. $F_\at{in}^v$) in $D_\at{out}$ (resp. $D_\at{in}$).
In particular, $F_\at{in}^v = \emptyset$
(resp. $F_\at{out}^v = \emptyset$) when
$v$ has no incoming edges (resp. outgoing
edges).

\eat{
\warn{
	The reference obtain function $\phi$ can then be
	established accordingly after all relations have been constructed.}
}

\sstab
(4) 
\revise{The pointers and reference function $\phi$ ensure the following:}
(a) for each $t \in D_\at{V}$, 
$t[\at{in_{L}}] = \phi(F_\at{in}^v)$
and $t[\at{out_{L}}] = \phi(F_\at{out}^v)$
where 
$t[\at{vid}] = v.\kid$;
(b) 
$s[\at{dst_{L}}] = \phi(t)$ (resp. $s[\at{in_{L}}] = \phi(t)$) for each $s$ in $D_\at{out}$ 
(resp. $D_\at{in}$) and $t$ in $D_\at{V}$ with
$t[\at{vid}] = v'.\kid$, and (c) 
$s[\at{eid}] = e.\kid$ and
$e = (v, v')$ (resp. $e = (v', v)$) is an edge in graph $G$.
\eat{the dereference function $\psi$ can also be
established accordingly. }

\vspace{1ex}
\revise{We refer to $\R^E_G$ = 
$\at{(V, V_A^l, E_{\kw{out}}, E_{\kw{in}}, E_A^{l'})}$
as {\em the \ModelName schema of graph $G$} 
(for \at{l} $\in$ \mbox{vertex types} and  \at{l'} $\in$ \mbox{edge types}),
and  $\D_G^E = $ $(D_\at{V}, D_\at{V_A}, D_\at{E_A},
D_\at{out}, D_\at{in}, \psi)$ as {\em the extended relation of $G$}.}

\eat{We refer to 
the combination of above extended relations
and reference function $\phi$
as the {\em extended database instance $\D_G^E$}~\wrt~{\em 
$G$}.
Note that labeled nulls
will fill certain fields when some vertex or edge
attributes 
do not exist in $G$.}
\looseness=-1

\eat{
\warn{It is stated above that \ModelName model can 
represent the topology of graph data, we then show
how it is possible by partitioning the relations
according to the pointer.}
}

\warn{
\begin{example}
\label{exa-regular}
$\D_G^E$ is in the regular form.
\end{example} 
}

\vspace{-1ex}
\etitle{Fragmentation}.
The subsets $F^v_\at{in}$
(resp. $F^v_\at{out}$)
for all vertices $v$ in $G$ \revise{induce} a partition
of the extended relation $D_\at{in}$ (resp. $D_\at{out}$),
\ie each tuple in $D_\at{in}$ (resp. $D_\at{out}$)
resides in
a subset linked by certain pointers.
We refer to each such 
subset as a {\em fragment}
of \revise{the} extended relation.
\looseness=-1

\eat{Duplicates? Each edge is stored at least twice}

\revise{Recall that the} conventional partitioning strategy
in \RDBMS relies on a large shallow 
index to access the fragments 
of a relation, 
\revise{which is costly when the number of fragments increases}.
In contrast, 
we use   heterogeneous memory management 
(Section~\ref{subsec-physical-implementation})
and pointer-based 
exploration (Section~\ref{sec-eval})
to efficiently  implement and fetch fragments.
Moreover, different from runtime partitioning
\cite{Jeffrey2, ambuj1994cache, stefan2002optimizing}, the
edges in graphs can be directly converted
into a set of fragments (in $D_\at{in}$
and $D_\at{out}$) 
based
on the \warn{topology}\eat{topological structure}.
\looseness=-1

\eat{Observe that 
the subsets $F^v_\at{in}$
(resp. $F^v_\at{out}$)
for all vertices $v$ in $G$
indeed form a partition
of the extended relation $D_\at{in}$ (resp. $D_\at{out}$),
\ie every tuple in $D_\at{in}$ (resp. $D_\at{out}$)
resides in
a subset linked by certain pointer.
In this case, we also refer to each such
subset as a {\em fragment}
of extended relation.

The conventional partitioning strategy
in \RDBMS relies on a large shallow 
index to access different fragments 
of a relation
(see Figure~\ref{fig-relation})
and incurs
extreme costs when a number of 
fragments exist.
In contrast, 
we use  
heterogeneous memory management 
(Section~\ref{subsec-physical-implementation})
and pointer-based 
exploration (Section~\ref{sec-eval})
to implement and fetch fragments
at the same scale efficiently 
in extended relations.
Moreover, different from runtime partitioning
\cite{Jeffrey2, ambuj1994cache, stefan2002optimizing}, the
edges in graphs can be directly converted
into a set of fragments (in $D_\at{in}$
and $D_\at{out}$) solely based
on the topological structure.}

\eat{
\warn{
\etitle{Disjoint pointed relation fragmentation (another name?)}.
\textbf{(simplify it here)} 
Given an {\em extended database instance} $\D^E = (D_1^E, \ldots, D_m^E, \psi)$,
and a disjoint pointed relation $D_d^E$ in $\D^E$.
We say that $\F^E = \{F_i^E | F_i^E \subseteq D_d^E,\; F_i^E \neq \emptyset \}$
is a disjoint pointed relation fragmentation of $D_d^E$, iff:
(1) $\bigcup F_i^E = D_d^E$;
(2) $F_i^E \cap F_j^E = \emptyset$ for $i \ne j$;
(3) For $\forall p \in \mathbb{P}$, 
if $\psi(p) \subseteq D_d^E$ then $\psi(p) \in \F^E$;
(4) Denotes the set of tuples in $D_d^E$ that are not 
referenced by any pointers
as: $T_n^E = D_d^E \setminus \bigcup_{p \in \mathbb{P}} \psi(p)$, 
if $T_n^E \neq \emptyset$ then $T_n^E \in \F^E$.

Intuitively, for each pointer points to the tuples in $\D^E$,
the process defined above would generate a fragment corresponds
to the set of tuples it points to,
and collect all the tuples not pointed by any pointers (if exists)
into a separated one.
It is well defined as $D_i^E$ is a disjoint pointed relation in $\D^E$.
}

\etitle{Remark}. It can be seen that after such a partition, 
all pointers point to a disjoint pointed relation can now 
directly point to a separated fragment of it,
which imply a more efficient physical design (Section~\ref{subsec-physical-implementation}).
}

\eat{
\vspace{0.36ex}
We refer to $(R_\at{V}, R_\at{V_A}, R_\at{E_A},
R_\at{out}^{v_1}, R_\at{in}^{v_1}, \ldots,
R_\at{out}^{v_o}, R_\at{in}^{v_o}, \psi)$
as the {\em extended database instance}~\wrt~{\em graph
$G$}. Here $o$ denotes the total number of vertices in
$G$. Intuitively, the topological structure
of $G$ is encoded by the pointers in its
corresponding extended database
instance, along which efficient exploration
of the graph $G$ can be achieved 
(see details in Section~\ref{sec-eval}).
\looseness=-1
}

\eat{
\warn{
	\etitle{Remark}.
	The progress described above would first represent the graph data in few relations 
	and then partition some of them,
	however, it can be easily seen that such a progress can be omitted and the graph data 
	can be converted to a set of fragments directly.
	Therefore, different from previous works~\cite{Jeffrey2, ambuj1994cache, stefan2002optimizing},
	\SystemName does not relay on runtime partition
	and the data can be pre-partitioned and stored according to the topology of graph data (Section~\ref{subsec-physical-implementation}).
} 
}

\begin{example}
\label{exa-conversion}
\warn{Figure~\ref{fig-RX-model} demonstrates 
a typical \ModelName-based representation of $D$ and $G$ in Figure~\ref{fig-example}
that eliminates the cross model access (details see~\cite{full}).
Taking tuple \eat{$t_{v_0}$}\warn{$t_{u2}$} for vertex \eat{$v_0$}\warn{$u_2$} as an example, 
all its incoming (resp. outgoing) edges 
are contained in a single fragment $F^{u2}_{in}$ (resp. $F^{u2}_{out}$) 
which is bound to its field $in_L$ (resp. $out_L$).\eat{ that can be iterated 
through the exploration.
Intuitively, the exploration on the graph structured data can be 
taken place on the \ModelName structured data cheaply 
through the pointers compared to the one demonstrated in Figure~\ref{fig-relation}~(a) or (b),
which would be detailed shown with the operator introduced in Section~\ref{sec-eval}.}
The exploration on graph can then be taken place\eat{through the pointers} (see Section~\ref*{sec-eval}).}
\looseness=-1
\end{example} 
\vspace{-2ex}

\warn{
\stitle{Discussion}.  
We collect a set of requirements for a {\em data model}~\cite{codd1980data}
to satisfy for uniformly supporting the hybrid query across graphs and relations,
which are:
(1) provide a uniform access and query interface for both relational and graph data; 
(2) inherit the relational query evaluation workflow to support declarative hybrid query; 
(3) achieve a comparable performance w.r.t. native graph-based methods on graph queries. 
The proposed \ModelName model achieves them all with minimal modifications
to the relational model
without further introducing various conceptions for graph~\cite{angles2008survey, stonebraker2005goes}
at the logical level.
Meanwhile, the \ModelName model can also achieve an equivalent expressiveness 
compared to previous graph algebras built atop of the relational model~\cite{Jeffrey1}.
The recursion feature~\cite{sakr2021future} is far beyond the scope of this paper but can be alternatively achieved as in~\cite{Verticagraph} (see~\cite{full}).
We defer a systemic study of this to future work.}
\looseness=-1

\eat{
\begin{example}
\label{exa-partition}
Figure~\ref{fig-relation} (b) shows how the edge relation can be 
partitioned, into fragments $F_E^0$ and $F_E^1$ here, and been access through the shallow index above
them.
\end{example} }

\eat{
\warn{
The process described above would first generate 
a database instance $\D_G^E$ and then partition it.
However, the graph data can be directly converted to the set of fragments 
as below:

\stitle{Directly convert graph data as fragments}. 
\dots
(briefly show how few steps in ``representing graphs''
can be replaced.)
}
}

\eat{
Generally speaking, the \ModelName model 
enhances the relational model on its 
``structural part''~\cite{codd1980data, codd1981relational}
to consider logical-level 
pointers and the induced exploration operations. 
Such mechanisms can be directly 
embedded in the logical query plan to
evaluate pattern queries without affecting
the foundations of relational model 
(see Section~\ref{sec-eval}).
This makes it possible to seamlessly manage relational
data and graph data together without cross-model access
\cite{vyawahare2018hybrid, hassan2018grfusion, gimadiev2021combined}, while
providing a larger space for optimizing hybrid queries, \eg
pushing relational join conditions down into 
the pattern matching.}

\eat{
\sstab
\warn{
(2) Different from
the conventional partitioning strategy
in \RDBMS, as demonstrated in Figure~\ref{fig-relation}, that relies on a large shallow 
index to access different fragments and thus yields
extreme costs when numerous partitions exist,
the efficient exploration supported by the pointers
can be applied to access the fragments. 
Meanwhile, the heterogeneous memory management strategy that would 
be introduced in Section~\ref{subsec-physical-implementation}
also makes it possible for \SystemName to manage this scale of fragments. 
}
(Seems cannot move forward, need to be listed after Example~\ref{exa-conversion}
to show there can actually be numerous fragments.)
}

\eat{
\sstab
(2) Such an abstraction of graph-structured
data in extended tables can eliminate various 
restrictions imposed by different native graph models,
\eg the property graph model adopted by
Neo4j~\cite{neo4j} 
treats all properties as key-value pairs~\cite{francis2018cypher}, but 
an additional lookup is required to access them.
\looseness=-1

\sstab
(3) While a number of extended relations (\ie partitions)
for storing edges are
built after the conversion process, we can apply
efficient exploration supported by 
the pointers. This is different from
the conventional partitioning strategy
in \RDBMS \warn{ as demonstrated in Figure~\ref{fig-relation}}, which relies on a large shallow 
index to access different fragments and yields
extreme costs when numerous partitions exist for,
\eg a single table containing the whole set 
of edges. Moreover, the pointer-based
exploration reduces the accesses incurred in
graph view~\cite{Kaskade} and 
join indices~\cite{jin2021making}
that are not memory-friendly.
}

\eat{
\warn{
\sstab
(4) Difference from the works with runtime partition~\cite{Jeffrey2, ambuj1994cache, stefan2002optimizing} for lowering the communication \& memory access cost?

\sstab
(5) \cite{seo2013socialite, seo2015socialite} introduced tail-nested table,
	somewhat a kind of hierarchical data model,
	which is a restricted version of our works that only allow the 
	pointer exists in the last column.
	It then prevents them to carry out new operators that 
	for seamless execution of exploration on graph data 
	and value-based join on relational data.
}
}

\vspace{-1ex}
\subsection{Physical Implementation}
\label{subsec-physical-implementation}
\vspace{-0.2ex}

\eat{We now show how we implement the \ModelName model,
especially its pointers and fragments,
with slight \revise{extension to the underlying \RDBMS}.}

\revise{Extending the classic}
relational model, the \ModelName model \warn{introduces} pointers to link to 
fragments in, \eg $D_\at{in}$
and $D_\at{out}$. 
\warn{As the introduced pointer can be implemented with a fixed-length field (see below) along with other type of value, 
	the rest of \RDBMS physical design can then be kept unchanged
	to~{\em uniformly} manage extended relations and the conventional ones.} 
\warn{We now show the slight extensions of the underlying \RDBMS physical design 
	to support the \ModelName model, especially its pointers and fragments.
A wide range of possible options supported by the \ModelName model are listed here, 
and the ones we implemented in \SystemName will be described in Section~\ref{subsec-system-imp}.}

\eat{
	highlight fragment before introduce heterogeneous staff
}

\eat{
\etitle{Horizontal fragment}. 
Given a relation $D$, a {\em horizontal fragment} of it 
is a subset of the tuples in $D$, denoted as $F \subseteq D$.
A set of horizontal fragments $\{F_i\}$ of $D$ is called a 
horizontal partitioning/fragmentation iff.
$\bigcup F_i = D$ and $F_i \cap F_j = \emptyset$ for $i \ne j$.
\looseness=-1

\vspace{0.36ex}
The horizontal fragmentation is frequently utilized in the 
distributed database~\cite{ozsu1999principles}
to store large relation in multiple distributed nodes 
where each of them holds only one or few fragments.
Other kinds of fragment, vertical or mixed~\cite{navathe1995mixed},
would not be further reviewed here.
All the ``fragment'' (resp. ``partition'') below reference to 
horizontal ``fragment'' (resp. ``partition'').

We now discuss how to implement the \ModelName model 
for maintaining graphs $G$ in \SystemName,
especially the pointers 
and fragments, 
by slightly extending the physical design of
\RDBMS.}

\eat{
Here we only provide a brief description
for managing extended relations in one practical way.
\warn{(unified address for fixed-length pointer; 
heterogeneous memory management for fragments)}
However, there also exist alternative solutions that
can implement the workflow of Section~\ref{sec-overview}  
in in-memory or
on-disk \RDBMS. 
\looseness=-1
}

\eat{
\warn{Here we only provide a relatively high-level description 
of how it is possible for the physical implementation to support 
the extended relation in a practical way
(, which does not restrict the workflow demonstrated in Section~\ref{sec-overview}
to be implemented as an in-memory or on-disk DBMS. ?)}
}

\eat{Compared to 
conventional relational model, 
our \ModelName model just 
introduces additional data types of
pointers linking to fragments (in $D_\at{in}$
and $D_\at{out}$) or single
tuples (in $D_V$). 
In fact, both kinds of pointers can be 
represented using a new fixed-length field  (see below),
while the rest of the physical design of \RDBMS
stays intact.
Hence \SystemName~{\em uniformly}
manages the extended relations
with conventional ones.

\vspace{0.36ex}
Since \ModelName model inherits 
the underlying features of relational model,
most of existing well-developed 
physical design techniques for \RDBMS, 
\eg~\cite{athanassoulis2019optimal,
arulraj2016bridging,stonebraker2018c,
alagiannis2014h2o,rasin2013automatic,
lightstone2010physical} can 
be directly utilized in \SystemName.
\eat{Thanks to the high degree of similarity 
between the two models, most of existing well-developed 
physical design techniques for \RDBMS, 
\eg~\cite{athanassoulis2019optimal,
arulraj2016bridging,stonebraker2018c,
alagiannis2014h2o,rasin2013automatic,
lightstone2010physical} can 
be directly inherited in \SystemName.}
As demonstrated in Figure~\ref{fig-RX-model}, 
\SystemName leverages different physical design
strategies in regards to the purposes. For instance, 
the column-store~\cite{stonebraker2018c} 
can be adopted to maintain vertex and edge attribute relations, 
to speed up the scanning of each specified column.
On the other hand, since the power-law degree distribution~\cite{clauset2009power} 
often results in a number of edges having the same
label and adjacent to the same vertex, 
column-oriented compression methods~\cite{daniel2006integrating}
can also be considered in \SystemName for storage and execution optimization.}

\eat{
\warn{
(Various index can also be constructed on each fragment separately if is needed,
especially for the field like ``label'',
that can equivalently achieve the secondary index.)
}
}

\eat{
\warn{
	\stitle{Heterogeneous memory management}. 
	Generally speaking, the conventional \RDBMS provides 
	two different memory management strategies:
	fixed-size block-based one for the (large) relations;
	and variable-length segment-based one for the variable 
	length fields.
	However, as the power-law degree distribution~\cite{clauset2009power} 
	would bring both numerous small fragments of the edge table, up to billions, 
	and few very large ones at the sametime,
	both strategies have their own limitations:
	Managing all the numerous fragments with block-based strategies would bring 
	unaffordable internal fragmentation~\cite{randell1969note} since most of them 
	would contain even only few tuples;
	Managing all fragments as segments would, 
	although can eliminate such a waste,
	enforce all of them to be stored in continuous memory space
	and thence bring difficulties for updating especially the large fragments.

	To avoid both these two limitations, \SystemName introduces a heterogeneous memory management strategy
	that combine both of them: 
	for the fragments larger than a threshold, divide them into blocks,
	does not enforce the entire fragment to be stored in a continues space
	to avoid reallocation during the update process;
	for the small ones that are cheap to copy, 
	regarding them as segments and tightly laying them in continues memory
	to avoid the waste of space.
	Each fragment, no matter stored in block or segment, 
	can then be carry out their physical design in their own memory space.

	(Meanwhile, as the conventional \RDBMS can easily manage the billions tuples
	with variable-length field in each of them, it can be proved that such a memory
	management strategy is capable for this scale of fragments.)
}
}

\eat{
\warn{In fact, it can be seen that comparing to the conventional relational model, 
\ModelName only allow additional value types, pointer-to-tuple and pointer-to-relation, 
to be considered for the fields. 
It would be shown latter in the following paragraphs \textbf{(pointers and memory space)} that both kinds of pointers can be represented 
with a fixed length field, which allows the extended relation to be {\em managed uniformly} with the conventional
ones.

\vspace{0.36ex}
Meanwhile, owing to such a similarity\eat{In addition, due to the high degree of similarity 
between the \ModelName model and relational 
model}, most of existing well-developed 
physical design techniques for \RDBMS, 
\eg~\cite{athanassoulis2019optimal,
arulraj2016bridging,stonebraker2018c,
alagiannis2014h2o,rasin2013automatic,
lightstone2010physical} can 
be directly inherited in \SystemName. 
As demonstrated in Figure~\ref{fig-RX-model}, 
\SystemName leverages different physical design
strategies in regards to the purposes. For instance, 
the column-store~\cite{stonebraker2018c} 
\textbf{can be adopted to maintain vertex (resp. edge) attribute relation $R_\at{V_A}$ (resp. $R_\at{E_A}$)}, 
for speeding up the scanning of each specified column;
\warn{
	Meanwhile, each vertex would typically have numerous inward/outward edges with the same label, 
	especially considering that some of them can have really large degree owing to the power-law degree distribution~\cite{clauset2009power},
	the column-oriented compression methods can then be considered~\cite{daniel2006integrating}
	for storage and execution optimization.
	}.
}
}

\eat{Following the conventional physical design for
\RDBMS~\cite{ozsu1999principles, kemper2011hyper, irving1982virtual}
and operating systems~\cite{silberschatz2018operating, stonebraker1981operating},
a {\em unified virtual address space} is utilized in \SystemName to 
uniformly hold all relations, both the conventional and
extended ones.
\eat{
Moreover, to avoid the unaffordable internal
fragmentation when maintaining
numerous extended relations with the 
traditional fixed-size block-based 
memory management~\cite{randell1969note},
\SystemName also adopts a segment-based management strategy,
similar to ones for handling 
variable-length data in \RDBMS.
This is effective in managing the large number
of small edge relations $R^v_\at{in}$ and
$R^v_\at{out}$ generated from the graphs, where
each small relation corresponds to a single segment
for having a tighter memory layout.}


\eat{
\stitle{Unified address space and pointer implementation}. 
Following the conventional physical design for
\RDBMS~\cite{ozsu1999principles, kemper2011hyper, irving1982virtual}
and operating systems~\cite{silberschatz2018operating, stonebraker1981operating},
a {\em unified virtual address space} is utilized in \SystemName to 
\warn{uniformly hold all the relations, both the conventional and the extended ones.} 
\warn{Meanwhile, it should be noticed that there is a challenge raised up from utilizing 
the \ModelName for graph data representation
to manage the enormous inward/outward edge \textbf{fragments/relations}
which can have a large number in total but each of them can be very small 
at the same time.
Keep using the conventional fixed size block-based management can lead to
unacceptable internal fragmentation~\cite{randell1969note}.
Therefore, in addition, the segment-based memory management strategies, 
similar to the ones for the variable-length data in conventional \RDBMS, 
can be utilized to manage all those small \textbf{fragments/relations} 
and each of them can then correspond to a single segment
to achieve a tighter memory layout.
}
}

\eat{Recall that each logical-level
pointer can link to either a single tuple or
an entire extended relation. We implement
the pointers in two ways, \ie
directly and indirectly, in \SystemName.
For direct way, each pointer is such implemented
that points to the starting address of the tuple 
or extended relation it refers to.
In indirect implementation, we follow the
technique that was developed for garbage
collection~\cite{piquer1991indirect}. 
\warn{For reference, it records the block index and 
offset of the  tuple or relation 
that the pointer links to, within
the unified virtual address space.
This method is more convenient for
data modification and materialization. Add more details.}}

\vspace{0.36ex}

\SystemName 
represents each pointer 
as a fixed length field, no matter it points to a single tuple or a fragment.
The pointers can directly 
record the starting address 
of the tuple or fragment they refer to, in the unified space.
Alternatively, 
following the technique developed for garbage collection~\cite{piquer1991indirect},
an indirect pointer is also supported
in conjunction with the segment-based management for segment,
which records only the segment id and an offset 
inside the segment.
\SystemName would first lookup the base address 
of the segment via its id in a segment table and then use the recorded 
offset to identify the tuple or fragment.
Such an indirect implementation 
can benefit the processing of graph updates
(Section~\ref{sec-update}). }

\vspace{-0.3ex}
\stitle{Heterogeneous 
fragment management}. 
\revise{\RDBMS supports two memory management strategies,
based on either fixed-size blocks or 
variable-length segment. Unfortunately,
\warn{neither} of them fits \warn{the} \ModelName well.
To model a real-life graph, it is common to find}
a number of small fragments and few large ones 
simultaneously in relations $D_\at{in}$ and $D_\at{out}$~\cite{clauset2009power}.
\revise{For such graphs, the block-based strategy often inflicts
unaffordable internal  fragmentation~\cite{randell1969note}
by small fragments, while the segment-based one requires
each fragment to be stored in a continuous memory space
and makes it difficult to handle updates.} 
\looseness=-1

\vspace{0.36ex}
\revise{In light of this}, 
\warn{we adopt a heterogeneous fragment management strategy
to support the \ModelName model}. 
\warn{For the fragments larger than a \revise{predefined} threshold,\eat{half of the block size by default,} 
they will be managed in a block-based manner,
and the rest of small fragments will be treated as segments~\cite{wilson1995dynamic} and 
stored in continuous memory.}
\revise{In this way, the cost of large-segment reallocation
	is reduced whilst the flexibility of
    segment-based strategy for small fragments is retained \warn{(see~\cite{full} for an evaluation)}.}
\looseness=-1

\begin{figure}[t]
\centerline{\includegraphics[scale=1]{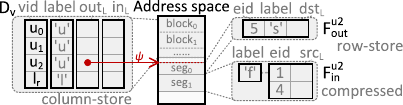}}
\centering
\vspace{-2.8ex}
\caption{Memory layout and physical design}
\label{fig-physical-design}
\vspace{-4.4ex}
\end{figure}

\vspace{0.36ex}
Moreover, \warn{such a flexibility allows the locality of graph data
to be taken} into account when managing memory. \revise{Observe
that graph computation often continuously accesses
both the incoming and outgoing 
edges of the same vertex, \eg graph pattern matching}.
Therefore, \warn{$F_\at{in}^v$ and $F_\at{out}^v$ 
can be organized} in a continuous memory address space
if possible, for 
better cache performance and memory locality 
(see Figure~\warn{\ref{fig-physical-design}}). 
Such flexible strategies are \revise{not supported by}, 
\eg~\cite{jin2021making,
mhedhbi2022modern,Kaskade}, 
which simply build \eat{graph }indices and views in \RDBMS.

\eat{\stitle{Heterogeneous 
fragment implementation}. 
As we may have a number of
small fragments and few large ones 
simultaneously in
relations $D_\at{in}$ and $D_\at{out}$,
both fixed-size block-based memory management
and variable-length segment-based one,
two memory management strategies in \RDBMS, 
have drawbacks
in 
implementing fragments. Block-based strategy
will result in 
unaffordable internal 
fragmentation~\cite{randell1969note}
\wrt small fragments. While 
segment-based one can
get rid of this problem, it enforces 
all fragments to be stored in 
continuous memory space, making it
harder to handle updates.

\vspace{0.36ex}
To this end, \SystemName adopts 
a heterogeneous memory management strategy
that combines the above two to implement
the fragments. 
For large fragments whose sizes are
above a threshold, 
it divides them into blocks and applies
block-based management.
In contrast, the rest of small fragments are 
treated as segments and 
placed in continuous memory. 
Then it avoids the costly reallocation of large fragments
while achieving relatively tight
memory allocation with the   heterogeneous strategy.

\vspace{0.36ex}
Moreover, the \ModelName model also allows 
\SystemName to take the locality 
of the topological structures of graphs into account
when conducting memory management.
In fact, it is 
common that the incoming and outgoing 
edges of the same vertex would be accessed 
continuously in graph computation, \eg pattern matching.
Therefore, \SystemName organizes $F_\at{in}^v$ and
$F_\at{out}^v$ in continuous memory address 
if possible, for 
better cache performance and memory locality 
(see Figure~\ref{fig-RX-model}). 
Such flexible strategies are beyond the capabilities
of \eg~\cite{jin2021making,
mhedhbi2022modern,Kaskade}, 
which simply build
graph indices and views in \RDBMS.
\looseness=-1
}

\eat{
Meanwhile, following the
conventional physical design for
\RDBMS~\cite{ozsu1999principles,kemper2011hyper}
and operating systems~\cite{silberschatz2018operating, stonebraker1981operating},
a {\em unified virtual address space} is utilized in \SystemName to hold 
all extended relations, and each extended 
relation corresponds 
to a block in it. 
Hence the extended relations are 
{\em managed uniformly} with conventional ones.

Following the
conventional physical design for
\RDBMS~\cite{ozsu1999principles,kemper2011hyper}
and operating systems~\cite{silberschatz2018operating, stonebraker1981operating},
a {\em unified virtual address space} is utilized in \SystemName to hold 
all extended relations, and each extended 
relation corresponds 
to a block in it. 
Hence the extended relations are 
{\em managed uniformly} with conventional ones.

\vspace{0.36ex}
In addition, due to the high degree of similarity 
between the \ModelName model and relational 
model, most of existing well-developed 
techniques for improving physical design of \RDBMS, 
\eg~\cite{athanassoulis2019optimal,
arulraj2016bridging,stonebraker2018c,
alagiannis2014h2o,rasin2013automatic,
lightstone2010physical} can 
be directly inherited in \SystemName. 
As demonstrated in Figure~\ref{fig-RX-model}, 
\SystemName leverages different physical design
strategies in regards to the purposes. For instance, 
the column-store~\cite{stonebraker2018c} is adopted 
to maintain vertex relation $R_\at{V}$, for
speeding up the scanning of vertex attributes;
the large number of small edge relations $R_\at{in}^v$
and $R_\at{out}^v$ are kept using the
\warn{efficient row-store~\cite{}}, because that
real-life graphs often have the 
power-law degree distribution~\cite{clauset2009power};
\warn{while compressed c-store~\cite{} add details}.

\stitle{Pointers}. Recall that each logical-level
pointer can link to either a single tuple or
an entire extended relation. We implement
the pointers in two ways, \ie
directly and indirectly, in \SystemName.
For direct way, each pointer is such implemented
that points to the starting address of the tuple 
or extended relation it refers to.
In indirect implementation, we follow the
technique that was developed for garbage
collection~\cite{piquer1991indirect}. 
\warn{For reference, it records the block index and 
offset of the  tuple or relation 
that the pointer links to, within
the unified virtual address space.
This method is more convenient for
data modification and materialization. Add more details.}

As will be seen in Section~\ref{sec-update}, these
implementation mechanisms can also easily
adapt to graph updates. 
}

\eat{
\stitle{Flexible physical storage}. Our 
proposed \ModelName model allows \SystemName to 
enforce flexible physical data organization when
storing extended relations.
As a simple example, \SystemName can take the locality 
of the topological structures of graphs into account
when conducting memory management within the
virtual address space. 

\vspace{0.36ex}
In fact, it is 
common that the incoming and outgoing 
edges of the same vertex would be accessed 
continuously in graph computation, \eg pattern matching.
Therefore, \SystemName organizes $R_\at{in}^v$ and
$R_\at{out}^v$ in continuous memory address for 
better cache performance and memory locality 
(see Figure~\ref{fig-RX-model}). 
Such flexible strategies are beyond the capabilities
of \eg~\cite{jin2021making,
mhedhbi2022modern,Kaskade}, 
which simply build
graph indices and views in \RDBMS.
\looseness=-1
}

\stitle{Unified address space and pointers}.
Following the conventional design for
\RDBMS~\cite{ozsu1999principles, kemper2011hyper, irving1982virtual}
and operating
systems~\cite{silberschatz2018operating, stonebraker1981operating},\eat{
	\SystemName utilizes a {\em unified virtual address space}
	for both the conventional relations. It
	represents each pointer as a fixed length field \revise{in the following
		two ways}}
\warn{a {\em unified virtual address space} is utilized for both the conventional and the extended relations.
	Each pointer can then be represented with a fixed length field in the following two ways.}
(1) The pointers can directly 
record the starting address 
of 
a fragment, in the unified space.
(2) Alternatively, 
an indirect pointer is supported
in conjunction with the segment-based management for segments,
which records only the segment id and an offset 
inside the segment, along the same lines as~\cite{piquer1991indirect}. 
\warn{The address of the fragment can then be identified by first lookup the base address 
	of the segment via its id in a segment table and then plus with recorded offset.}\eat{
	\SystemName would first lookup the base address 
	of the segment via its id in a segment table, and then use the recorded 
	offset to identify the tuple or fragment.}
Such an indirect implementation simplifies the processing of graph updates
(Section~\ref{sec-update}).\looseness=-1
\eat{In particular, if a pointer in $\D_G^E$ 
	links to a relation that is not fragmented accordingly, 
	it can only be a single tuple in relation $D_V$
	and can simply record the tuple id accordingly.\looseness=-1}

\etitle{Remark}. \revise{Observe the following.
(1) In addition to \warn{a} better memory locality, 
the unified virtual address space \warn{unifies the} in-memory and on-disk graph data representation,
and hence, \eat{extend \SystemName with}\warn{makes it possible to incorporate} 
disk-based \eg~\cite{fan2022hierarchical},
and even distributed \eg~\cite{bouhenni2021survey} graph processing 
(2) Since the \ModelName model is a mild extension of
the relational model,}
\warn{most of existing physical design techniques for \RDBMS, 
\eg~\cite{athanassoulis2019optimal,
arulraj2016bridging,stonebraker2018c,
alagiannis2014h2o,rasin2013automatic,
lightstone2010physical}, 
can be leveraged as showcased in Figure~\warn{\ref{fig-physical-design}}}.
\eat{As \revise{showcased} in Figure~\warn{\ref{fig-physical-design}}, 
\SystemName can leverage different physical design
strategies.}For instance, 
column-store~\cite{stonebraker2018c} 
can be adopted to maintain vertex and edge attribute relations 
to speed up the scanning of each specified column.
\revise{Moreover, since the power-law degree distribution
of real-life graphs~\cite{clauset2009power}
often yields} a number of edges that share the same
label and are adjacent to the same vertex, 
column-oriented compression methods~\cite{daniel2006integrating, lang2016data}
can also be adopted\eat{ by \SystemName} for storage and execution optimization.

\eat{\etitle{Remark}. Besides achieving a
better memory locality, 
the unified virtual address space also 
enables \SystemName to unify in-memory and
on-disk graph data representation. Moreover, 
a further extension of physical design 
\wrt secondary storage
is doable by incorporating disk-based graph
processing approaches \eg~\cite{fan2021making}.
\looseness=-1

\vspace{0.36ex}
Since \ModelName model inherits 
the underlying features of relational model,
most of existing well-developed 
physical design techniques for \RDBMS, 
\eg~\cite{athanassoulis2019optimal,
arulraj2016bridging,stonebraker2018c,
alagiannis2014h2o,rasin2013automatic,
lightstone2010physical} can 
be directly utilized in \SystemName.
\eat{Thanks to the high degree of similarity 
between the two models, most of existing well-developed 
physical design techniques for \RDBMS, 
\eg~\cite{athanassoulis2019optimal,
arulraj2016bridging,stonebraker2018c,
alagiannis2014h2o,rasin2013automatic,
lightstone2010physical} can 
be directly inherited in \SystemName.}
As demonstrated in Figure~\ref{fig-RX-model}, 
\SystemName leverages different physical design
strategies in regards to the purposes. For instance, 
the column-store~\cite{stonebraker2018c} 
can be adopted to maintain vertex and edge attribute relations, 
to speed up the scanning of each specified column.
On the other hand, since the power-law degree distribution~\cite{clauset2009power} 
often results in a number of edges having the same
label and adjacent to the same vertex, 
column-oriented compression methods~\cite{daniel2006integrating}
can also be considered in \SystemName for storage and execution optimization.}

\eat{
\warn{
\sstab
\warn{(3) 
The introduced pointers directly connect different tuples
and thus the exploration based on it reduces the accesses incurred in
graph view~\cite{Kaskade} and 
join indices~\cite{jin2021making}
that are not memory-friendly and also hard to consider secondary index.
Moreover, 
different from previous works~\cite{Jeffrey2, ambuj1994cache, stefan2002optimizing},
\SystemName does not require runtime partition 
for better memory locality or lower communication cost
as the data has already been pre-partitioned according to the topology of graph data.
}
(move backward)
}
}

\vspace{-0.7ex}
\subsection{Updating and Browsing Graphs}
\label{sec-update}
\vspace{-0.4ex}

\eat{
\warn{To do:
\mbi
\item browsing graphs: lossless
\item updating graphs: necessary
\item implementation
\item summarize properties
\mei
}
}

\eat{Based on the \ModelName model, \SystemName allows
users to \revise{browse and update graphs}, while
guaranteeing \revise{the} physical data
independence
\cite{stonebraker1975implementation}.}

\warn{The logical conversion process from graph to the \ModelName model shown in Section~\ref{subsec-define-model}
can be reversed and incrementalized for users to browse and update graph respectively,
regardless of how fragments and pointers are physically implemented
to not break the physical data independence~\cite{stonebraker1975implementation}\eat{,
and differs from 
the maintenance of, \eg ``network model''~\cite{stonebraker2005goes}}.}

\begin{figure*}[t]
	\vspace{-5.4ex}
	\centerline{\includegraphics[scale=1.25]{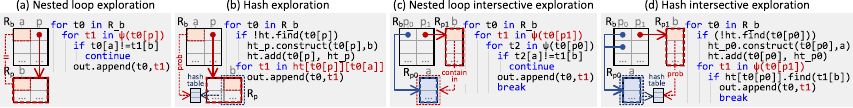}}
	\centering
	\vspace{-2.4ex}
	\caption{A demonstration of the operators introduced in the \ModelName model}
	\label{fig-operator}
	\vspace{-3.4ex}
\end{figure*}

\vspace{-0.5ex}
\stitle{Browsing graphs}.\eat{ \SystemName provides an 
interface for users to browse \revise{graphs} stored
in extended relations with 
explicit topology, as depicted in Figure~\ref{fig-overview}.}
\warn{The \revise{graphs} stored in extended relations can be browsed with 
explicit topology, as depicted in Figure~\ref{fig-overview}, by
simply reversing the graph conversion process to 
restore the topological structure of the graphs, in which
pointers are used to fetch relevant data.}
\warn{Users may opt to pick entities of their interests, 
for the browsing to be conducted lazily and expand localized subgraphs only upon their request.}
Moreover, \warn{common labels and attributes of \revise{entities
(vertices) and relationships (edges)}
can also be extracted from its extended
relations}, by referencing the \ModelName schema
for a  graph $G$. \revise{It can then facilitate the creation of an
ontology~\cite{matthews2005semantic} for such entities and links,
such that users can easily} write pattern queries over $G$.
\looseness=-1

\stitle{Incremental maintenance}. \revise{Real-life graphs
are often updated}.
Consider batch updates  $\Delta G = (\Delta V, \Delta E)$ to a graph $G$,
which consist of a set $\Delta V$ of vertex insertions and deletions,
and \revise{a set} $\Delta E$ of edge insertions and deletions.  
Updates to vertex (resp. edge) attributes can be simulated by \revise{a}
sequences of vertex (resp. edge) deletions and insertions, which
first delete the old  vertices (resp. edges),  followed by 
inserting  the new ones with updated attributes.

\vspace{0.36ex}
Given the batch updates $\Delta G$,\eat{ \revise{\SystemName maintains}
the extended relations \revise{of $G$} by}
\warn{$\D_G^E$ can be maintained by}
(a) appending newly inserted vertices to
$D_\at{V}$ along the same lines
as in the graph conversion process in
Section~\ref{subsec-define-model};
(b) allocating new vertex and edge attribute values 
to $D_\at{V_A}^{l_i}$ and $D_\at{E_A}^{l_j}$;
(c) sorting the new edges by source vertices 
$v_i$ and destination
vertices $v_j$, and appending them to
fragments $F_\at{out}^{v_i}$ and 
$F_\at{in}^{v_j}$ accordingly;
(d) identifying and removing \revise{the data about the} deleted 
vertices and edges based on the value and reference function; and 
(e) \revise{revising the reference function accordingly, as described
in} Section~\ref{subsec-define-model}.

\etitle{Implementation}. \revise{We implement the procedure based on
	the physical design of Section~\ref{subsec-physical-implementation} as follows}.
Since the small fragments are \revise{stored as segments
	in a} continuous 
memory space (Section~\ref{subsec-physical-implementation}),
\warn{the reallocation will be conducted when the inserted tuples make
the fragments exceed the capacity of their underlying segments.}\eat{\SystemName conducts  segment reallocation when
	the inserted tuples make fragments larger 
	than the capacity of their segments.}
\revise{In particular}, 
\warn{if the size of an enlarged fragment reaches the threshold, 
a block will be allocated for the block-based management
to be applied on it thence,
which makes the reallocation cost under control \warn{(see~\cite{full} for an evaluation)}.}\eat{\SystemName 
splits it into blocks and applies block-based 
management instead. \revise{This strategy ensures} that
only small enough fragments are managed as 
segments, making the reallocation cost
under control.}
\warn{Once a segment has been reallocated, the {\em direct} pointers point to 
	it will need to be reassigned to tract it while the {\em indirect} pointers need not to be updated and save such a cost.}
\looseness=-1

\eat{\vspace{0.36ex}
The procedure above works regardless of how fragments and
pointers are physically implemented. \revise{It 
processes graph updates  without
breaking the physical data independence~\cite{stonebraker1975implementation},
and differs from 
the maintenance of, \eg ``network model''~\cite{stonebraker2005goes}}.}

\eat{We consider the batch graph updates
as  $\Delta G = (\Delta V, \Delta E)$,
which consists of a set $\Delta V$ of vertex insertions and deletions
and another set $\Delta E$ of edge insertions and deletions.  
Updates to vertex (resp. edge) attributes can be simulated by
sequences of vertex (resp. edge) deletions and insertions, which
first delete the old  vertices (resp. edges),  followed by 
inserting  the new ones with updated attributes.

\vspace{0.36ex}
Given the batch updates $\Delta G$, \SystemName manipulates the extended
relations by
(a) appending newly inserted vertices to
$D_\at{V}$ along the same lines
as in the graph conversion process in
Section~\ref{subsec-define-model};
(b) allocating new vertex and edge attribute values 
to $D_\at{V_A}^{l_i}$ and $D_\at{E_A}^{l_j}$;
(c) sorting the new edges by source vertices 
$v_i$ and destination
vertices $v_j$, and appending them to
fragments $F_\at{out}^{v_i}$ and 
$F_\at{in}^{v_j}$ accordingly;
(d) identifying and removing information \wrt deleted 
vertices and edges based on the value and pointer 
assignment semantics; and 
(e) establishing new pointer relationship
as in Section~\ref{subsec-define-model}.

\eat{
\sstab
(1) For each newly inserted vertex $v$ in $\Delta V$, 
\SystemName appends it to
the vertex relation $R_\at{V}$ along the same lines
as in the graph conversion process described in
Section~\ref{subsec-define-model}; and the attributes of $v$ are
allocated to the vertex attribute relation $R_\at{V_A}$.
Similarly, attributes of new edges in $\Delta E$ are also 
included in the edge attribute relation $R_\at{E_A}$.

\sstab
(2) Sort the new edges by their source vertices 
$v_i$ and destination
vertices $v_j$; and append them to $R_\at{out}^{v_i}$ and
$R_\at{in}^{v_j}$ accordingly.
Besides, establish the new pointer relationship, \ie filling in the fields of
$\at{in_L}$, $\at{out_L}$, $\at{src_L}$ and $\at{dst_L}$
as in Section~\ref{subsec-define-model}. 
Note that some pointers may 
replace the labeled nulls, 
\eg when a vertex  
has no outgoing or incoming edges at all
in the original graph. 

\sstab
(3) Capitalizing on the value and pointer 
allocation process
in graph conversion, 
\SystemName identifies information 
\wrt deleted 
vertices and edges in extended relations. 
They will be removed directly.
}

\vspace{0.36ex}
Note that the above procedure does not change the 
the schema of extended relations, 
and it does not rely on how the fragments and
pointers are physically managed and 
implemented either. 
Thus, different from maintaining the data 
in ``network model''~\cite{stonebraker2005goes},
here the graph updates are processed without 
breaking the physical data independence~\cite{stonebraker1975implementation}.
\looseness=-1

\vspace{0.6ex}
Below we highlight the key
features in implementing this logical
process with the physical design in Section~\ref{subsec-physical-implementation}. 

\etitle{Implementation}. Since the small fragments are managed as segments and stored in continuous 
memory space (Section~\ref{subsec-physical-implementation}),
\SystemName enforces segment reallocation when 
the inserted tuples make fragments larger 
than the capacity of their 	underlying segments.
However, if the size of an enlarged fragment
reaches the threshold, \SystemName further
splits it into blocks and applies block-based 
management instead. Such a strategy can ensure that
only small enough fragments are managed as 
segments, making the reallocation cost
under control in response to the updates.

\eat{
\warn{
	Since the small fragments are managed as segment 
	and are forced to be stored in continuous memory space in \SystemName
	as described in Section~\ref{subsec-physical-implementation},
	comparing to the conventional \RDBMS,
	there is a challenge to manage them when accepting the updates.
	For a fragment managed as segment,
	once the tuples inserted in it exceed the capacity of its 
	underlying segment, such a segment might require to be reallocated
	and the entire fragment would need to be moved to another position. 
}

\warn{
	In light of these, we first assign a larger-than-need
	reserved space for each segment for holding the
	extended relation, 
	to avoid reallocating segment when only few tuples
	are inserted to it.
	Meanwhile, once the insertion make the size of one segment reaches 
	the threshold, it would be divided into blocks and then be managed 
	as blocks thence to avoid further reallocation.
	Such a management strategy can ensure only small enough fragments
	need to be regarded as segment
	which then make the reallocation cost under control.
	Together, both the required times of reallocation and the cost for 
	each single reallocate can be controlled.
}
}

\eat{
There are two major
challenges in physical implementation.
(1) Inserted vertices and edges can make
the size of some extended relation beyond the 
capacity of the segment it resides in, hence another
larger segment in the address space is demanded
for reallocating this segment. 
(2) When a segment is moved to another location
in the address space, 
all existing pointers linked to the data within this
segment would be invalid, \ie the connection to
relevant tuples or relations is lost.

\vspace{0.36ex}
\eat{
In light of these, we assign a larger-than-need
reserved space for each segment for holding the
extended relation. Then segment reallocation
is done only if a number of tuples
are inserted to the extended relations.
This is analogous to the 
conventional block-based memory management
which typically has 
in-block free space to 
place inserted tuples.
The experimental results 
in Section~\ref{sec-expt} show that 
$\tbf\times$ speedup is achieved
in processing graph updates when
$\tbf\%$ extra space is reserved for each
segment. }
}

When a fragment has been reallocated
to another position,
\SystemName just updates its base
address in the segment table  for
indirect pointer implementation.
Then the segment address can still
be obtained by the old id and offset.
In contrast, the recorded starting address must be
adjusted in response to every
segment reallocation for indirect
pointer implementation.
This is more costly when multiple pointers
link to the same segment. 
Nevertheless, direct pointers
can be coupled with the indirect counterpart 
in \SystemName to strike a balance between 
memory usage and efficiency.
}

\eat{
\warn{
	Moreover, as the updated base address of a segment can still be looked-up 
	in the segment table using its id after it has been moved to another position.
	The indirect pointer (Section~\ref{subsec-physical-implementation}) 
	can thus still keep tracking the tuple/fragment it points to after accept 
	the updates that potentially require the segment reallocation.
	In contrast, although is more efficient to dereference, 
	the direct pointers that directly record the address must be updated
	in response to every reallocated segment,
	which is more costly when multiple pointers
	link to the same segment. 
	Nevertheless, direct pointers
	can be coupled with the indirect counterpart 
	in \SystemName to strike a balance between 
	memory usage and efficiency.
}
}

\eat{
Moreover, we assure that the id of each segment
keeps intact even if the segment has moved
to another position. By doing so, the
recorded segment id's also remain the same in the
pointers \wrt indirect implementation
(Section~\ref{subsec-physical-implementation})
when processing updates; and we only need to
adjust the mappings from the id's to addresses in
\eg an array.
In contrast, for pointers implemented in the direct
way, the recorded addresses must be updated
in response to every reallocated segment,
which is more costly when multiple pointers
link to the same segment. 
Nevertheless, direct implementation of pointers
is coupled with the indirect counterpart 
in \SystemName to strike a balance between 
memory usage and efficiency.}

\eat{
\stitle{Browsing graphs}. \SystemName provides an 
interface for users to browse data stored
in extended relations with 
explicit  topology, as briefly depicted in Figure~\ref{fig-overview}. This is achieved by 
simply reversing the graph conversion process
in Section~\ref{subsec-define-model} to 
restore topological structure of the graphs, in which
pointers are used for fetching relevant data.
Users can also specify
entities of their interests, and \SystemName
just returns localized subgraphs surrounding them.
\looseness=-1

\vspace{0.36ex}
On the other hand, \SystemName can extract common
labels and attributes of
the vertices (resp. edges) from  extended
relations, by referencing the \ModelName schema
that corresponds to data graph $G$.
It organizes them in the form of 
ontology~\cite{matthews2005semantic}, which facilitates
the users to write pattern queries over $G$.}

\vspace{-0.3ex}
\section{Hybrid Query Evaluation}
\label{sec-eval}
\vspace{-0.4ex}

\eat{
\warn{To do: efficiency
\mbi
\item strategy for seamless query evaluation, inside \RDBMS
\item pattern queries: no slower than native graph engines, vs.~\cite{GSQL};
\item hybrid queries: RDBMS query optimizer
\item examples: to illustrate how hybrid queries are optimized
\item I/O cost when processing large graphs
\item Justification: properties and advantages, versus prior systems
\mei
}

\etitle{Implementation of $\delta$ function}. \warn{Show to how
implement the heterogeneous entity resolution function.}
}

\warn{In this section, we present the entire
\ModelName-based hybrid query evaluation workflow.}
We first introduce the new
exploration operator 
induced by the \ModelName model
that \warn{facilitates} the pattern matching
(Section~\ref{subsec-RX-model-operator}).
We then review the 
general strategy of query optimization
in \RDBMS  (Section~\ref{subsec-query-graph}) 
and show how 
to extend it by incorporating exploration to
evaluate hybrid queries  (Section~\ref{subsec-query-optimization-with-exploration}).


\subsection{Exploration Operator}
\label{subsec-RX-model-operator}

Based on the pointers introduced in the \ModelName
model, we define the exploration operator, which 
can be used to ``explore'' the data in extended
relations through those pointers. 
\eat{
We define the exploration operator, which 
can be used to ``explore'' data in the extended
relations through the pointers introduced in \ModelName
model, }

\eat{
\stitle{Exploration}.
The {\em exploration operator}, denoted as $\explore{A}$,
is applied on an extended relation $D$ with one of its
column $A_p$ whose domain consists of the pointers. 
More specifically, 
$D \explore{A} D[A_p]$ {\em joins} 
every tuple 
$t$ in $D$ with each one of the tuples in $\psi(t[A_p])$.}

\stitle{Exploration}.
The {\em exploration operator}, denoted as $\cexplore{A}$, 
is applied on an extended relation $S_1$, 
\warn{regarding one of its field $A$ as pointers, toward} 
another extended relation $S_2$.
Here $\theta$ is condition set  that can be empty.
That is, $S_1 \cexplore{A} S_2$ {\em joins} 
every tuple $t \in S_1$ 
with each one of the tuples 
$t'  \in \psi(t[A]) \subseteq S_2$ together
{\em only if} $t$ and $t'$ satisfy condition $\theta$,
\eg comparison among multiple attributes
on $t$ and $t'$.\eat{
This is similar to the 
conventional $\theta$-join supported by \RDBMS, which
combines two tables based on the condition $\theta$.
That is, $S_1 \cexplore{A} S_2$ {\em joins} 
every tuple 
$t$ in $S_1$ with each one of the tuples in $\psi(t[A]) \subseteq S_2$
to form the result relation.}
\looseness=-1

Intuitively,\eat{
$S_1 \cexplore{A} S_2$ combines
every tuple $t$ from $S_1$ and 
$t'$ from $S_2$ %
s.t. $t'$ is linked by the 
pointer $t[A]$ and they satisfy conditions $\theta$.}
if a graph $G$ is stored with the \ModelName model as in Section~\ref{sec-model},
then the exploration
operation is indeed analogous to exploring $G$
by, \eg iterating over \warn{all} adjacent edges of each vertex
in $G$, 
a common step in graph algorithms, 
\warn{instead of a single tuple~\cite{shekita1990performance}}.\eat{
Therefore, employing the 
exploration operator, most of the strategies in 
exploration-based native graph algorithms, 
\eg~\cite{yang2021huge,bhattarai2019ceci,han2019efficient,
huang2014query,cordella2004sub} can
be equivalently achieved here.}
\looseness=-1

\eat{
\begin{example}
\label{exa-exploration}
	Considering extended relation $\D_G^E$
	of data graph $G$, exploring the outgoing edges from a 
	vertex $v \in G$ can be described with the exploration
	operator as: $D_V[out_L] \explore{A} D_{out}$.
	To consider the	opposite vertex pointed by 
	the outgoing edges:
	$D_{V}[out_L] \explore{A} D_{out}[dst_L] \explore{A} D_{V}$.
\end{example}}

\eat{
\stitle{Incorporating conditions}. We also allow 
conditions $\theta$ to be enforced 
on $\explore{A}$, yielding {\em conditional exploration}
$S_1 \cexplore{A} S_2$. That is, 
a tuple $t \in S_1$ and another tuple 
$t' \in S_2$ from the 
subset pointed by $t[A]$ are joined together
{\em only if} $t$ and $t'$ satisfy condition $\theta$,
\eg comparison among multiple attributes
on $t$ and $t'$. 
This is similar to the 
conventional $\theta$-join supported by \RDBMS, which
combines two tables based on the condition $\theta$.}

\eat{
However, besides simply verifying the condition 
over the plain values in $S_1$ and data pointed by \warn{$D_B$},
here $\theta$ can also inspect other 
data fetched by the pointers. 
}

\etitle{Explorative condition}.
Besides simply verifying it over plain values,
the data \warn{linked} by the pointers can also be inspected 
in the \warn{join condition} here.
Considering a condition $\theta^I$
defined as $\psi(A)[B] \ni C$.
For each tuple $t \in S_1$
and $t' \in S_2$, 
$S_1 \ijoin S_2$ checks whether there 
exists a tuple $t'' \in \psi(t[A])$ such that $t''[B]$ matches the value
of  $t'[C]$; if so, $t$ and $t'$ are joined and
returned in the result relation.
As the evaluation process of such a condition \warn{will} also 
explore the tuples pointed by $t[A]$,
we refer to $\theta^I$ as an {\em explorative condition}.

\vspace{0.36ex}
By incorporating such a condition into an exploration operator, 
$S_1 \iexplore S_2$ 
\warn{can then} pairwisely check the tuples linked by 
pointers in {\em different fields}, \ie $A$ and $A'$ 
of the same tuple $t \in S_1$ when enforcing $\theta^I$, just like computing the
intersection of $B$-values and 
$C$-values \wrt those linked tuples.\eat{
These will be joined with $S_1$.} Hence we refer to $S_1 \iexplore S_2$ as the
{\em intersective exploration}.
\looseness=-1

In fact, the intersective exploration can serve as
a basic component, checking whether the 
matches of two connected pattern edges
intersect at a same vertex in the data graph,
\ie an essential step in matching
cyclic patterns.\eat{
Moreover, this can be done without physically
joining the data as required by conventional
relational operators.} As will be seen shortly,
the (intersective) exploration can be 
executed seamlessly along with other conventional
operators in a 
pipelined manner~\cite{boncz2005monetdb} \warn{without performance degradation}.
\warn{The comparison to the
	native graph-based pattern matching algorithms in Section~\ref*{subsec-query-optimization-with-exploration}
	will further show how most of their strategies 
	can then be achieved equivalently.}

\eat{
\begin{example}
\label{exa-intersection}
\warn{
	A possible execution process of the intersective exploration
	$R_b \explore{p_1}_{\psi(R_b[p_0])[c] \ni b}  \;  R_b$
	is demonstrated in
	Figure~\ref{fig-operator} (b).
	For each tuple $t_0 \in R_b$ and all tuples $t_1 \in \psi(t_0[p_1])$,
	export tuple pair $(t_0, t_1)$ 
	iff $\exists t_2 \in \psi(t_0[p_0])$ \st $t_1[b] = t_2[c]$.
}
\end{example}
}

\vspace{-0.2ex}
\warn{
\begin{example}
	\label{exa-intersective-exploration}
	A possible logical plan to match pattern $P$ in Figure~\ref{fig-example} on graph stored as $\D_G^E$, 
	omit label checking, can be described as:

	\centering{
	$(((D_{V}^2 
	\explore{out_L} D_{o}^1 )
	\explore{dst_L} D_{V}^1 )
	\explore{out_L}_{\psi(D_{V}^2.in_L)[src_L] \ni dst_L} D_{o}^2 )
	\explore{dst_L} D_{V}^0$
	}

	It represents matching the pattern with the order: $v_2$, $v_1$, $v_0$,
	where $D_{V}^0$, $D_{V}^1$ and $D_{V}^2$ (resp. $D_{o}^1$, $D_{o}^2$) are all aliases of $D_{V}$ 
	(resp. $D_{out}$).
\end{example}}
\vspace{-0.4ex}
\looseness=-1

\stitle{Physical implementations}.
Same as the standard operators supported by \RDBMS, 
each logical exploration operator
can also correspond to multiple
physical operators \warn{in} different scenarios.
We first use a logical exploration:  
$R_b \explore{p} _{a = b} R_p$,
where \warn{field $R_b[p]$ consists of pointers toward fragments},
as an example to show how various physical operators can 
correspond to it, and then consider the explorative condition.\eat{
Further, as the explorative condition is not that straightforward 
to be considered as the conditions with plain value.
We also utilize a logical intersective exploration:  
$R_{b} \explore{p_1}_{\psi(R_{b}[p_0])[a] \ni b} R_{p_1}$,
to show how various physical implementation
can be considered for it.}

\warn{
\etitle{Nested-loop exploration}.
As the most native implementation,
once a tuple $t_0 \in R_{b}$ has been enumerated,
$t_0[p]$ \warn{will} be dereferenced to locate the fragment it points to.\eat{, 
thence all tuples in it can be accessed
An additional lookup of the address as described in Section~\ref*{subsec-physical-implementation} would be required here
if the pointer is indirected.}
Thence all tuples
$t_1 \in \psi(t_0[p])$ can then be accessed and iterated in a nested loop manner to directly verify equalities as shown 
in Figure~\ref{fig-operator}(a).
The $t_0, t_1$ pairs can then be exported if there are $t_0[a] \text{=} t_1[b]$.
Such an implementation will be very efficient if all $t_0[p]$ point
to relatively small fragments.}
\looseness=-1

\warn{
\etitle{Hash exploration}.
Alternatively,
the hash-based methods 
can also be considered to speed up the equality checking by 
constructing a hash table for the data to be probed.
That is, as shown in Figure~\ref{fig-operator}(b), 
a large hash table $ht$ will be maintained.\eat{
to map the value of the pointer in field $R_b[p]$ to their own separated hash table.
 that contains the value of field $[b]$ of the tuple they point to.}
For each enumerated tuple $t_0 \in R_{b}$, 
it will be first checked whether 
there already exists a hash table $ht_{p}$ in $ht$ \warn{corresponding} to the value in $t_0[p]$.
If not, such a $ht_{p}$ will be constructed,
to hold all tuples $t_1 \in \psi(t_{0}[p])$ and the value of their field $t_1[b]$,
and be stored in $ht$ \warn{corresponding} to $t_0[p]$;
otherwise, such a pre-constructed $ht_{p}$ can be directly reused 
to find all tuples $t_1 \in \psi(t_{0}[p])$ s.t. $t_0[a] \text{=} t_1[b]$.
\looseness=-1

Note that those hash tables stored in $ht$ can 
be reused for different tuples in $R_b$ that have the same value in their pointer field,
and thus eliminate the need of duplicate construction.
A performance improvement can then be achieved if the pointers
in column $R_b[p]$ all point to relatively large fragments
and a large portion of them have the same value, 
\ie toward the same set of tuples.
This can happen when $R_b$ is the intermediate result of the pattern matching
where the same vertex from data graph can appear multiple times
in different partial matches.
Other typical physical implementations for \warn{the} join operator, 
\eg index, merge sort etc.~\cite{lightstone2010physical},
can also be considered for exploration but will not be further discussed here.}

\warn{
\etitle{Explorative condition}.
As the explorative condition cannot be simply evaluated 
by verifying the plain value of the two input tuples,
we consider a logical intersective exploration:  
$R_{b} \explore{p_1}_{\psi(p_0)[a] \ni b} R_{p_1}$,
and demonstrate
the nested loop and hash-based physical operators \warn{corresponding} to it
in Figure~\ref{fig-operator}(c) and \ref{fig-operator}(d) respectively.}

\eat{
\etitle{Explorative condition}.
As the explorative condition cannot be simply evaluated 
by comparing the plain value,
we then utilize a logical intersective exploration:  
$R_{b} \explore{p_1}_{\psi(R_{b}[p_0])[a] \ni b} R_{p_1}$,
to show how various physical implementations
can also be considered for it.

Figure~\ref{fig-operator}(c) demonstrates 
a nested loop implementation.
For each tuple $t_0\in R_b$, all tuples $t_2 \in t[p_0]$ and $t_1 \in t[p_1]$
will be iterated for verifying equality.
$t_0, t_1$ pairs will be exported once we find a
$t_2 \in t[p_0]$ s.t. $t_2[a] \text{=} t_1[b]$.
Meanwhile, as shown in Figure~\ref{fig-operator}(d),
the hash method can also be considered.
For each $t_0\in R_b$,
separated hash tables $ht_{t_0}$ will be constructed and reused
corresponding to its pointer field $t_0[p_0]$ 
for the value of field $[a]$ it points to. 
Then for each tuple $t_1 \in \psi(t_0[p_1])$, 
there will only need to verify whether value $t_1[b]$ exists in $ht_{t_0}$
to decide whether to export the tuple pair $t_0, t_1$.}

\eat{
\etitle{Nested loop intersective exploration}.
\warn{Similar to the nested loop exploration with plain value join conditions,
once a tuple $t_0 \in R_{b}$ has been enumerated,
both $t_0[p_0]$ and $t_0[p_1]$ would be dereferenced to locate 
the fragment they point to for all tuples in both fragments to be accessed.}
After that, for each $t$, we can iterate over all tuples 
$t_2 \in t[p_0]$ and $t_1 \in t[p_1]$
in a nested loop manner to directly verify equalities as shown 
in Figure~\ref{fig-operator}(c).
$t_0, t_1$ pairs would be exported once we find a  
$t_2 \in t[p_0]$ s.t. $t_2[a] = t_1[b]$.

\etitle{Hash intersective exploration}.
As shown in Figure~\ref{fig-operator}(d), 
we would also maintain a large hash table $ht$ that maps the value of 
the field of pointer in $R_b[p_0]$ to their own separated
hash table for the value of field $[a]$ they point to.
For each enumerated tuple $t_0 \in R_{b}$, we would first find whether 
there already exists a hash table stored in $ht$ \warn{corresponding to the value
of its pointer field $t_0[p_0]$}.
If so we can directly use the 
existing one to find whether there are $t_1[b] \in \psi(t_0[p_0])[a]$;
if not, we would create such a hash table $ht_{p_0}$ 
for all values in $\psi(t_0[p_0])[a]$ and store it in $ht$.
\warn{Such a hash table can also be reused for the duplicated value in field $R_b[p_0]$.}
}


\eat{
For instance, given a logical 
intersective exploration: 
\warn{$R_{b} \explore{p_1}_{\psi(R_{b}[p_0])[c] \ni b} R_{p_1}$
while both $p_0$ and $p_1$ are all pointers toward fragment,
once a tuple $t \in R_{b}$ has been enumerated,
$t[p_0]$ and $t[p_1]$ would be dereferenced for the
tuples in both of them to be iterated.
After that,} we can use nested loop to iterate over tuples
in the data linked by different pointers
from \warn{both $t[p_0]$ and $t[p_1]$ can then} 
directly verify equalities, as depicted in 
Figure~\ref{fig-operator}. Alternatively,
a hash-based implementation of intersective exploration 
can speed up the equality checking by using a hash table
constructed for the data to be probed 
(see Figure~\ref{fig-operator}).
Note that the hash table can 
even be reused when different pointers link 
to the same data, without the need of duplicate
construction. This 
is helpful 
for handling intermediate results
created during the pipelined evaluation of 
pattern queries, where the same vertex 
from data graph often appears multiple times
in different partial matches.}

\looseness=-1

\eat{
\etitle{Remark}. In principle, enforcing multiple
conditional exploration operators can equivalently
accomplish the standard multi-way joins.
As a special case, triangle counting in
graphs $G$ can be 
done by hash-based 
conditional explorations with a worst-case complexity
of $O(|V||E|)$, which equals the ones achieved
by sophisticated worst-case optimal join algorithms~\cite{mhedhbi2019optimizing}.
}

\vspace{-0.8ex}
\subsection{Relational Query Optimization Overview}
\label{subsec-query-graph}

	We then show how only \warn{a} few minor extensions, 
	without \warn{modifying the overall and the various sophisticated detailed technologies}, 
	\warn{will be} required for the conventional \RDBMS to take the exploration into consideration 
	to \warn{facilitate the evaluation of} both pattern query and hybrid query.
	Before showing the extensions required in each step, 
	the hypergraph-based conventional relational query optimization 
	workflow~\cite{AbHuVi1995, chaudhuri1998overview}\eat{, 
	briefly demonstrated in Figure~\ref{fig-overview},}
	\warn{will first be} reviewed here.

	Generally speaking, 
	in \warn{a} conventional \RDBMS optimizer, the input \SQL query \warn{will first be}  
	parsed to generate an unoptimized logical query plan in a canonical form~\cite{joseph2007architecture} accordingly.
	After that, a query graph~\cite{bernstein1981using} \warn{will} be constructed correspondingly 
	where each relation (resp. constant) 
	in the input logical plan \warn{will} be modelled as a node 
	and each join or selection condition \warn{will} be represented as an edge.
	Each edge connects multiple \warn{nodes for relation or constant} 
	involved in the corresponding condition~\cite{AbHuVi1995}, 
	making those edges {\em hyperedges} and the query graph a {\em hypergraph}.
	
	\eat{
		Multiple nodes for all the relations and constants involved in the corresponding condition
		will be connect by a single edge~\cite{AbHuVi1995},
		which thus makes them {\em hyperedges} and the query graph a {\em hypergraph}.
		\looseness=-1
		
	Each of those edges 
	connects multiple \warn{nodes for relation or constant} 
	involved in the corresponding condition~\cite{AbHuVi1995}, 
	which thus makes them {\em hyperedges} and the query graph a {\em hypergraph}.
		
	Those edges can be {\em \warn{hyperedges}} that \warn{connect} multiple \warn{nodes for relation or constant} 
	involved in the corresponding condition~\cite{AbHuVi1995}, 
	which thus makes the query graph a \emph{hypergraph}.}

	Such a hypergraph equivalently represents the input 
	declarative query while providing an elegant perspective
	for it to become the input of most query optimization algorithms
	especially for join reordering~\cite{fender2013counter, bhargava1995hypergraph, moerkotte2008dynamic}.
	Each CSG (Connected SubGraph) in the query graph 
	denotes a possible subquery,
	and each partition of a CSG into a CMP (connected complement) pair,
	regarding all edges \warn{that} connect them as join conditions,
	can then be resolved \warn{as} a join operator.
	That is, by enumerating all possible top-down partitions from the entire query graph 
	down to the CSGs with only one node,
	all possible join \warn{orders} can then be considered.
	Heuristic methods~\cite{steinbrunn1997heuristic, neumann2009query} 
	or several restrictions on the join tree topology~\cite{ahmed2014snowstorms}, \eg to be left-deep~\cite{Ioannidis1991LeftDeep},
	can be utilized in such an enumeration process as a trade-off to reduce the plan space.

	The cost of each enumerated join tree \warn{will then} be evaluated for the
	optimal query plan \warn{among them} to be selected accordingly.
	The cardinality of each CSG in the join tree 
	\warn{will first be} estimated~\cite{leis2017cardinality, Heule2013HyperLogLog, lu2009new, Poosala1996Improved, Lipton1990Practical, Muller2018Improved, perron2019learned},
	and a cost model~\cite{manegold2002generic, bellatreche2013exploit,SAC79:sigmod} \warn{will} be further utilized to consider the physical execution process,
	to \warn{generate an optimal} physical plan for final execution.
	If \warn{the pipelined execution~\cite{boncz2005monetdb} is utilized}, 
	the generated physical query plan will be further divided into pipelines according to the 
	operators that \warn{can} break the pipeline execution\eat{,
	and a meta pipeline will be generated \warn{that} consists all these pipelines}.

	\begin{figure*}[t]
		\vspace{-5.4ex}
		\centerline{\includegraphics[scale=1.3]{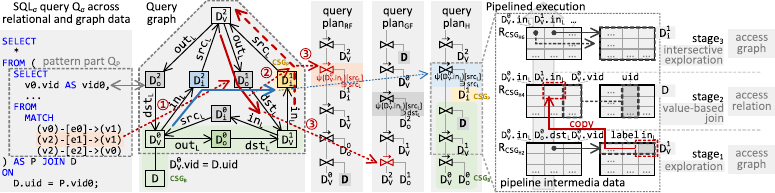}}
		\centering
		\vspace{-2.4ex}
		\caption{A demonstration of the hybrid query evaluation process with the \ModelName model}
		\label{fig-hybrid-query}
		\vspace{-3.4ex}
	\end{figure*}

	\stitle{Extension}.
	It \warn{will} be shown that the above workflow only needs to be extended at three
	minor points to incorporate the exploration and the $\delta$-join 
	to support the graph pattern
	and hybrid query:
	
	\etitle{(1) Exploration representation\eat{ and resolution}}. 
	We introduce a new kind of \warn{{\em exploration edge}} into the query graph.
	\warn{Using this} revised query graph as the input,
	the existing optimization strategies can then be directly
	\warn{applied on it to support} hybrid query \warn{with slight revises}.
	
	\etitle{(2) Exploration cost estimation}. 
	We extend the cost estimation methods for conventional relational query
	\warn{to consider} the \warn{query plan that} contains \warn{the} exploration operator.
	
	\etitle{(3) Dynamic entity resolution}. 
	The $\delta$-join \warn{will} be introduced as an operator for dynamic entity resolution
	according to the query from the user,
	which \warn{will} not \warn{increase} difficulty compared to existing relational operators.
	
	Those \warn{modifications are all minor} without the need to 
	touch the well-developed sophisticated optimization techniques
	from plan enumeration\eat{fender2013counter, moerkotte2008dynamic, steinbrunn1997heuristic, neumann2009query}
	to cost estimation\eat{leis2017cardinality,perron2019learned, park2020g, kim2021combining, leis2015good}.

\vspace{-0.8ex}
\subsection{
Hybrid Query Optimization}
\label{subsec-query-optimization-with-exploration}

	\warn{We now present the extensions on the query optimization workflow 
	enabled by the \ModelName model,
	which are all minor and allow most of the 
	well-developed sophisticated optimization techniques
	to be directly inherited.
	The hybrid query can then be optimized in an enlarged
	unified space and evaluated without performance degradation\eat{,
	and the dynamic entity resolution can also be integrated}.}
	\eat{We would also compare such an extended optimization workflow
	with the native graph-based algorithms, to show the equality and the strengths to them.}

	\stitle{Optimization workflow with exploration}. 
	\warn{We first show how the conventional query graph can be extended to uniformly
		represent the hybrid query, 
		followed by the modifications required on the existing optimization methods 
		for them to be applied on such an extended query graph.}
	\eat{We first show how few minor revises on several procedures 
	allow the conventional query optimization workflow 
	to integrate the introduced exploration operator 
	for hybrid query.}

	\etitle{Exploration representation}.
	We introduce {\em exploration edge} to represent the exploration through pointers in the query graph along with the hyperedges for join and select conditions.
	An exploration edge has the form $(S_1, A, S_2)$
	that directly links the extended \warn{relation} $S_1$ to $S_2$
	and is labelled by the field of \warn{the} pointer \warn{to be explored},
	which can be resolved into an exploration operator: $S_1 \explore{A} S_2$.
	The hybrid query can then be encoded with such a query graph 
	uniformly \warn{to provide an enlarged unified plan space for optimization}.
	\eat{
	The hybrid query can then be represented uniformly after the graph has been represented in the extended relations $\D_G^E$ as in Section~\ref{sec-model}.}
	
	\eat{
	The pattern part $Q_P$ of the hybrid query can then be represented uniformly  provide a unified representation of the hybrid query.
	
	We then show how the hybrid query can be 
	
	As the query graph is considered as the input of
	most of the mainstream optimization methods, 
	to generally inherit the existing workflow,
	it needs to be first extended to represent the introduced exploration.
	In addition to the types of edge for join and selection conditions in the conventional query graph, 
	we introduce another kind of \warn{directed} {\em exploration edge} to the query graph 
	in the form of $(S_1, A, S_2)$.
	Such an exploration edge directly links extended relations
	$S_1$ to $S_2$ and is labelled by
	the name of attribute that contains pointers,
	which can be resolved into an exploration operator: $S_1 \explore{A} S_2$.

	After the data graph has already been transformed into the extended
	relations $\D_G^E$ as in Section~\ref{sec-model},
	a unified representation of the hybrid query for plan enumeration can be generated.
	To simplify the description below, we skip the trivial parsing step to generate 
	the canonical logical plan,
	and directly show how query graph can be constructed 
	corresponding to \warn{$Q_P$}.
	
	A unified representation of the hybrid query can then be 
	}
	
	We then show how the pattern part \warn{$Q_P$} of an input \SQLd query can be encoded with such 
	a query graph, while the rest part can \warn{still} be constructed as it used to be.
	To simplify the description below, we skip the trivial parsing step \warn{for generating} the canonical logical plan 
	and directly show how \warn{the extended} query graph can be constructed corresponding to \warn{$Q_P$}.
	\warn{Assume the data graph is stored in the extended
	relations $\D_G^E$ as described in Section~\ref{sec-model},
	for each pattern vertex $v_i$ in $Q_P$, 
	we add a node $D_\at{V}^i$ into the query graph as an alias of
	the extended relation $D_\at{V}$.
	A pulled down filter condition for vertex label checking, \ie to be $L_P(v_i)$,
	will then be contained in such a node
	in the query graph and associated with its field \emph{``label''}.}

	For each edge $e_i = (v_j, v_k)$ in $Q_P$, 
	two nodes $D_\at{out}^i$ and $D_\at{in}^i$\warn{, alias of $D_\at{out}$ and $D_\at{in}$ respectively,} 
	\warn{will} be included into the query graph and also be associated with a filter condition to check
	the \emph{``label''} to be $L_P(e_i)$.
	After that, four exploration edges $(D_\at{V}^j, \at{out_L}, 
	D_\at{out}^i)$, $(D_\at{out}^i, \at{dst_L},
	D_\at{V}^k)$, $(D_\at{V}^k, \at{in_L},
	D_\at{in}^i)$ and $(D_\at{in}^i, \at{src_L},
	D_\at{V}^j)$ will be incorporated into the query graph,
	which form two opposite paths and enable two alternative
	exploration directions induced \wrt the
	same pattern edge $e_i$.\eat{In fact, 
	the path formed by the 
	latter two corresponding to the opposite
	of the first two hyperedges, enabling two alternative
	exploration directions induced \wrt the
	same pattern edge $e_i$.}
	Such a redundancy 
	introduced on purpose enlarges the optimization space
	for enumerating query plans.
	Finally, ``same-pointer conditions'' \warn{will} be generated among different 
	exploration edges toward the same node in the query graph,
	directly comparing the value of the pointer to 
	restrict them \warn{to} point to the same tuple.
	It then allows the explorative condition to be resolved 
	during the query plan generation (see below),
	and helps to compute
	the intersection of matched vertices, 
	an essential step to optimize matching the cyclic patterns.
	(see~\cite{full} for a formal definition of query graph).
	\looseness=-1

\begin{example}
\label{exa-query-graph}
	Figure~\ref{fig-hybrid-query} shows the query graph 
	corresponding to an example \SQLd query simplified from the one in Example~\ref{exa-SQL}, 
	where the labels of both vertices and edges in its pattern part $Q_P$ are all omitted.
	Taking edge $e_1$ as an example, it corresponds to two opposite paths with edges:
	$(D_V^2, out_L, D_o^1)$, $(D_o^1, dst_L, D_V^1)$; and
	$(D_V^1, in_L, D_i^1)$, $(D_i^1, src_L, D_V^2)$, respectively.
	This denotes that the edge can be explored from two directions.
	Meanwhile, the value-based join across relation and graph data 
	is also represented as an edge \warn{connecting} the nodes \warn{$D$} and $D_V^0$ uniformly. 
	The same-pointer conditions, \eg $D_i^2[src_L] = D_i^1[src_L]$, are not shown in Figure~\ref{fig-hybrid-query} as a simplification \warn{(see~\cite{full} for a step-by-step explanation)}.
\end{example}

	\etitle{Hybrid query plan enumeration}.
		The enlarged unified plan space for hybrid queries, \warn{facilitated} by the extended query graph, 
		allows the access on graph and relational data to be considered in any order.
		The value-based join with low selectivity can then be pulled down into the pattern matching process in the generated plan for optimization according to the estimated cost (see below).
		\eat{
		As the need of cross model access has been eliminated 
		by modeling the graph and relational data uniformly with the \ModelName model
		at the very beginning,
		the query plan can then be executed	without bringing performance degradation.}


		By accepting the extended query graph as input,
		\warn{CSG-CMP pairs can still be exported from it using}
		existing mainstream plan enumeration methods~\cite{fender2013counter, moerkotte2008dynamic, steinbrunn1997heuristic, neumann2009query}.
		There are only a few differences on the following steps to resolve them into exploration operators:\eat{
		and operators can then be generated accordingly
		by resolving all edges connect them as join conditions. 
		As another kind of edge for exploration
		is introduced into the query graph, 
		there will be only few difference on the following steps to resolve 
		them into exploration operators:}

		\eat{
	\etitle{Hybrid query evaluation}.
		As described previously, the exploration can be 
		represented in the query graph and then be resolved 
		into the query plan along with other conventional
		relational operators uniformly.
		Therefore, using such an extended query graph for the hybrid query
		as the input,
		the well-developed methods, 
		exhaustively~\cite{fender2013counter, moerkotte2008dynamic}
		or heuristically~\cite{steinbrunn1997heuristic, neumann2009query},
		can then be utilized to optimize the hybrid query
		in a larger unified space.
		The access of graph and relational data can be taken place in any order.
		That is, if a join condition to a relation has very low selectivity,
		it will be pulled down to the pattern matching process
		in the generated optimized query plan according to the estimated cost for early pruning.}
	
	\sstab
	(1)	If an enumerated CSG-CMP pair is connected by an exploration edge,
	then we add the exploration operator it encodes to the 
	query plan according to the edge's
	direction.
	\warn{The prob-side CSG, \ie the one pointed by the exploration edge, 
	is required to contain only one node in the query graph,
	to ensure the pointer will not be used to track the data 
	that have already been moved into the intermediate result.}
	This restriction is \warn{not tighter than that of} the left-deep trees,
	which are frequently 
	enforced to achieve pipelined execution in \RDBMS~\cite{boncz2005monetdb},
	and one can easily verify that such
	a legal query plan always exists.
	\looseness=-1
	
	\sstab
	(2) \warn{A node in the extended query graph can be resolved
	as an explorative condition incorporated into other operators.
	That is, once a CSG-CMP pair, $CSG_0$ and $CSG_1$, is enumerated, an operator $op_i$ will be
	resolved according to the edges that directly connect them.
	After that, it will be checked whether there exist another node $n_i$ in the query graph,
	such that: 
	(a) there exists an exploration edge from a node in $CSG_0$ (resp. $CSG_1$) toward $n_i$, and
	(b) there exists an edge for plain-value join connects $n_i$ and a node in $CSG_1$ (resp. $CSG_0$).
	For each node $n_i$ \warn{that} satisfies these two requirements,
	it will be resolved as an explorative condition and be incorporated into $op_i$,
	which does not need to be further contained in other CSGs.} 

	
	\sstab
	(3) \warn{The nodes that are duplicated to each other in the query graph
	will not be both contained in the generated query plan. 
	That is, when enumerating the query plan, 
	it will be guaranteed that only one path 
	among each duplicate 
	path pairs in the query graph is inspected \wrt 
	the corresponding pattern edge. 
	Hence, the space for planning is enlarged without
	introducing duplicated computation into the generated query plan}
	\warn{(see~\cite{full} for an overall description).}
	\looseness=-1

	\eat{
	\etitle{Hybrid query evaluation}.
		As described previously, the exploration can be 
		represented in the query graph and then be resolved 
		into the query plan along with other conventional
		relational operators uniformly.
		Therefore, using such an extended query graph for the hybrid query
		as the input,
		the well-developed methods, 
		exhaustively~\cite{fender2013counter, moerkotte2008dynamic}
		or heuristically~\cite{steinbrunn1997heuristic, neumann2009query},
		can then be utilized to optimize the hybrid query
		in a larger unified space.
		The access of graph and relational data can be taken place in any order.
		That is, if a join condition to a relation has very low selectivity,
		it would be pulled down to the pattern matching process
		in the generated optimized query plan according to the estimated cost for early pruning.
		As can also be seen in the following example, the generated hybrid 
		plan can still be executed in a pipelined manner without bring
		any performance degradation:}





    
    \etitle{Exploration cost estimation}.
    We also present a method for estimating the 
    cost of  query plans \wrt hybrid queries,
	\warn{for the optimal plan to be selected from all those enumerated accordingly.}
    Here we opt to extend the previous cost estimation
    strategies for \RDBMS, which are based on (1) cardinality
    estimation, and (2) a cost model to evaluate each operator
    in the query plan. 
    
    \eat{All existing estimation
   	approaches that are designed \wrt  either the 
   	general relational queries in 
   	\RDBMS~\eg~\cite{leis2017cardinality,perron2019learned},
   	or the specific pattern matching 
   	task~\eg~\cite{park2020g, kim2021combining}, and even using some AI-based methods~\cite{Zhou2022DatabaseAI}.}
    
    To estimate the cardinality of intermediate
    results,\eat{ especially the ones deduced by pattern matching, }
    we inherit the existing estimation
    approaches that are designed \wrt  either the 
    general relational queries in 
    \RDBMS~\eg~\cite{leis2017cardinality,perron2019learned},
    or the specific pattern matching 
    task~\eg~\cite{park2020g, kim2021combining}, and even using some AI-based methods~\cite{Zhou2022DatabaseAI}.
    This is inspired by the fact that the same number of matched
    results will be deduced when matching a
    given pattern on the data graph,
    \warn{no matter how the data is modelled}.
    
    Since all operators in the conventional relational query plan remain
    intact in the query plan for hybrid queries, we bring a slight revision 
    of the existing cost models for \RDBMS to consider the introduced operators, 
    instead of developing a new one from scratch. 
    As an example, below we show 
    that the simple cost function $C_{mm}$~\cite{leis2015good},
    which is tailored for the main-memory setting,
    can be extended to an other model, denoted 
    as $C_{emm}$, as follows.
    

    \begin{align*}
    	C_{emm}(T) = 
    	\begin{cases}
    		C_{emm}(T_1) + \tau |T_1[A]|              \quad \text{(if $T = T_1 \cexplore{A} T_2$, nested loop)}  \\
    		C_{emm}(T_1) + \tau |\Pi_{A}(T_1)[A]| + |T| \ \ \text{(if $T = T_1 \cexplore{A} T_2$, hash)}\\
    		\intertext{$C_{emm}(T_1) + \tau |T_1[A]| \times \frac{1}{2} |T_1[A']| / |T_1| $}
    		\vspace{-2.5ex}
    		\intertext{\qquad \qquad \qquad \text{(if $T = T_1 \explore{A}_{\psi(A')[a] \ni b} T_2$, nested loop)}} 
    		\vspace{-2.5ex}
    		\intertext{$C_{emm}(T_1) + \tau |T_1[A]| + \tau |\Pi_{A'}(T_1)[A']| +  |T|$}
    		\vspace{-2.5ex}
    		\intertext{\qquad \qquad \qquad \text{(if $T = T_1 \explore{A}_{\psi(A')[a] \ni b} T_2$, hash)}} 
    	\end{cases}
    \end{align*}
	\vspace{-1ex}
    \looseness=-1
    
    Here only the estimated cost 
    for the physical \warn{(intersective)} exploration operator 
    is shown, \ie using nested loop or  hash table;
    $|S|$ (resp. $|S[A]|$) denotes the cardinality
    estimated for relation $S$ (resp. data linked by the
    pointer in the field $A$ of all tuples in $S$);
    and $\tau \leq 1$ represents the cost of table 
    scan compared to join~\cite{leis2015good}.
    \warn{Intuitively, the nested-loop methods will iterate over all tuples pointed by the pointer; 
    	while the hash-based methods allow the pointer with same value to be reused
    	and thus only need to consider the cardinality after projection.}
    The cost for other conventional relational operators follows
    from the counterparts in~\cite{leis2015good}; while the cost
    for $\delta$-join is estimated by the complexity
    of the ER method applied.

The following example provides an overall demonstration
of how various hybrid query plans
that access graph and relational data in different orders
can be enumerated and then executed without performance degradation:

\begin{example}
	\label{exa-query-plan}
		Figure~\ref{fig-hybrid-query} shows three representative
		query plans corresponding to the
		\SQLd query $Q_\delta$: 
		$\text{Plan}_\text{RF}$, 
		$\text{Plan}_\text{GF}$, 
		and $\text{Plan}_\text{H}$, that access the relational data
		at first, at last and during the pattern matching process respectively.
		$\text{Plan}_\text{H}$ is highlighted as a typical example to show 
		how a hybrid query plan can be generated through CSG-CMP pair enumeration and 
		then be executed with a seamlessly access on graph and relational data 
		without switching to another relational/native graph-based engine
		\warn{(see~\cite{full} for details)}.
\end{example}

\eat{
\begin{example}
\label{exp-cost-model}
\warn{We estimate the cost of the three query plan, 
$\text{Plan}_\text{RF}$, 
$\text{Plan}_\text{GF}$ and 
$\text{Plan}_\text{H}$, 
listed in Figure~\ref{fig-hybrid-query}, assuming that
they all utilizing the hash intersective exploration.
We then have:\\
$C_{emm}(\text{Plan}_\text{RF}) = $  \\
$C_{emm}(\text{Plan}_\text{GF}) = $  \\
$C_{emm}(\text{Plan}_\text{H})  = $  

We then provides an set of assumption about the cardinality appeared in 
above formulas to actually compute the cost of them: 
\warn{$|...| = ...$ ...}.
One can easily verify that these cardinalities do not conflict with 
each other and there exists an actual data set that satisfy all them
above.
Under this setting, it can be calculated that $C_{emm}(P_{H}) < C_{emm}(P_{GF}) < C_{emm}(P_{RF})$
which show that accessing the relational data in the middle 
is more efficient than considering these two part of data separately.}
\end{example}}

\warn{Intuitively, if the join on \warn{$D$} and $D_V^0$ in 
Figure~\ref*{fig-hybrid-query} has a low selectivity
but is not lower than the partial match of vertices $v_0$ and $v_1$,
$\text{Plan}_\text{H}$ will then have the lowest cost among the three of them.
The experimental results in Section~\ref{sec-expt} show 
that optimizing the hybrid query in such an enlarged unified plan space and pulling the join conditions 
down to the pattern matching process in such a way can bring a
$10.1\times$ speedup on average.}

\eat{
	\etitle{Equivalently consider the graph algorithms}.
	This query plan enumeration strategy
	can also integrate the candidate
	matching sets for pattern vertices.
	In practice, computing candidate sets in
	advance is widely adopted in the state-of-the-art pattern matching 
	algorithms
	\cite{shixuan2020inmemory}.
	Here one can apply any existing
	method to derive the 
	candidate matching set for each 
	pattern vertex $v_i$, maintain them
	in extended relation and use a   
	vertex extended relation node
	in the query graph to encode. This
	indeed replaces node $D_\at{V}^i$
	which only verifies matched labels. 

	\warn{
		The candidate set for each vertex can be stored in a separated relation that contains 
		only one column for vertex id.
		To make a use of these candidate set, 
		only need to join the id field of each vertex with 
		their corresponding relation for candidate vertex id,
		in a form of hybrid query.
		Considering that some candidate set can have a very limited
		selectivity but the native graph-based algorithms would 
		consider all of them equivalently, that might actually bring
		negative impact to the performance.
		The declarative query optimization workflow allow 
		us to make a use of them according to the detailed data and query.
		If a candidate set cannot bring much efficiency improvement,
		it would be left on the top of the query tree after optimization.
		We can then remove them as they cannot be pull down to the 
		pattern matching process for efficiency improvement.
	}

	\etitle{Remark}.
	The exploration process can be equality performed with the introduced operator.
	The candidate set can be considered with the hybrid execution with the tables storing the vertex id of the candidate set (see Section~\ref{subsec-query-graph}).
	Other tech for graph pattern matching can also find their \RDBMS alternative:
	Matching order selection \& join reordering.
	Failing set~\cite{han2019efficient} \& sideway-information-passing~\cite{neumann2009scalable}.
	Adaptive query order~\cite{han2019efficient, TurboIso} \& Adaptive execution\cite{Amol2007Adaptive}.
}

\stitle{Dynamic entity resolution with $\delta$-join}.
As the ER algorithm is regarded as a black box here,
each $\delta$-join $Q_1 \join_\delta Q_2$
\warn{will} be constructed as a hyperedge in the query graph
that links to all nodes involved in subqueries $Q_1$ and $Q_2$, along the same lines of handling the union and set difference operators.
As for the query plan generation, 
optimized plans for its subqueries \warn{will first be}
generated separately
and the result relation of the $\delta$-join \warn{will} be 
treated as a base relation for applying
global optimizations further,
in a way similar to that for aggregation operators in conventional
relational query plans~\cite{galindo2001orthogonal}
without bringing more difficulty.
\looseness=-1

During the execution process, the ER algorithm \warn{will} be
conducted once all results of its subqueries have been obtained.\eat{, 
	and the result relation of $\delta$-join would be 
	treated as a base relation for applying
	global optimizations further
	Such an operator should also be able to move across the plan tree 
	if some assumptions can be made about the ER methods as in ~\cite{Leone2022Critical}
	instead of treating it as a black box,
	we defer them to further work.}
	\warn{Such an operator will also be able to move across the plan tree 
		and the required space for intermediate result can thence be reduced 
		if it can be assumed that the utilized ER algorithm will always join or not join two tuples independently
	instead of being treated as a black box.
	We defer the study on this to \warn{future} work.}

\stitle{Comparison to native graph-based algorithms}.
	The introduced exploration operator provides the most fundamental component
	in the native graph-based algorithms,
	while the revised \RDBMS optimizer can equivalently perform the matching 
	order selection.\eat{
	Moreover, the capability for hybrid query plan also allows 
	\SystemName to integrate the candidate sets for optimization,}
	\warn{Moreover, the capability for hybrid query also allows 
	the \emph{candidate sets} to be integrated for optimization}
	which \warn{are} widely adopted in the state-of-the-art pattern matching 
	algorithms~\cite{shixuan2020inmemory}.
	That is, 
	candidate set $C(v_i)$ can be generated for each pattern vertex
	$v_i$ using any of existing methods\eat{through the imperative programming interface demonstrated in Figure~\ref{fig-overview},
	and then be}
	and stored in a separated relation $D^C_{v_i}$ with one column $vid$ only for the vertex id.
	The generated candidate set can then be used for vertex filtering
	as a join: $D^C_{v_i} \join_{vid = vid} D_\at{V}^i$.

	The query-related data-driven declarative optimization workflow 
	actually allows \warn{the candidate sets to be better utilized}, 
	where only the ones that
	can bring more benefits than cost \warn{will} be considered instead
	of all of them.
	That is, as those additional joins on the candidate sets
	\warn{will} not differ the query result,
	if the join tree with the lowest cost does not contain all these 
	joins on the candidate set, it can still be confidently selected to 
	generate the query plan.
	It can thus avoid considering the candidate set with high selectivity, 
	\ie cannot prune \warn{many} vertices, for performance improvement.
	
	\eat{The optimizations from database community on query execution~\cite{boncz2005monetdb,menon2017relaxed,neumann2011efficiently}, 
	a stage rarely considered by the research of graph pattern matching algorithms, 
	can also bring advantages for \SystemName to exceed them.
	As the pipelined execution shown in Example~\ref{fig-hybrid-query},
	the edges connects to a vertex can be iterated continuously
	to further take an advantages of the locality of graph data compared to the native graph	pattern matching
	algorithms~\cite{yang2021huge,bhattarai2019ceci,han2019efficient,
	huang2014query,cordella2004sub} whose workflow basically coincides with 
	the volcano model~\cite{graefe1993volcano} for query
	evaluation in \RDBMS.}

	\etitle{Remark}.
	By considering the exploration as an operator, 
	most of the other native graph-based pattern matching 
	optimizations can also find their \RDBMS alternatives.
	As an example, sideway information passing~\cite{neumann2009scalable} holds a similar idea as the failing set~\cite{han2019efficient};
	The adaptive execution~\cite{Amol2007Adaptive}, well-studied in the database community, 
	can somewhat be regarded as a more advanced adaptive query ordering~\cite{han2019efficient, TurboIso}.
	\warn{Meanwhile, optimizations from the database community on query execution~\cite{boncz2005monetdb,menon2017relaxed,neumann2011efficiently,li2013bitweaving, zhou2002implementing,schuh2016experimental}, 
	a stage that is rarely studied in the research of graph pattern matching algorithms~\cite{yang2021huge,bhattarai2019ceci,han2019efficient,huang2014query,cordella2004sub} 
	whose workflow \warn{essentially} coincide with 
	the volcano model~\cite{graefe1993volcano},
	can also bring advantages to exceed them.}
	\looseness=-1

\eat{
To enumerate logical and physical 
query plans for hybrid queries,
\SystemName extends the 
hypergraph-based 
methods~\cite{moerkotte2008dynamic,bhargava1995hypergraph,
AbHuVi1995}, 
which are adopted in most 
of mainstream \RDBMS 
query optimizers.
Here unified query graphs (hypergraphs) are proposed 
to take all operators into consideration, 
including the newly introduced exploration.
They also enlarge the space for optimizing
hybrid queries. 

\eat{
\warn{The hybrid query plan can then be enumerated 
	from a larger optimization space.
}
}
\eat{
	By traversing such graphs,  // the query plan enumeration process cannot be called as ``traverse''
	\SystemName generates 
	the query plans with a larger optimization space.
}

\stitle{Query graph construction}.
Similar to building the query graph for  conventional \SQL queries,
each relation (resp. constant) 
in the input \SQLd query $Q$
is modeled as a node in the query graph, \ie hypergraph, while
the join and selection conditions
in $Q$ are represented as hyperedges
with labels.
Note that
every hyperedge is connected to all relation and constant nodes
involved in the condition, not limited to
two nodes as in standard simple graphs.

\vspace{0.36ex}
In addition, we introduce another
kind of {\em exploration edges}
to the query graph. 
Each exploration edge is in the
form of $(S_1, A, S_2)$, for encoding
exploration operator of $S_1 \explore{A} S_2$. 
Hence it directly links extended relations
$S_1$ to $S_2$ and is labeled by
the name of attribute that contains pointers.

After transforming graphs into
extended relations as in Section~\ref{sec-model},
we encode the  pattern query part
in $Q$ by using extended relations 
and exploration edges.
More specifically, for each pattern vertex $v_i$ in $Q$, 
we add a vertex extended 
relation node $D_\at{V}^i$
to the query graph, which represents
the subset of extended
relation $D_\at{V}$ that matches
the pattern vertex label $L_P(v_i)$ of $v_i$,
and it is also associated with
the attributes from $D_\at{V_A}^{L_P(v_i)}$.
For each pattern edge $e_i = (v_j, v_k)$, 
two edge 
extended relation nodes $D_\at{out}^i$ and $D_\at{in}^i$ are included, 
indicating subsets that match edge
label $L_P(e_i)$ and have attributes from
$D_\at{E_A}^{L_P(e_i)}$.
Moreover, we incorporate four exploration
edges $(D_\at{V}^j, \at{out_L}, 
D_\at{out}^i)$, $(D_\at{out}^i, \at{dst_L},
D_\at{V}^k)$, $(D_\at{V}^k, \at{in_L},
D_\at{in}^i)$ and $(D_\at{in}^i, \at{src_L},
D_\at{V}^j)$ in the query graph. In fact, 
the path formed by the 
latter two corresponds to the opposite
of the first two hyperedges, enabling two alternative
exploration directions induced \wrt the
same pattern edge $e_i$.
Such redundancy 
introduced can enlarge the optimization space
for enumerating query plans.

\eat{
\warn{
\stitle{Query graphs construction}.
	Same to constructing the query graph for the conventional \SQL,
	each relation appeared in the input \SQLd query $Q$, 
	with the select conditions pull down to it,
	would be modeled as a node in the query graph.
	The join conditions among different relations in $Q$
	would then be represented as hyperedges that connect different nodes.
	We extend the query graph to incorporate
	the introduced exploration operator,
	by considering a new kind of {\em directed exploration edge}: $(D_A)-[A_p]->(D_B)$
	that denotes the operation: $D_A[A_p] \explore{A} D_B$.
}
}
\eat{
Given an \SQL query $Q$, 
the conventional query graph (hypergraph) models
each relation and constant that appears in $Q$
as a node, while the hyperedges in this graph represent 
selection or join conditions of $Q$. Note that
each hyperedge is connected to all relation and constant nodes
involved in the condition, not limited to
two nodes as in standard simple graphs.}

\eat{
\vspace{0.36ex}
\warn{
	Such an edge adopted in the query graph then allows us 
	to describe the graph pattern matching with the exploration operator
	through the introduced pointers.
	That is, for each pattern vertex $v_i$,  we add a 
	vertex relation node $D_\at{V}^i$, 
	with a select condition for its label $L_P(v_i)$ on it; 
	and for each pattern edge $e_i = (v_j, v_k)$, 
	two edge relation nodes $D_\at{out}^i$ and $D_\at{in}^i$
	are included, 
	with select condition for its label $L_P(e_i)$ on both of them.
	After that, directed exploration edges would be added to connect 
	all those nodes. 
	That is, for each pattern edge $e_i = (v_j, v_k)$,
	connects relevant vertices with paths:
	
	$(D_\at{V}^j)-[out_L]->(D_\at{out}^i)-[dst_L]->(D_\at{V}^k)$; and
	
	$(D_\at{V}^k)-[in_L]->(D_\at{in}^i)-[src_L]->(D_\at{V}^j)$.
	\\
	It can be seen that the above two paths are 
	actually semantically duplicate to each other that
	correspond to the two alternative exploration directions of the same edge $e_i$. 
	Node $D_\at{out}^i$ and $D_\at{in}^i$ would be marked as duplicated with each other,
	while such a redundancy introduced on purpose can enlarge the optimization space
	for enumerating the query plan (see below).
}
}

\eat{
We extend the conventional query graph by incorporating the 
pattern queries. That is, for each pattern vertex $v_i$,  we add a 
vertex relation node $D_\at{V}^i$; 
and for each pattern edge $e_i = (v_j,
v_k)$, two edge relation nodes $D_\at{out}^i$ and $D_\at{in}^i$
are included, together with the hyperedges linking
to vertex relation nodes $D_\at{V}^j$ and $D_\at{V}^k$.
To simplify the 
discussion, here each vertex relation node (resp. edge relation node)
indicates the set of tuples extracted from $D_\at{V}$ (resp.
union of $F^v_\at{in}$ or $F^v_\at{out}$ for
all $v$ in graph $G$) by matching the label of pattern vertex (resp.
pattern edge), and they are also associated with the 
attributes from $D^l_\at{V_A}$ (resp. $D^l_\at{E_A}$). 
 
Note that such new hyperedges are ``directed'',
encoding the edge connections $(v_j, v_k)$ 
in the pattern query and their opposites $(v_k, v_j)$.
They are represented by pointer fields, \ie
\at{in_L}, \at{out_L}, \at{src_L} and \at{dst_L}.
For instance, the hyperedge from $D_\at{V}^j$ to $D_\at{out}^i$
(resp. $D_\at{out}^i$ to $D_\at{V}^k$) is attached
with \at{out_L} (resp. \at{dst_L}), while the others
that correspond to the opposite direction of pattern edge $(v_j, v_k)$
are given \at{in_L} and \at{src_L}. 
This duplication can enlarge the optimization space
for hybrid queries (see below).
}

\eat{
\warn{
	The \emph{``hybrid''} join conditions, which across the relations and the result of pattern query,
	can then be considered as the hyperedge in query graph,
	between the nodes representing the ordinary relations
	and the nodes representing the relations for graph data, 
	\ie $D_\at{V}^i$, $D_\at{in}^i$ and $D_\at{out}^i$.

	Finally, a ``same-pointer condition'' can be resolved among the 
	exploration edges toward the same vertex in the query graph,
	\eg $E_{2b}[src_L] = E_{1b}[src_L]$ in Figure~\ref{fig-hybrid-query} (not shown).
	Such a same-pointer condition allows the intersective exploration to be resolved 
	during the query plan generation (see below).
	This helps compute
	the intersection of matched vertices, which is an essential
	step to optimize matching the cyclic patterns.
}
}

\vspace{0.6ex}
For each $\delta$-join $Q_1 \join_\delta Q_2$ in $Q$, 
a hyperedge is also included, linking to all nodes
involved in subqueries $Q_1$ and $Q_2$. 
The union and set difference
operators are handled along the same lines. 
\looseness=-1

\eat{
	Following the dynamic programming
	approach~\cite{moerkotte2008dynamic}, \SystemName
	traverses the query graph recursively to enumerate
	CSG-CMP (connected subgraph-connected complement) 
	pairs~\cite{moerkotte2008dynamic} and generates
	the logical and physical query plans based on
	the traversal order accordingly. In particular, the following
	operations are invoked regarding the pattern queries
	and $\delta$-join.
}

\eat{
\stitle{Plan generation}.\warn{
	Following the dynamic programming
	approach from~\cite{moerkotte2008dynamic}, \SystemName
	would enumerate all CSG-CMPs (connected subgraph-connected complement pair) 
	and generate the logical and physical query plans based on them. 
	Here, we highlight its difference on 
	resolving the (intersective) exploration and $\delta$-join compared to the plan 
	generation process in the conventional relational query.
}
}

\stitle{Plan generation}.
Following the dynamic programming
approach ~\cite{moerkotte2008dynamic}, \SystemName traverses the
query graph to enumerate 
all \kw{CSG} (connected subgraph) pairs 
and generates the logical and 
physical query plans according to  
the traversal order. 
Below we highlight the
difference in resolving (intersective) explorations from the query
graph and processing  $\delta$-join.

\eat{
\sstab
(1) If the traversal accesses a directed hyperedge that is attached
with a pointer field, then an exploration operator
is added to the plan following the hyperedge's direction.
This is equivalent to expanding the partial matches
in native graph pattern matching
algorithms~\cite{yang2021huge,bhattarai2019ceci,han2019efficient,
huang2014query,cordella2004sub}, which verifies the edge connections.

\sstab
(2) The intersective exploration is incorporated whenever possible.
That is, if the hyperedge currently being inspected
directs to node $D$,
and $D$ has a field whose values may intersect with the ones
in the data linked by other pointers from the existing CSG 
enumerated, then we include an intersective exploration in
the plan. This helps compute
the intersection of matched vertices, which is an essential
step when matching cyclic patterns.

\sstab
(3) To ensure that the evaluation is not affected by the
inconsistent data and pointers which may 
appear in the joined result after exploration,  we enforce the query
plans to be in the form of left-deep tree. 
By doing so, each tuple or relation linked by pointers will be involved
in at most one exploration operation.
}

\sstab
(1)	If the enumerated two \kw{CSGs} are connected by an exploration edge,
then we add the 
exploration operator it encodes to the 
query plan following the hyperedge's
direction.
To keep the consistency of pointers,
\ie data can still be tracked by pointers
after applying exploration, we also ensure that
the  right-hand-side operand of the exploration is resolved from a single 
node in the query graph.
This restriction is not tighter 
than the left-deep trees,
which are frequently 
enforced to achieve pipelined execution in \RDBMS~\cite{boncz2005monetdb},
and one can easily verify that such
a legal query plan always exists.
\looseness=-1

\sstab
(2) \SystemName resolves 
intersective exploration 
from the query graph whenever possible.
That is, when an exploration operator
is to be added regarding two \kw{CSGs},
\SystemName checks whether these
two \kw{CSGs} 
also connect to another same node in
the query graph within two hops via
different exploration edges. Such
connections and the specific
node will be resolved as the 
condition of an intersective exploration
operator,
which replaces the exploration operator
that yet to be included.
Moreover, these elements will no 
longer be accessed for further resolving
other operators or conditions. Hence
intersective exploration indeed
combines multiple atomic
operations into one,
reducing the overall cost.
\looseness=-1

\sstab
(3) When enumerating each query plan,
\SystemName also 
guarantees that only one path 
among each duplicate 
path pairs in the query graph is inspected \wrt 
the corresponding pattern edge. 
Hence the exploration 
can be incorporated in the query plan
regarding either  direction of the pattern 
edge.

\eat{
\warn{
	\sstab
	(1)	If the enumerated two CSGs are connected by an exploration edge,
		then an exploration operator would be added to the 
		plan according to the direction of such an edge 
		between two CSGs.
		As the right-hand-side operand of the exploration operator cannot be subquery
		(Section~\ref{subsec-RX-model-operator}),
		the prob-side CSG should contain only one vertex.
		Such a restriction is not tighter than the left-deep tree,
		which would be frequently considered to ensure the pipelined execution~\cite{boncz2005monetdb},
		and it can then be guaranteed that legal plan can always be generated.
		
	\sstab
	(2) The intersective exploration would be incorporated whenever possible.
		That is, when a pair of CSGs is enumerated, 
		\SystemName would explore all the vertices
		through the out going exploration edges from one of the CSGs, 
		to find whether there are conditions, \eg \emph{``same-pointer condition''}, 
		connect to the other CSG.
		Those connections would be resolved as the conditions of 
		the intersective exploration.


	\sstab
	(3) For the vertices in the query graph 
		that have already been considered as conditions
		in the intersective exploration,
		while there are no fields need to be grabbed from it in the result.
		There would be no need to further resolve another operator to explore it, 
		as the constraints have already been considered equivalently. 
		Meanwhile, for the nodes in query graph that are marked as duplicated with each other, 
		\ie $D_\at{out}^i$ and $D_\at{in}^i$,
		it would also be restricted to consider only one in the generated query plan,
		which allows any of the exploration directions to be considered in the query plan.
}
}

\eat{
\sstab
\warn{
	(2) The intersective exploration can be resolved into the generated query plan 
		without actually involving the vertex into the query graph.
		That is, when enumerating two CSGs, \SystemName would exploring the out going 
		exploration edges from one of them, to find whether there are connections toward
		the other CSG.
		If so, such a vertex not contained in both CSGs would be considered as a explorative
		condition in the exploration join.
}
}

\sstab
(4) \SystemName handles the $\delta$-join operator  in a way 
similar to that for aggregation operators in conventional
relational query plans~\cite{galindo2001orthogonal}.
Optimized query plans would be first generated for 
each subquery of the $\delta$-join separately.
Then JedAI~\cite{papadakis2020three} is called for
conducting ER, and the result relation of $\delta$-join is 
treated as a base relation for applying
global optimizations further.

\begin{example}
\label{exa-query-plan}
\warn{
	Figure~\ref{fig-hybrid-query} demonstrates three possible query plans
	generated from the query graph, that join the relational data in the 
	head (RF), in the end (GF) and in middle (H) respectively.

	The resolution process of the intersective exploration in $query\;tree_{RF}$ 
	is especially highlighted here 
	as a representative step to explain the plan generation process:
	Once the CSG pair: $CSG_B = \{R, V_0, E_{0a}, V_1\}$ and $CSG_P = \{E_{1b}\}$ 
	have been enumerated,
	according to the edge $in_L$ between them,
	an exploration operator would be resolved from $CSG_B$ toward $CSG_P$.
	After that,
	the \emph{same-pointer condition} $E_{2b}[src_L] = E_{1b}[src_L]$
	would be considered to generate the intersective exploration,
	as $E_{2b}$ is connected with an outgoing exploration edge from $CSG_B$
	and $E_{1b}$ is in $CSG_P$.

	As $E_{2b}$ has already been considered in the intersective exploration,
	it would not be further contained in the generated query plan.
	The node $E_{0a}$, $E_{1a}$ and $E_{2a}$ would also not be considered
	as they are duplicated with the ones that have already been contained in 
	the generated query plan,
	$E_{0b}$, $E_{1b}$ and $E_{2b}$ respectively.
}
\end{example}

\etitle{Enlarged optimization space}. As demonstrated in 
Example~\ref{exa-query-plan}, joining attributes
from relations and graphs can take place at any time,
\ie ahead of, after, or even in the middle of the pattern matching
process. These result in a larger space for enumerating
query plans and selecting the one with optimal cost
estimated by the cost model (Section~\ref{subsec-optimization}).
Moreover,  the redundancy introduced in the query graphs on purpose reflects the 
two alternative exploration 
directions \wrt each pattern 
edge, enlarging the optimization space as well.
\looseness=-1

\etitle{Remark}.
This query plan enumeration strategy
can also integrate the candidate
matching sets for pattern vertices.
In practice, computing candidate sets in
advance is widely adopted in the state-of-the-art pattern matching 
algorithms
\cite{shixuan2020inmemory}.
Here one can apply any existing
method to derive the 
candidate matching set for each 
pattern vertex $v_i$, maintain them
in extended relation and use a   
vertex extended relation node
in the query graph to encode. This
indeed replaces node $D_\at{V}^i$
which only verifies matched labels. 


\eat{
	\warn{
		\etitle{Remark}. 
		Moving a tuples from their original relation to the 
		pipeline intermediate result~\cite{boncz2005monetdb} would 
		make them failed to be tracked by the pointers point to them,
		therefore, the right-hand-side operand of the exploration cannot be subquery.
		The plan generation process in Section~\ref{subsec-query-graph}
		can guarantee that legal plan can always be generated.}
}

\eat{
\warn{
	\etitle{Remark}.
	The exploration process can be equality performed with the introduced operator.
	The candidate set can be considered with the hybrid execution with the tables storing the vertex id of the candidate set (see Section~\ref{subsec-query-graph}).
	Other tech for graph pattern matching can also find their \RDBMS alternative:
	Matching order selection \& join reordering.
	Failing set \& sideway-information-passing.
	Adaptive query order \& Adaptive execution.
}
}

\subsection{Query Execution}
\label{subsec-hybrid-query-execution}

We now show that \SystemName can execute the 
query plans seamlessly over the extended relations,
by pipelining~\cite{boncz2005monetdb}.

\stitle{Pipelined execution}. Pipelined
execution~\cite{boncz2005monetdb}
has been adopted in \RDBMS to process conventional
relational queries. Similarly, the exploration 
operators can also be executed in a pipelined manner, because of
their semantics. In fact, given an exploration $S_1 \explore{A} S_2$,
we iterate over tuples in the data pointed by $t[A]$ for
each $t \in S_1$, and join them with $t$ to form the intermediate 
result (relation).
After that, the next exploration operator
can readily inspect the plain values and pointers within
the joined tuples of this intermediate result, \ie
multiple exploration operators constitute a pipeline,
without the need of materializing the entire 
temporary  intermediate relation which
incurs excessive I/O cost. Intersective exploration
works in the same way. In fact, the pointers 
in intermediate result still links to the original tuple
or extended relation, hence subsequent exploration
operators can safely employ them to fetch linked data. 

\etitle{Seamless integration}. \SystemName uniformly integrates the
exploration and conventional join operators together 
in the pipelined execution, as guided by the query plan.
Moreover, the tuples are passed in batches from one
operator to another, which is different from the native graph
pattern matching
algorithms~\cite{yang2021huge,bhattarai2019ceci,han2019efficient,
huang2014query,cordella2004sub} whose workflow coincides with 
the volcano model~\cite{graefe1993volcano} for query
evaluation in \RDBMS.
\looseness=-1

\begin{example}
\label{exa-pipeline}
\warn{
	Figure~\ref{fig-hybrid-query} demonstrates the pipelined execution process of three different operators: 
	exploration, conventional value-based join and intersective exploration
	in the hybrid execution plan (H).
	The tuple of vertex pointed by the pointer in field $e_{2a}[dst]$ would be
	first explored, and its fields $label$ and $out$ would be copied
	into the next stage of the pipeline intermediate result, for joining the relation $R$
	and intersective exploration respectively.
	Since such a vertex can appear in the intermediate result for multiple times,
	it would lead to the duplicated pointers in column $v_0._{out}$ as described in Section~\ref{subsec-RX-model-operator}.
	The value-based join and the intersective exploration can then be executed
	without breaking the pipeline.
	}
\end{example}

We remark that other well-developed methods for optimizing query execution,
\eg vectorization and efficient compilation~\cite{menon2017relaxed,neumann2011efficiently}
can also be inherited in \SystemName. 
But native graph pattern matching algorithms rarely consider such
aspects.

\subsection{Cost Estimation}
\label{subsec-optimization}

We finally present a method for estimating the 
cost of  query plans \wrt hybrid queries, based on
which the optimal one is selected by \SystemName.
Here we opt to extend the previous cost estimation
strategies for \RDBMS, which are based on (1) cardinality
estimation, and (2) a cost model to evaluate each operator
in the query plan.

\etitle{Cardinality estimation}. 
To estimate the cardinality of intermediate
results, especially the ones deduced by pattern matching, 
we inherit the existing estimation
approaches that are designed \wrt  either the 
general relational queries in 
\RDBMS~\eg~\cite{leis2017cardinality,perron2019learned},
or the specific pattern matching 
task~\eg~\cite{park2020g, kim2021combining}.
This is inspired by the fact that the same number of matched
results will be deduced when matching a
given pattern on the data graph,
no matter whether this is achieved by using the relational 
join-based methods or native graph pattern matching algorithms.

\etitle{Cost model}.
Since all operators for conventional relational query plan remain
intact in the query plan for hybrid queries, we extend the 
the existing cost models developed for \RDBMS, instead
of developing a new one starting from scratch. 
The extension should cope with the new operators
introduced, \eg exploration. Then the cost 
of a query plan is the sum of all 
operators involved. As an example, below we show 
that the simple cost function $C_{mm}$~\cite{leis2015good},
which is tailored for the main-memory setting,
can be extended to an other model, denoted 
as $C_{emm}$, as follows.

\vspace{-2ex}
\begin{align*}
C_{emm}(T) = 
\begin{cases}
\tau |S_1|  + \Sigma_{t \in S_1} |\psi(t[A])| & \text{(if $T = S_1 \explore{A} S_2$)}  \\
		\intertext{$\tau |S_1| + \tau \Sigma_{t \in S_1} (|\psi(t[A])| \times ( 1 + 0.5 |\psi(t[A'])|))$}
		& \text{(if $T = S_1 \iexplore S_2$,
		nested loop)}  \\
		\intertext{$\tau |S_1| + \Sigma_{t \in S_1}( (1 + \tau) |\psi(t[A])| + \tau |\psi(t[A'])|$)}
		& \text{(if $T = S_1 \iexplore S_2$,
			hash table)} 
	\end{cases}    
\end{align*}

Here only the estimated cost 
for exploration and intersective exploration
are shown, together with the physical implementation
strategies, \ie using nested loop or  hash table;
$|S|$ (resp. $|\psi(t[A])|$) denotes the cardinality
estimated for relation $S$ (resp. data linked by the
pointer in the field $A$ of tuple $t$);
and $\tau \leq 1$ represents the cost of table 
scan compared to the lookup in hash table~\cite{leis2015good}.
The cost for other conventional relational operators follows
from the counterparts in~\cite{leis2015good}; while the cost
for $\delta$-join is estimated by the complexity
of the ER method applied.

\begin{example}
\label{exp-cost-model}
\warn{We estimate the cost of the three query plan, RF, GF and H, 
	listed in Figure~\ref{fig-hybrid-query}, assuming that
	they all utilizing the hash intersective exploration.
	We then have:\\
	$C_{emm}(P_{RF}) = $  \\
	$C_{emm}(P_{GF}) = $  \\
	$C_{emm}(P_{H}) = $  

	We then provides an set of assumption about the cardinality appeared in 
	above formulas to actually compute the cost of them: 
	\warn{$|...| = ...$ ...}.
	One can easily verify that these cardinalities do not conflict with 
	each other and there exists an actual data set that satisfy all them
	above.
	Under this setting, it can be calculated that $C_{emm}(P_{H}) < C_{emm}(P_{GF}) < C_{emm}(P_{RF})$
	which show that accessing the relational data in the middle 
	is more efficient than considering these two part of data separately.
}
\end{example}

\eat{
\warn{
	(It can also be verified through such a cost model,
	the intersective exploration allows the \ModelName-based
	system to out-perform the conventional \RDBMS
	which can consider query plans with lower cost.)
}
}

The above example shows that pushing the access of relational
data down to the pattern matching process can be cheaper
according to the cost model.
The experimental results in Section~\ref{sec-expt} also show 
	that $\tbf\%$ of the hybrid queries can benefit from such a
	seamless execution and can achieve $\tbf\times$ speedup on average. }
\vspace{-0.3ex}
\section{System and evaluation}
\label{sec-expt}

\warn{We carry out \SystemName, an implementation of the proposed \ModelName model, 
	to evaluate the benefits it can bring on real-world data.}

\subsection{System Implementation}
\label{subsec-system-imp}

\warn{We implemented the entire \SystemName 
extended from DuckDB~\cite{raasveldt2019duckdb}.
We introduced the exploration operator and the exploration condition into both the logical and physical plans 
as in Section~\ref{subsec-RX-model-operator},
and then integrated them into the query optimization workflow as in Section~\ref{subsec-query-optimization-with-exploration}.
We added a filter atop of the plan enumeration method~\cite{moerkotte2008dynamic} adopted in DuckDB
to resolve exploration operator from CSG-CMP pairs enumerated from the extended query graph as described
in Section~\ref{subsec-query-optimization-with-exploration}
(see~\cite{full} for the details).
We used a simple histogram to estimate the {\em ``selectivity''} of the exploration
in the optimizer, and the experiment results
show it works sufficiently well.

For physical plan generation, we utilized a simple heuristic approach 
which will generate hash (intersective) exploration operator only if the base-side 
relation is a pipeline intermediate result and the explored pointers are not 
point to single tuple;
the nested-loop (intersective) exploration will be considered otherwise.
We integrated the introduced operators into the pipelined execution engine~\cite{boncz2005monetdb}
as shown in Figure~\ref{fig-hybrid-query} 
with the size of the pipeline intermediate result set to $2048$ as in DuckDB~\cite{Raasveldt_DuckDB}.
For all extended relations, 
we uniformly considered the row-store as the physical implementation.
Meanwhile, we opted to divide the first input relation of the pipeline into several subsets
and duplicate the pipeline into multiple copies to process them in parallel to support intra-query parallelism.\eat{~\cite{schuh2016experimental}}
\looseness=-1

For entity resolution,
we integrated JedAI~\cite{papadakis2018jedai}, one of the most advanced open-sourced ER implementation we can find, using their official implementation with Java and called it from the C++-side through the JNI (Java Native Interface)~\cite{gabrilovich2001jni}
for dynamic entity resolution.
We used group linkage in the {\em entity matching}
step and unique mapping clustering in the {\em entity clustering} step
according to its characteristics following~\cite{papadakis2020three}.
\SystemName also provides a native graph interface to support imperative programming as 
shown in Figure~\ref{fig-overview}.
Based on it, we implemented the candidate set generation method in~\cite{han2019efficient}
to optimize the pattern-part of \SQLd query as
in Section~\ref{subsec-query-optimization-with-exploration}.}
\looseness=-1

\eat{
The optimizations from database community on query execution~\cite{boncz2005monetdb,menon2017relaxed,neumann2011efficiently}, 
	a stage rarely considered by the research of graph pattern matching algorithms, 
	can also bring advantages for \SystemName to exceed them.
	As the pipelined execution shown in Example~\ref{fig-hybrid-query},
	the edges connects to a vertex can be iterated continuously
	to further take an advantages of the locality of graph data compared to the native graph	pattern matching
	algorithms~\cite{yang2021huge,bhattarai2019ceci,han2019efficient,
		huang2014query,cordella2004sub} whose workflow basically coincides with 
	the volcano model~\cite{graefe1993volcano} for query
	evaluation in \RDBMS.

\vspace{0.6ex}
Below we highlight the key
features in implementing this logical
process with the physical design in Section~\ref{subsec-physical-implementation}. 

\etitle{Implementation}. \revise{We implement the procedure based on
	the physical design of Section~\ref{subsec-physical-implementation} as follows}.
Since the small fragments are \revise{stored as segments
	in a} continuous 
memory space (Section~\ref{subsec-physical-implementation}),
\revise{\SystemName conducts  segment reallocation when
	the inserted tuples make fragments larger 
	than the capacity of their} segments.
\revise{In particular}, 
if the size of an enlarged fragment
reaches the threshold, \SystemName 
splits it into blocks and applies block-based 
management instead. \revise{This strategy ensures} that
only small enough fragments are managed as 
segments, making the reallocation cost
under control.

When a fragment \revise{is} reallocated to another position,
\SystemName updates its base
address in the segment table  for
indirect pointer implementation.
The segment address can still
be obtained by the old id and offset.
In contrast, the recorded starting address must be
adjusted in response to every
segment reallocation for indirect
pointer implementation. \revise{To this end, 
	direct pointers
	can be coupled with indirect ones
	to strike a balance between 
	memory usage and cost}. \looseness=-1}

\vspace{-1.6ex}
\subsection{Experimental Study}
\label{subsec-expt-study}

\vspace{-0.4ex}
    
\eat{
Using real-life and synthetic data, we empirically evaluated \SystemName for
its (a) efficiency, (b) scalability and (c) effectiveness
in enrichment.} 
\warn{
Using real-life datasets\eat{ and massive queries}, we empirically evaluated \SystemName for
its (a) efficiency, (b) scalability and (c) effectiveness
in enrichment.}\looseness=-1

\vspace{-0.8ex}
\stitle{Experimental setting}. We start with the experimental settings.

\vspace{-0.4ex}
\etitle{Datasets}.
\warn{We used three real-life graphs:
(1) \dblp~\cite{DBLP}, a real-life citation network
with $0.2$M vertices and $0.3$M edges, 
constituting bibliographic records of research papers in computer science;
(2) \yago~\cite{mahdisoltani2013yago3}, a knowledge graph with
$3.5$M vertices and $7.4$M edges; 
(3) \imdb~\cite{imdb}, a graph database  
that includes attributes for the information of movies, directors
and actors, having $5.1$M vertices and $5.2$M edges.
(4) \movie~\cite{movie}, a rating network with $25$M of 
0.0\textasciitilde5.0 rating from $0.16$M\eat{162541} users to $59$K\eat{59047} movies,
each \emph{user} vertex holds a unique $uid$
and each \emph{movie} vertex holds its $title$ and $genres$.
And two real-life relations:
(1) \dbpedia relations~\cite{dbpediag2t}, 
    where the \at{Celebrity}~\cite{dbpediaAthelete,dbpediaPolitician} intersects with the peoples in \yago 
    and the \at{Journal}~\cite{dbpediaAcademicJournal} intersects with the journals in \dblp;
(2) The structuralized nationality of movies from \linkedmdb~\cite{hassanzadeh2009linked},
    that intersects with the movies in \imdb.

\vspace{-0.2ex}
\etitle{Baselines}.
We considered four types of baselines to evaluate the performance of \SystemName:
(1) Relational systems, we utilized PostgreSQL (v13.5)~\cite{stonebraker1990implementation} and DuckDB~\cite{raasveldt2019duckdb};
(2) Native graph-based systems, we utilized Neo4J (v5.6.0)~\cite{neo4j} and GraphFlow~\cite{kankanamge2017graphflow};
(3) Native graph-based algorithms, we utilized CECI~\cite{bhattarai2019ceci}, DP-iso~\cite{han2019efficient}, 
    TurboFlux~\cite{Kim2018TurboFlux}, 
    IEDyn~\cite{Idris2017Dynamic} and 
    SymBi~\cite{Min2021Symmetric}.
    Following the recommendation in~\cite{sun2020memory}, 
    we also combined the candidate vertex computation method of 
    GraphQL~\cite{GQ} with the ordering methods and auxiliary data structures of DP-iso as another baseline,
    denoted as GQL-DPiso;
(4) Relational system with pattern matching-oriented optimization, 
    we utilized GRainDB~\cite{jin2021making}
    which is not merely acting as an interface atop of \RDBMS for query mapping.
    
For DuckDB, we used their official open-source C++ implementation~\cite{Raasveldt_DuckDB}.
The rest baselines, GraphFlow, CECI, DP-iso, GQL-DPiso, TurboFlux, IEDyn and SymBi,
we directly utilized the C++ implementations in~\cite{sun2022depth, sun2020memory}
as a fair comparison.

\vspace{-0.2ex}
\etitle{Configuration}.
The \SystemName and other baselines implemented in C++ are all compiled with GCC-12
and are set to use single thread if not specially mentioned. 
The experiments were run on a desktop with 3.2GHz processor and 16 GB memory.}
\looseness=-1

\vspace{-0.8ex}
\stitle{Experimental results}. We next report our findings.

\begin{figure}[t]
\centerline{\includegraphics[scale=0.875]{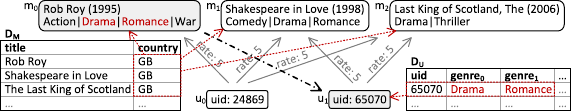}}
\centering
\vspace{-2.4ex}
\caption{A case study \warn{on} the movie recommendation task}
\label{fig-case-study}
\vspace{-4ex}
\end{figure}

\vspace{-0.4ex}\warn{
\stitle{Exp-1: Case study}.
We first validated the need 
to enrich relational data (resp. graph) with graph properties (resp. relational data)
all in \RDBMS.
Considering movie recommendation over datasets:
(1) Graph data \movie, denoted as $G_M$.
(2) User portrait $D_U(uid, genre_0, \dots, genre_2)$ of their top three interesting genres.
    Its field $uid$ aligns with the $uid$ of user vertices in $G_M$
    and is not merely an equivalent relational representation of $G_M$.
(3) $D_M(title, country)$ from \linkedmdb of the title and country of each movie, 
    the value of title here cannot directly match with the $title$ property 
    of movie vertex in $G_M$, as they can be stored in different forms and errors often occur.
 
As shown in Figure~\ref{fig-case-study}, 
\SystemName can recommend movie $m_0$\eat{(Rob Roy (1995))} to user $u_1$\eat{(uid: 65070)} according to: 
(1) $u_0$ (resp. $u_1$) rates $m_0$, $m_1$ and $m_2$ (resp. $m_1$ and $m_2$) as $5.0$;
(2) $m_0$, $m_1$ and $m_2$ all from the same country\eat{(GB)}, according to the data enhanced from $D_M$;
(3) $m_0$ is marked with two genres, \emph{Drama} and \emph{Romance}, two of the $u_1$ favourites
    according to the portrait of it from $D_U$.
\warn{This} task cannot be fairly achieved by:
(1) Various data enrichment or graph data extraction methods, 
    they are not developed to extract data from graph strictly 
    according to the pattern specified by users;
(2) \RDBMS or other relation-based polyglot systems, as their performance does not allow
    them to query graph data with such a complicated pattern \warn{(>10hr in DuckDB)};
(3) Native graph-based systems, the result of graph query \warn{(>$10^{10}$ matches)}
    needs to be exported into the \RDBMS for further relational processing
    with the data in it;
(4) Static offline entity resolution, since both $D_M$ and $G_M$ are updated frequently, 
	which can invalidate the precomputed result\eat{
    and much information users are not interested about would also be involved};
(5) Combinations of the above methods, 
    which fail to optimize the hybrid query in a unified space,
    not even to mention the cost of data movement and format conversion.}
\looseness=-1

\begin{figure*}[tb!]
\vspace{-4ex}
\begin{center}
	\vspace{-6ex}
	\begin{minipage}[t]{\textwidth}
		\subfigure[\small \dblp: Pattern matching speedup]{\label{fig-dblp-pattern-all-baselines}
			{\includegraphics[width=6.1cm]{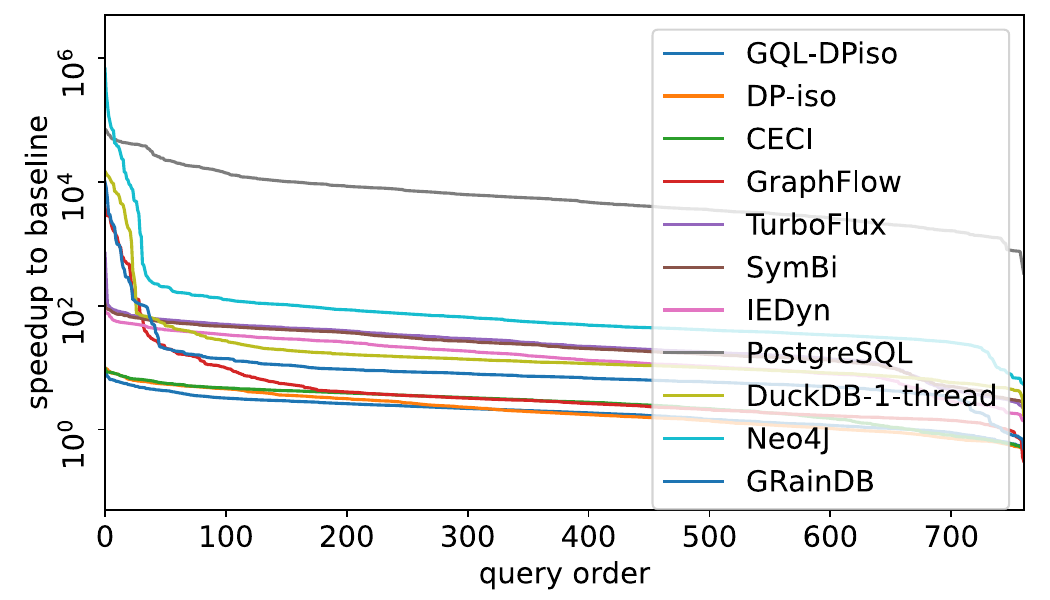}}}
		\hspace{-0.3cm} 
		\subfigure[\small \imdb: Pattern matching speedup]{\label{fig-imdb-pattern-all-baselines} 
			{\includegraphics[width=6.1cm]{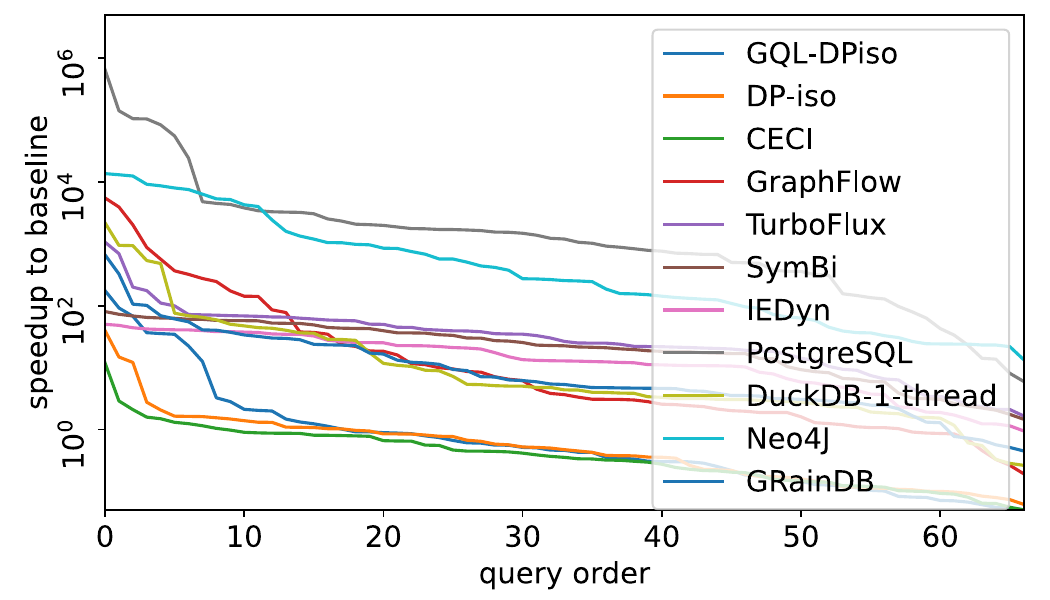}}}
		\hspace{-0.3cm} 
		\subfigure[\small \yago: Pattern matching speedup]{\label{fig-yago-pattern-all-baselines}
			{\includegraphics[width=6.1cm]{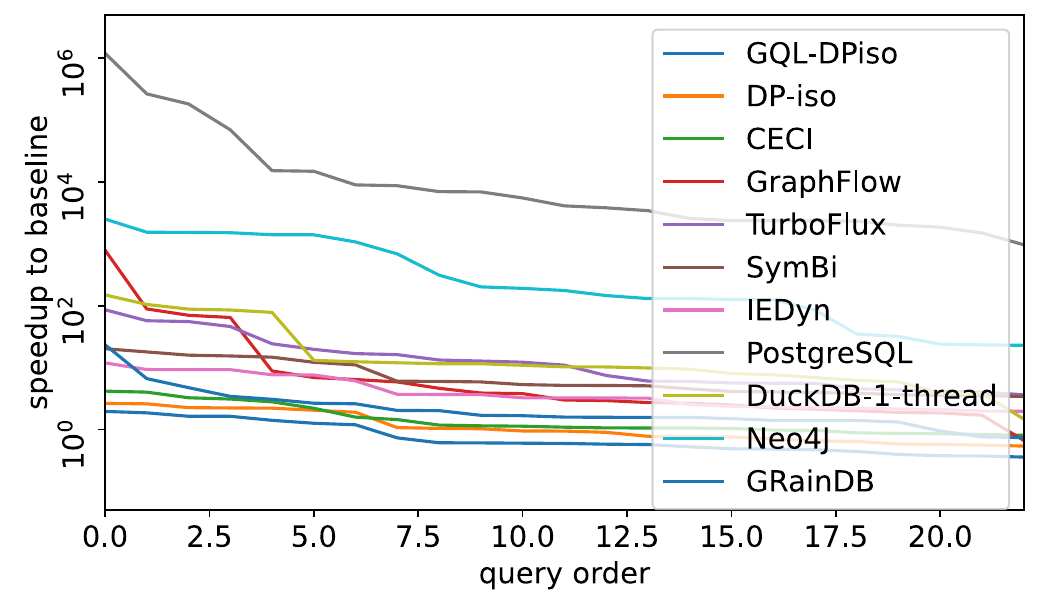}}}
	\end{minipage}
\end{center}
\vspace{-4.6ex}
\caption{Pattern query performance evaluation}
\label{fig-exp}
\vspace{-3ex}
\end{figure*}

\vspace{-0.2ex}
\warn{\stitle{Exp-2: Pattern query efficiency}.
    The most significant difference for \SystemName to depart from previous relational systems is 
    the efficiency on pattern queries by utilizing the more efficient operators
    induced by the \ModelName model.
    We first evaluated the performance of \SystemName on pattern queries compared to 
    various baselines, to prove that it can bring orders of magnitude speedup compared to 
    other relational systems and is also comparable to the most efficient
    native graph-based pattern matching algorithms.
    The patterns \warn{used} here are all discovered from their underlying data graphs as in~\cite{fan2022discovering}.

    \etitle{\warn{Comparison} to all baselines}.
    We first carried out an experiment to compare \SystemName to all kinds of baselines 
    we considered, to evaluate the efficiency on pattern query.
    Restricted by the performance of some baselines, especially the relational ones, 
    we only used a small number of simple patterns here:
    761 (8 cyclic), 
     68 (0 cyclic) and 
     23 (1 cyclic) patterns on \dblp, \imdb and \yago respectively. 
    Figures~\ref{fig-dblp-pattern-all-baselines} 
         to~\ref{fig-yago-pattern-all-baselines},
    show the speedup \SystemName brings compared to all baselines
    on all these patterns, ordered in a descending order.

    The experimental results show that \SystemName is 
    orders of magnitude more efficient than the relational baselines:
    32500x\eat{dblp: 8248.280588593952, imdb: 18395.561058897063, yago: 79231.2038182609} to PostgreSQL and 
    112x\eat{dblp: 218.91749496846282, imdb: 87.38631455882353, yago: 28.77046834782609}  to DuckDB with one thread 
    (121x 
    when both used 4-threads).
    Meanwhile, the \warn{enormous} size of intermediate result, >$100GB$ in some cases, 
    required by these relational baselines
    much beyond the typically capability of main-memory, which brought an 
    unaffordable cost for them to be written on disk and thence forbidden a large portion of queries here
    to be executed.
    \SystemName also achieved a
    34.0x\eat{dblp: 69.26740743101185, imdb: 29.70092431343285, yago: 3.0355782173913046} 
    speedup compared to GRainDB,
    as it only optimizes pattern query with sideway-information-passing.
    Although such a relatively less aggressive way optimization 
    allows it to achieve a visible performance improvement compared to DuckDB, 
    it still fails to compare with the most efficient algorithms.

    \SystemName achieved an efficiency improvement 
    compared to the native graph-based systems, 
    1640x\eat{dblp: 2601.2746619106447, imdb: 1740.2886428970585, yago: 584.3012781304349} to Neo4J and 
    106x\eat{dblp: 47.884401501971126, imdb: 222.54781610294117, yago: 47.17015482608696} to GraphFlow.
    It should be admitted that a large portion of the speedup 
    compared to Neo4J actually came from the efficiency improvement our C++-based implementation 
    brought compared to their Java-based one.
    However, for GraphFlow, although the C++-based implementation we utilized~\cite{sun2022depth} 
    eliminates such a gap, it still fails to consider the pipelined execution
    and utilize any of the candidate set selection methods.

    \SystemName achieved an efficiency improvement compared to several native graph-based algorithms: 
    36.7x\eat{dblp: 29.26887962023654,  imdb: 62.23542313235295,  yago: 18.62679030434783} to TurboFlux, 
    40.9x\eat{dblp: 18.3601074454665,   imdb: 18.03528322058824,  yago: 4.567213043478261} to IEDyn and
    20.9x\eat{dblp: 26.124743164257584, imdb: 28.477289882352938, yago: 7.857778913043477} to SymBi.
    As only a limited number of simple patterns were used here, 
    it can only be roughly seen that \SystemName has a similar performance
    compared to the most efficient baselines,
    \ie GQL-DPiso, DP-iso and CECI,
    which will be compared in details in the following experiments with a larger number of patterns.
}

\begin{figure}[tb!]
\vspace{-0.8ex}
\begin{center}
	\vspace{-4.3ex}
	\begin{minipage}[t]{\textwidth / 2}
		\subfigure[\small \dblp]{\label{fig-dblp-pattern-native-algorithms}
			{\includegraphics[width=2.9cm]{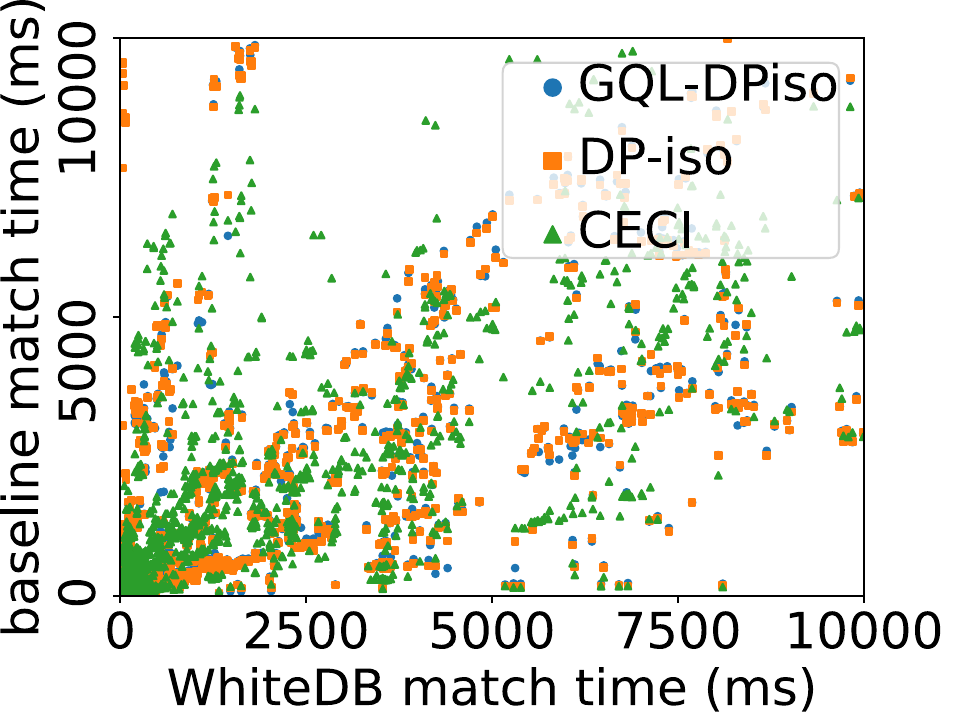}}}
		\hspace{-0.3cm} 
		\subfigure[\small \imdb]{\label{fig-imdb-pattern-native-algorithms} 
			{\includegraphics[width=2.9cm]{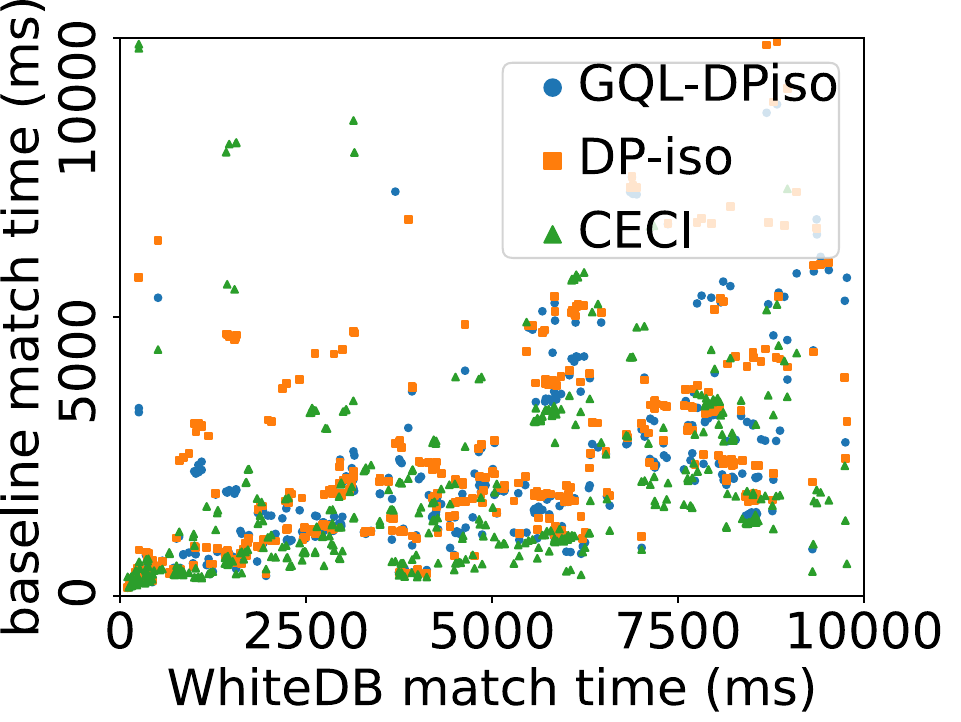}}}
		\hspace{-0.3cm} 
		\subfigure[\small \yago]{\label{fig-yago-pattern-native-algorithms}
			{\includegraphics[width=2.9cm]{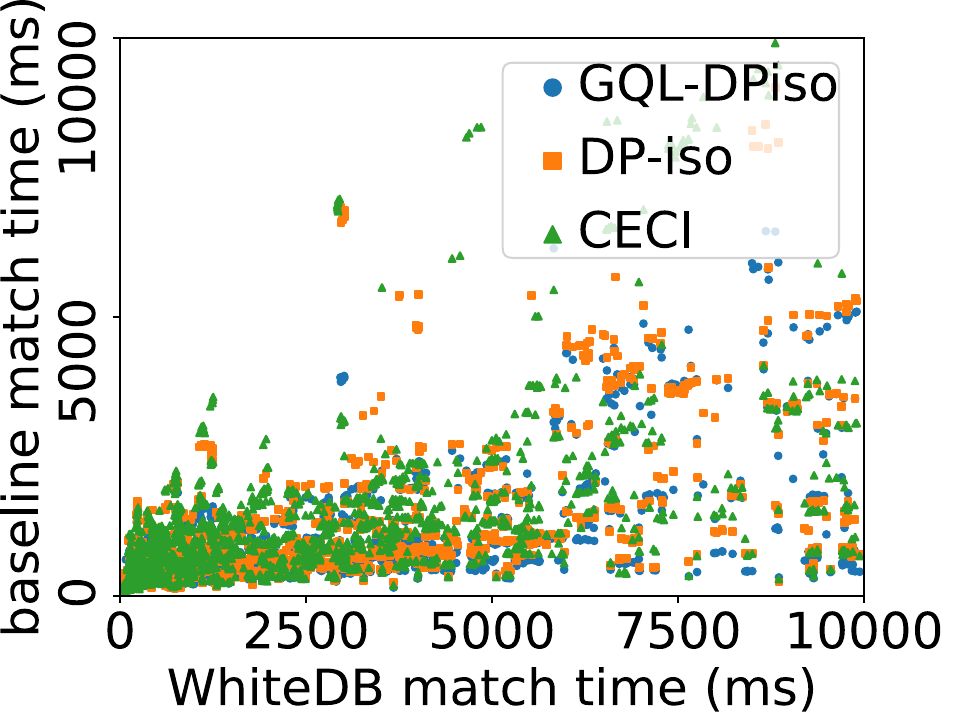}}}
		\vspace{-2ex}
	\end{minipage}
\end{center}
\vspace{-2.4ex}
\caption{Pattern matching time of \SystemName and baselines}
\label{fig-native-algo-exp}
\vspace{-4.8ex}
\end{figure}

\warn{
    \etitle{\warn{Comparison} to the most efficient baselines}.
    After proving that the implemented \SystemName can bring orders of magnitude improvement
    compared to other baselines, 
    we used a large amount of patterns to empirically evaluate its efficiency compared to 
    the most efficient baselines:
    3333 (1908 cyclic), 
    417 (257 cyclic) and 
    2485 (1445 cyclic) patterns
    on \dblp, \imdb and \yago respectively.
    Figures~\ref{fig-dblp-pattern-native-algorithms} to \ref{fig-yago-pattern-native-algorithms} 
    visually present the performance of \SystemName and the baselines on these patterns,
    where the X-axis (resp. Y-axis) represents the execution time on \SystemName (resp. baseline).
    \looseness=-1


    \warn{Generally} speaking, the experimental results 
    prove that the \ModelName supports \SystemName 
    to achieve a comparable performance compared to these most efficient native graph-based 
    pattern matching algorithms:
    1.01x\eat{dblp: 1.6162571217534185, imdb: 0.6465179632925603, yago: 0.7665271053112139} to GQL-DPiso, 
    1.20x\eat{dblp: 1.8576056508978713, imdb: 0.8266663895163926, yago: 0.9440862569400488} to DP-iso and 
    1.41x\eat{dblp: 2.4277911568346076, imdb: 0.7185907523983109, yago: 1.0818651517275377} to CECI, 
    which is even less significant than the variation among different native graph-based algorithms.
    In particular, \SystemName achieved a visible speedup
    on a portion of queries and can be very significant in some cases, that might own to: 
    (1) \warn{t}he pipelined execution method fairly deploys the locality of the graph structure 
        and is more efficient than the tuple-by-tuple one;
    (2) \warn{t}he heuristic ordering methods miss the optimal plan.
    (3) \warn{u}niformly considering the candidate set for all vertices in all queries 
        lowers the performance in some cases.
    \looseness=-1

    \SystemName failed to achieve the same performance on some patterns, which might be due to:
    (1) \warn{t}he requirement of runtime resolving the type of data and operator in \SystemName, as in other RDBMSs;
    (2) \warn{s}ome optimizations utilized in native graph-based algorithms 
    	cannot be considered in \RDBMS developed for general tasks, 
        like pre-sorting the adjacent vertex id to directly utilize the merge intersection.
    (3) \warn{t}he current implementation of \SystemName does not 
        support adaptive query execution to consider different 
        matching order in different portions of the data graph.
        We defer introducing adaptive query execution into \SystemName to future work.}

\begin{figure}[tb!]
\vspace{-0.8ex}
\begin{center}
	\vspace{-4.3ex}
	\begin{minipage}[t]{\textwidth / 2}
		\subfigure[\small \dblp $\times$ \dbpedia]{\label{fig-dblp-hybrid}
			{\includegraphics[width=2.9cm]{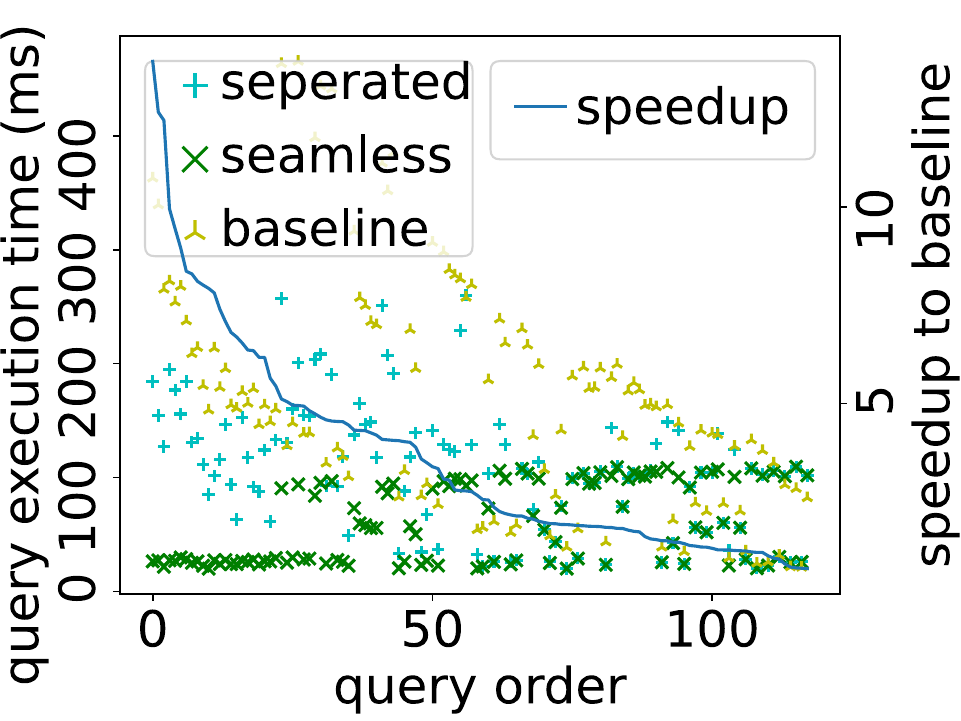}}}
		\hspace{-0.2cm} 
		\subfigure[\small \imdb $\times$ \linkedmdb]{\label{fig-imdb-hybrid}
			{\includegraphics[width=2.9cm]{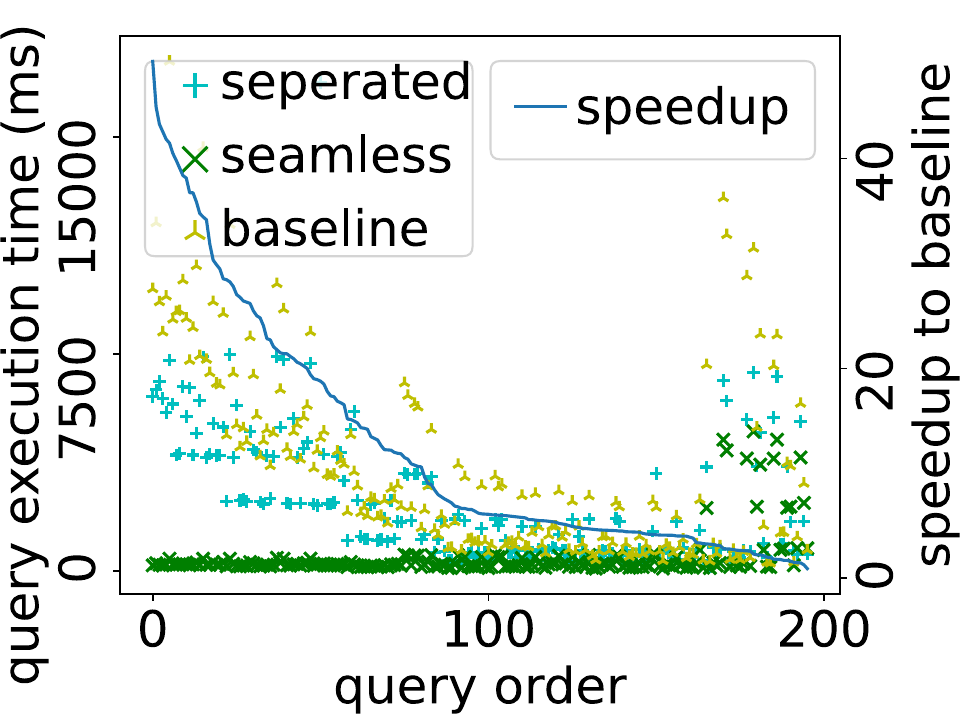}}}
		\hspace{-0.2cm} 
		\subfigure[\small \yago $\times$ \dbpedia]{\label{fig-yago-hybrid}
			{\includegraphics[width=2.9cm]{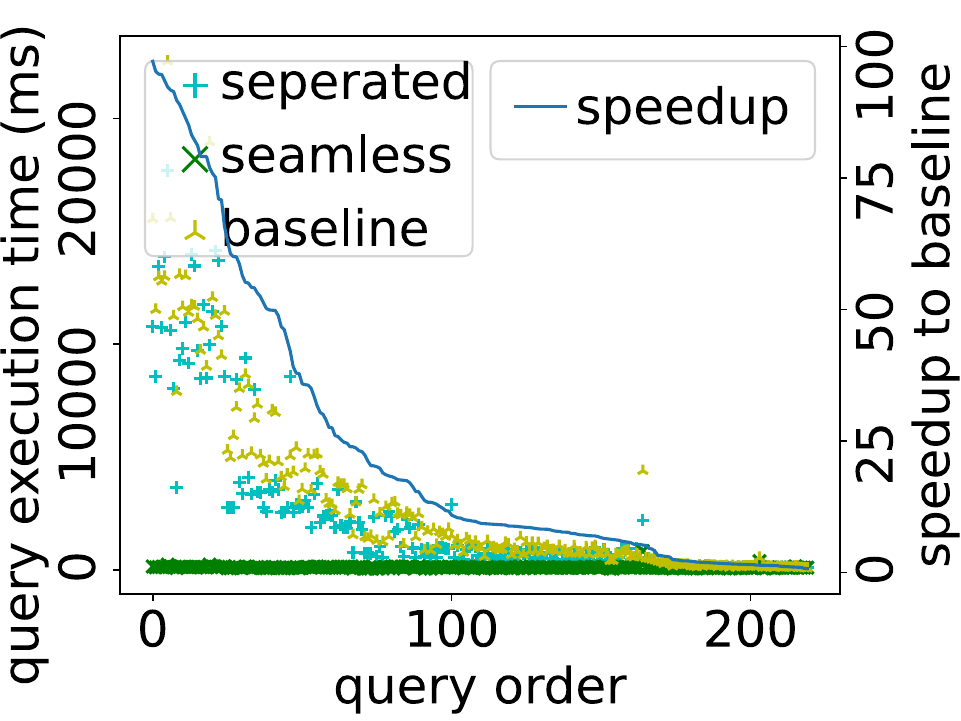}}}
		\vspace{-2ex}
	\end{minipage}
\end{center}
\vspace{-2.4ex}
\caption{Hybrid query performance}
\label{fig-hybrid-exp}
\vspace{-5.3ex}
\end{figure}

\begin{figure}[tb!]
\begin{center}
	\vspace{-3ex}
	\begin{minipage}[t]{\textwidth / 2}
		\subfigure[\small \dblp]{\label{fig-dblp-scalability}
			{\includegraphics[width=2.9cm]{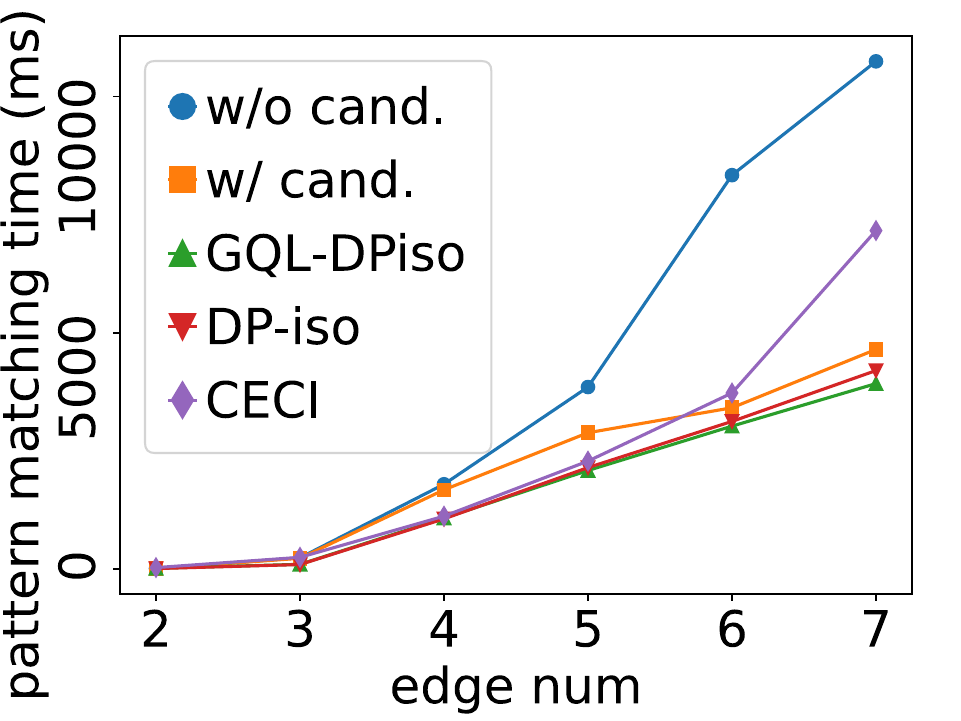}}}
		\hspace{-0.2cm} 
		\subfigure[\small \imdb]{\label{fig-imdb-scalability}
			{\includegraphics[width=2.9cm]{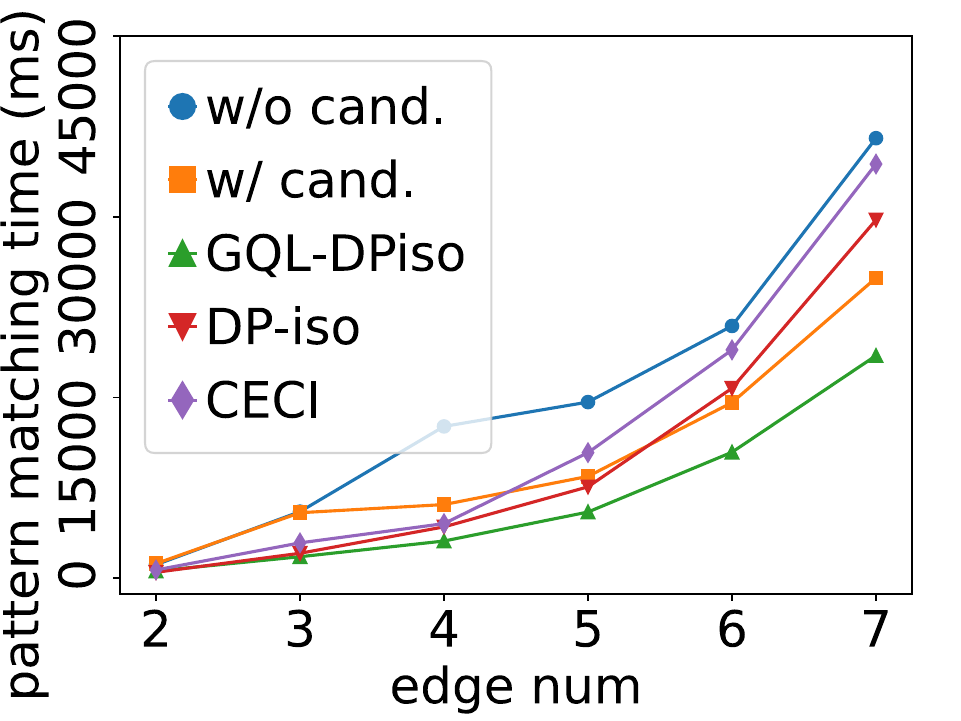}}}
		\hspace{-0.2cm} 
		\subfigure[\small \yago]{\label{fig-yago-scalability}
			{\includegraphics[width=2.9cm]{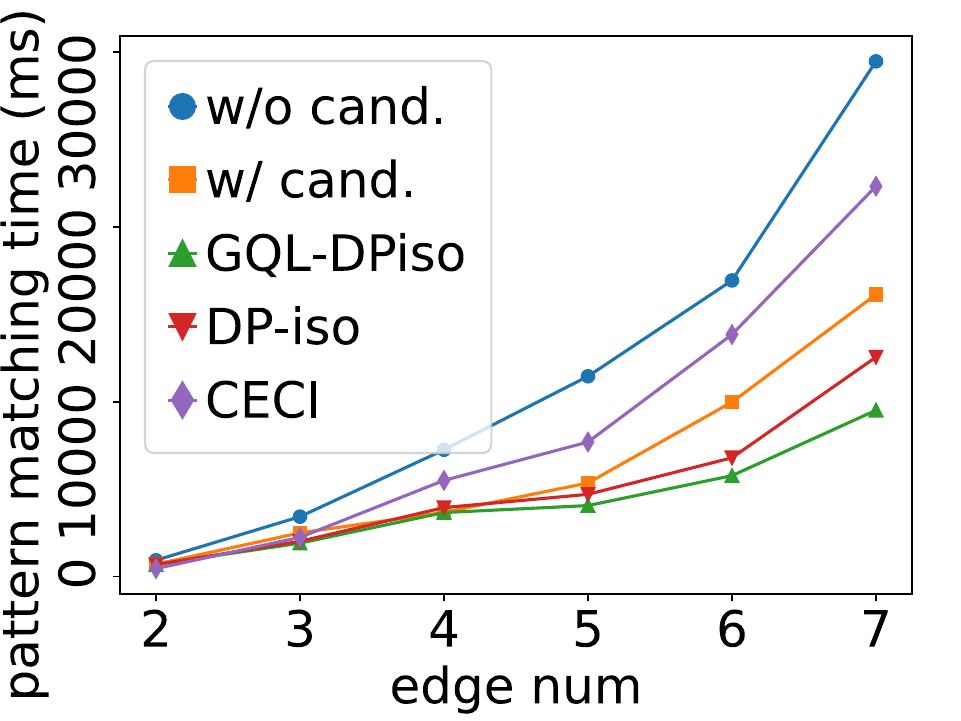}}}
		\vspace{-2ex}
	\end{minipage}
\end{center}
\vspace{-2.4ex}
\caption{Scalability of \SystemName}
\label{fig-scalability-exp}
\vspace{-4.4ex}
\end{figure}

\warn{
    \stitle{Exp-3: Hybrid query}.
    We evaluated the performance improvement \SystemName~\warn{achieved} on hybrid queries.
    For each pattern, we selected two vertices from it for them 
    to join on the corresponding relational data,
    and thence \warn{resulting in} 184, 254 and 207 hybrid queries on 
    $\text{\dblp}\times$\dbpedia,
    $\text{\imdb}\times$\linkedmdb and 
    $\text{\yago}\times$\dbpedia respectively.
    We considered the baseline as:
    (1) utilize GQL-DPiso to estimate the cost of pattern-part of the \SQLd query;
    (2) store the entire match results into DuckDB;
    (3) complete the rest relational processing,
    and added the time they consume together.
    Further, we generated two kinds of query plans for hybrid query in \SystemName:
    (a) \emph{Separated}, we first generated the optimized plan for the pattern part
    of the \SQLd query and then added additional joins on the external relations 
    on top of it. 
    Such a plan can be executed all inside \SystemName
    without storing the intermediate result into another separated engine.
    However, it still needs to first complete the entire pattern matching process
    and failed to make a full use of the capability for hybrid query optimization in \SystemName;
    (b) \emph{Seamless}, we further optimized the \emph{``separated''} hybrid query 
    plan within the unified enlarged plan space to pull the low selectivity joins down
    into the pattern matching process.
    
    Figures~\ref{fig-dblp-hybrid} to \ref{fig-yago-hybrid} 
    demonstrate their performance on those hybrid queries, 
    sorted in a descending order of the speedup 
    \emph{``seamless''} achieved compared to the baseline.
    The experimental result shows that \SystemName 
    achieved a visible speedup (>10x) on
    7.38\eat{9/122}\%, 39.8\eat{80/201}\% and 51.4\eat{111/216}\%
    of the queries and on average achieved
    3.62x, 12.9x and 23.9x speedup on 
    $\text{\dblp}\times$\dbpedia,
    $\text{\imdb}\times$\linkedmdb and 
    $\text{\yago}\times$\dbpedia respectively.
    The cost to hold the entire match result, 
    which can easily reach billions, 
    and store them into a separated engine largely 
    slows down the performance of the baseline. 
    Among the dataset we used, the speedup it achieved on \dblp is relatively less significant,
   	which is due to the cost to complete the entire pattern matching is relatively cheap\eat{on it }
   	and the match count is also relatively small.
    More specifically, \emph{``seamless''} also achieved 
    10.1\eat{dblp: 2.2947447749843968x, imdb: 8.774702905377154x, yago 19.056762521679147x}x speedup
    compared to \emph{``separated''}
    by pulling the low selectivity join conditions
    down for early pruning on average of all data graphs,
    which shows that optimizing hybrid query in enlarged unified plan space
    contributes a large portion of the speedup. 
}

\warn{
    \stitle{Exp-4: Scalability}.
    We evaluated \SystemName with larger patterns
    and \warn{showed} the average execution time for the patterns with different number of edges
    in
    Figures~\ref{fig-dblp-scalability} 
    to \ref{fig-yago-scalability},
    where \emph{``w/o cand.''} (resp. \emph{``w/ cand.''}) 
    represents \SystemName without (resp. with) considering candidate set.
    The result shows the performance of \SystemName will not degrade
    and is still comparable to the baselines when the patterns scale to 7 edges,
    0.91\eat{dblp: 1.464846063269822, imdb: 0.8154555338114161, yago: 0.46376028563188615}x,
    1.15\eat{dblp: 1.1919907371904022, imdb: 1.5299148788491967, yago: 0.7211785712945361}x and
    1.68\eat{dblp: 2.1415563016757564, imdb: 2.110509773179692, yago: 0.7809599096145141}x
    compared to GQL-DPiso, DP-iso and CECI respectively on 
    average of all data graphs.
    Moreover, the capability for \SystemName 
    to consider the candidate set brings visible improvement for large patterns, 
    3.17\eat{dblp: 6.412197354324131, imdb: 1.0988944703810073, yago: 2.006977459721716}x
    for the patterns with 7 edges on average of all data graphs.
    It fairly proved that the proposed evaluation workflow is not only suitable to 
    small patterns.}
    \looseness=-1

\begin{table}
	\vspace{1.3ex}
	\begin{small}
		\centering
		\begin{tabular}{|c|c|c|c|c|} 
			\hline
			{\bf Graph} & {\bf Candidate set} & {\bf Hash operator} & {\bf Pipeline} & {\bf Optimizer}  \\ 
			\hline
			\dblp		&      4.10x          &        4.90x        &      3.82x     &  3.79x \\
			\hline
			\imdb	    &      1.66x          &        4.25x        &      5.64x     &  11.9x \\ 
			\hline
			\yago       &      2.33x          &        5.98x        &      4.50x      & 13.3x \\
			\hline
		\end{tabular}
	\end{small}
	\caption{Ablation study on the efficiency of pattern query}\label{tab-ablation-study}
	\vspace{-7.7ex}
\end{table}

\warn{\stitle{Exp-5: Ablation study}. 
Finally, we performed an ablation study to investigate 
how much efficiency improvement each component contributes,
with the same set of queries used in Exp-2.
We omitted the four components one-by-one to evaluate the speedup they brought.
As shown in Table~\ref{tab-ablation-study},
the candidate set, the hash operators, the pipelined execution and the optimizer 
contribute
2.70x, 
5.04x, 
4.65x  
and 
9.66x 
efficiency improvement on average of all data graphs respectively.
It can be seen that 
it is the advantage brought by the pipelined execution 
that makes up the benefits that can be brought by other optimization strategies in various native graph-based 
algorithms not considered here.}
\looseness=-1

\eat{\warn{\stitle{Exp-6: Incremental maintenance}.
The data can be maintained with the data set .}
\looseness=-1

\warn{\stitle{Exp-7: Impact of dynamic entity resolution}.}
\looseness=-1
    }






\stitle{Summary}. We find the following.
(1) Modelling graph data with the \ModelName model induces 
    an exploration operator into the relational query evaluation 
    workflow, which allows \SystemName to outperform
    other relational systems with orders of magnitude on pattern query, 
    32500x to PostgreSQL and 112x to DuckDB.
    A comprehensive experiment with \warn{a large number of} patterns also proved
    \warn{that it allows} \SystemName to achieve a comparable performance to the most efficient graph-based
    algorithms, 1.01x to GQL-DPiso, 1.20x to DP-iso and 1.41x to CECI.
(2) By optimizing the hybrid query in an enlarged unified plan space 
    and evaluate it without involving cross model access, 
    \SystemName achieves a
    13.5x\eat{3.62 + 12.9 + 23.9} speedup.
(3) \SystemName still works for large patterns, 
    0.91x
    compared to the most efficient native graph baselines as the patterns scale to 7 edges.
\vspace{-0.3ex}
\section{Related Work}
\label{sec-related}
\vspace{-0.4ex}

We categorize the related work as follows.

\vspace{-0.2ex}
\etitle{Polyglot systems}.
The performance on pattern query of the relation-based polyglot systems~\cite{Verticagraph, Vertexica, Cytosm, Jeffrey1,Jeffrey2}
\warn{is} \warn{essentially} all relatively limited 
as the topology of graph is not preserved \warn{to support an efficient exploration}\eat{,
which forbids them to consider complicated patterns}.
Some of them are actually built atop of existing relational system\warn{s} 
and provide only an intuitive query interface~\cite{DaveJLXGZ16, Vertexica, Verticagraph}
instead of bringing any performance improvement for querying graph-structured data.
Among them, GRainDB~\cite{jin2021making,Jin2022GRainDB} utilizes sideway-information-passing
to optimize pattern query but still fail\warn{s} to achieve a comparable 
performance to the most efficient native graph-based \warn{algorithms}.
The polyglot systems with the native graph-core
also somewhat \warn{rely} on runtime graph view extraction from the
relational data~\cite{NScaleSpark, GraphGen, Kaskade, DuckPGQ, DuckPGQ2023vldb}
instead of in-situ processing all in \RDBMS.
As a comparison, \SystemName evaluates both the pattern query
and the hybrid query inside the \RDBMS with minor modifications
for it to achieve comparable performance \warn{to} the 
native graph engines for pattern query.

\etitle{Multi-model stores}.
Instead of providing an \emph{``one size fit all''}~\cite{stonebraker2005one} solution,
both the multistores systems~\cite{AbouzeidBARS09,ZhuR11,KepnerABBBBCGHKMMPRRY12}
and the polystores systems~\cite{DugganESBHKMMMZ15,HalperinACCKMORWWXBHS14,KolevBVJPP16} 
\warn{store} data in different models separately \warn{to deploy the advantages they provide}.
\warn{However}, they \warn{are typically restricted with} costly cross model access,
\eg relying on \emph{``cut-set join''} to exchange intermediate results from different engines~\cite{GSQL};
restricted semantics and optimization to query graph-side data~\cite{GRFusion};
and lack of an enlarged unified plan space to optimize hybrid query.


\etitle{Data enrichment}. 
\warn{Enriching data in different formats from diverse sources is considered to be a crucial need~\cite{DE2,DE5,DE4, DE6, DE7, DE1, DE3, DatabaseBasedIntegrationTools}.
Various technologies are developed
to link aligned entities~\cite{KB1, ahmad2010link},
or to extract information from graph-side data~\cite{nestorov1998extracting, vcebiric2019summarizing}, 
from simple label of category~\cite{torisawa2007exploiting}
to structured set of data~\cite{AT1, AT2}.
Meanwhile, the structured data is also utilized to enhance graph data,
as in graph neural network feature propagation~\cite{wang2019knowledge, wang2019www},
or embedding graph entities with literals~\cite{kristiadi2019incorporating}.

In the field of data enrichment, it is also recognized as important
to provide a unified embedding space for heterogeneous entities~\cite{toutanova2015representing, pezeshkpour2018embedding, trisedya2019entity}.
However, these methods all concentrate on the semantic connection
and can lead to unnecessary ambiguousness and cost
when the user can clearly and explicitly express their queries with a \SQL-like query language.}

\eat{
\cite{pezeshkpour2018embedding} embed entities from relational and graph in a unified space,
to gain link-prediction accuracy. enrichment

\cite{trisedya2019entity}  embedding for align, uses an embedding-based model
as illustrated in Fig 1 unified vector space. alignment

\cite{toutanova2015representing} we propose a model that cap-
tures the compositional structure of tex-
tual relations, and jointly optimizes entity

\cite{KB1} Entity linking

Extract category from knowledge graph~\cite{torisawa2007exploiting}
or structured data~\cite{AT1, AT2} 

Accurate Text-Enhanced Knowledge~\cite{an2018accurate} 

\cite{kristiadi2019incorporating} combines literals, but for embedding

\cite{wang2019knowledge, wang2019www} structured feature propagation using a graph neural network

\cite{ahmad2010link} Inter-network link-prediction

\etitle{Data enrichment}. 
Various data extraction strategies from knowledge bases
can retrieve information from different data sources~\cite{AT1,AT2,SheikhpourSGC17,
	KB3,hu2018leveraging,wang2019knowledge,wang2019www,wang2020ckan,
	gesese2021survey,kristiadi2019incorporating,KB1,
	ahmad2010link,an2018accurate,toutanova2015representing,
	pezeshkpour2018embedding,torisawa2007exploiting,trisedya2019entity}
\eat{AT1,AT2,SheikhpourSGC17,
chandrashekar2014survey,
KB3,hu2018leveraging,wang2019knowledge,wang2019www,wang2020ckan,
xie2016image,gesese2021survey,kristiadi2019incorporating,KB1,
ahmad2010link,an2018accurate,toutanova2015representing,
pezeshkpour2018embedding,torisawa2007exploiting,trisedya2019entity}.
However, some of them would only extract data from graph as a black box instead of 
strictly according to the users specification
when they can exactly describe the data they want.
More importantly, they cannot consider dynamic resolution 
driven by the query from the users.
They also fail to be directly evaluated inside the \RDBMS
not to mention enriching graph pattern matching with relational predicates.
}


\vspace{-0.3ex}
\section{Conclusion}
\label{sec-conclude}
\vspace{-0.4ex}

\warn{
    We have introduced minor extensions to \RDBMS
    to efficiently support pattern query and hybrid query,
    consisting of: 
    (1) an extended relational model to represent graph in relations
        while preserving the topology of graph data
        without breaking the physical data independence,
    (2) an introduced operator for exploring through the graph data, and
    (3) minor extensions to the \RDBMS optimizer for hybrid 
        query evaluation without modifying the overall workflow.
    We have experimentally verified that those extensions
    allow \SystemName to achieve a comparable performance compared to 
    the state-of-the-art native graph-based algorithms
    and can pull the join on relations down to the pattern matching process
    for further accelerating the hybrid query.
}

One topic for future work is to extend \SystemName to support
graph traversal queries, \warn{adaptive query execution and efficient transection processing}.\eat{Another topic is to revisit
entity resolution and conflict resolution by enriching
relational data with graph properties.}
\warn{Another topic is to consider some assumptions of the entity resolution 
for it to move across the query tree.}

\balance

\clearpage
\balance
%
{
\bibliographystyle{ACM-Reference-Format}
\bibliography{paper}
}

\balance
\clearpage
\balance

\section*{Appendix}

\renewcommand{\thesubsection}{\Alph{subsection}}

\subsection{Extended Query Graph Definition}

We provide a formal definition of the extended query graph here,
based on its counterpart in~\cite{moerkotte2008dynamic}.

\stitle{Extended query graph}.
Given a {\em \ModelName schema} $\R^E = (R_1^E, \ldots, R_m^E)$.
The extended query graph is defined as $Q = (N, E, L, F)$, where:
(1) $N \neq \emptyset$ denotes the set of nodes in it;
(2) $E \subseteq  \mathcal{P}(N)  \times \mathcal{P}(N)$ denotes the set of hyperedges, 
	($\mathcal{P}$ for power set), 
	where $\forall e = \langle n_s, n_d \rangle \in E$, 
	there are $n_s, n_d \subseteq N$ and $n_s \cap n_d = \emptyset$;
(3) $L : E \ra \{\kw{explore},  \kw{value}, \kw{duplciate}\} \times \{ A_{ij}\}$ 
	that associates each edge with a label to denote its type and attribute (after rename) of the vertices it connects;
(4) $F : E \ra N \times N$ that maps each exploration edge to a pair of vertices
	$\langle n_s, n_d \rangle$ that denotes the source and destination vertex respectively
	where $n_s \neq n_d$.

Intuitively, along with the set of node and hyperedge in~\cite{moerkotte2008dynamic},
each hyperedge here is associated with a label with $L$ in addition.
$L$ marks the type of the edge, \ie to be an exploration, plain value condition, or a duplicated (``same-pointer'') condition;
and an attribute for the join condition.
Meanwhile, although the exploration edge is conceptually defined as directed edge,
such a {\em ``direction''} is actually equivalently represented by a function $F$.
Therefore, the query graph for the conventional relational query is only extended with 
two functions here and thus retains the definition of CSG and CSG-CMP pair in~\cite{moerkotte2008dynamic}.

As the all plan enumeration methods 
considering only the topology of the hypergraph,
they can then be directly adopted here to enumerate 
hybrid query plan from the extended query graph.
Meanwhile, as a simplification of the description,
we also utilized a specified notation for the exploration edge below:
\vspace{-0.4ex}

\stitle{Directed edge notation}.
Given a {\em \ModelName schema} $\R^E = (R_1^E, \ldots, R_m^E)$
and an extended query graph $Q = (N, E, L, F)$.
For an attribute $A_p$ in $\R^E$, node $n_s, n_d \in N$ and $n_s \neq n_d$,
we say there is an edge $( n_s,  A_p, n_d )$ in $Q$,
it means there exists $e = \{ n_s \} \times \{ n_d \} \in E$, such that:
(1) $L(e) = \langle \kw{explore},  A_p \rangle$;
(2) $F(e) = \langle  n_s, n_d \rangle $.

Informally, for a node $n$ in the query graph, we use $N(n)$ and $N_E(n)$
to denote its neighbouring nodes and edges respectively;
for an edge $e$, we use $N(e)$ to denote its neighbouring vertices.
As a simplification,
we consider a query graph-based representation
of intermediate result based on 
the notation of subgraph in~\cite{moerkotte2008dynamic}.

\vspace{-0.4ex}

\stitle{Intermediate result}.
Given a query graph $Q = (N, E, L, F)$
and a subset of vertices $N' \subseteq N$.
We use $R_{N'}$ to denote the intermediate join result 
corresponds to the sub-query graph $Q|_{N'}$.
\vspace{-0.4ex}

\subsection{Logical Plan Parsing}

We provide a detailed demonstration of the 
parsing process in Figure~\ref{fig-hybrid-query-parse} from \SQLd query $\kw{Q_{\delta}}$
to a logical plan $\text{Plan}_\text{P}$,
which was omitted in Section~\ref{sec-eval}:

\begin{enumerate}[leftmargin=4ex]
\item 
Parse the edge: $(v2)-[e2]\to(v0)$ into two subsequent exploration operators:
$D_V^2 \explore{out_L} D_o^2$ and $R_{CSG_{C0}} \explore{D_o^2.dst_L} D_V^0$,
where $CSG_{C0} = \{D_V^2, D_o^2\}$.

\item 
Parse the edge: $(v2)-[e1]\to(v1)$ into two subsequent exploration operators:
\\
$R_{CSG_{C1}} \explore{D_V^2.out_L} D_o^1$ and 
$R_{CSG_{C2}} \explore{D_o^1.dst_L} D_V^1$\\
where $CSG_{C1} = \{D_V^2, D_o^2, D_V^0\}$
and   $CSG_{C2} = \{D_V^2, D_o^2, D_V^0, D_o^1\}$.
\vspace{0.5ex}

\begin{figure}[H]
	\vspace{-2ex}
	\centerline{\includegraphics[scale=1.0]{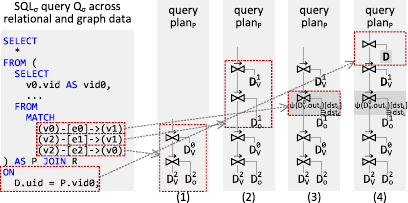}}
	\centering
	\vspace{-2.4ex}
	\caption{A step-by-step demonstration of the parsing process from input \SQLd query to a logical query plan}
	\label{fig-hybrid-query-parse}
	\vspace{-3ex}
\end{figure}

\begin{figure}[H]
\vspace{-1.4ex}
\centerline{\includegraphics[scale=1.0]{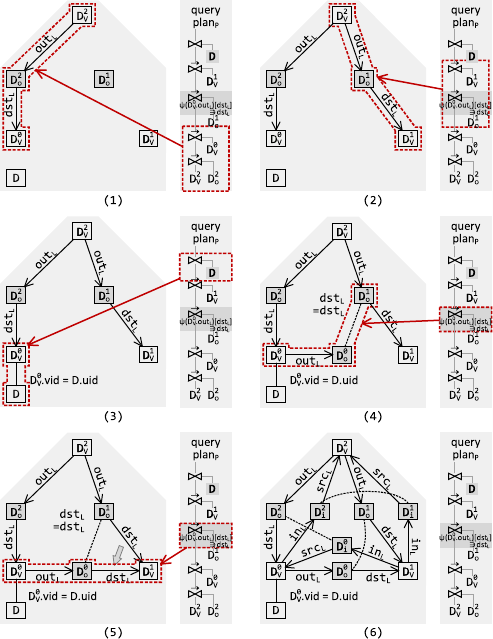}}
\centering
\vspace{-2.4ex}
\caption{A step-by-step demonstration of the query graph construction process from an input logical query plan}
\label{fig-hybrid-query-construct}
\vspace{-2.4ex}
\end{figure}

\item 
As both vertex $v0$ and $v1$ have already been considered in the query plan,
the edge: $(v0)-[e0]\to(v1)$ will be parsed as 
an exploration condition:
$\psi (D_V^0.out_L)[dst_L] \ni dst_L$
and be incorporated into the operator:
$R_{CSG_{C1}} \explore{D_V^2.out_L}_{\psi (D_V^0.out_L)[dst_L] \ni dst_L} D_o^1$.

\item 
Parse the join on the relation $D$ as $R_{CSG_{C3}} \bowtie_{D_V^0.vid = uid} D$
on the top of the query plan generated so far, were
$CSG_{C3} = \{D_V^2, D_o^2, D_V^0, D_o^1, D_V^1\}$.
\end{enumerate}

\vspace{-0.4ex}
\subsection{Query Graph Construction}
\label{appendix-query-graph-construct}

We provide a detailed, step-by-step demonstration in Figure~\ref{fig-hybrid-query-construct} 
of the query graph construction process from $\text{Plan}_\text{P}$:

{
\begin{enumerate}[leftmargin=4ex]
\setcounter{enumi}{-1}
\item 
Each leaf node in the query plan will be constructed as 
a node in the query graph:
$D$, $D_V^1$, $D_o^1$, $D_V^0$ $D_o^2$ and $D_V^2$.

\item 
An edge $(D_V^2, out_L, D_o^2)$ will be constructed 
corresponding to the exploration operator: $D_V^2 \explore{out_L} D_o^2$.
Following that, an edge $(D_o^2, dst_L, D_V^0)$ will be constructed 
corresponding to the exploration operator: $R_{CSG_{C0}} \explore{D_o^2.dst_L} D_V^0$.

\item 
An edge $(D_V^2, out_L, D_o^1)$ will be constructed 
corresponding to the exploration operator: 
$R_{CSG_{C1}} \explore{D_V^2.out_L} D_o^1$.
Following that, an edge $(D_o^1, dst_L, D_V^1)$ will be constructed 
corresponding to the exploration operator: 
$R_{CSG_{C2}} \explore{D_o^1.dst_L} D_V^1$.

\item 
$D_V^0$ and $D$ will be connected with an edge that denotes the join:
$D_V^0 \bowtie_{vid = uid} D$.

\item 
Corresponding to the condition: 
$\psi (D_V^0.out_L)[dst_L] \ni dst_L$,
a node $D_o^0$ will be first inserted into the query graph
and an edge $(D_V^0, out_L, D_o^0)$ will be constructed.
As $\D_G^E$ is in the regular form, it can be looked up that $D_o^0$
is an alias of $D_o$ as it is explored from the field $D_V^0[out_L]$.
$D_o^0$ and $D_o^1$ will be further connected by an edge 
denoting the ``same-pointer condition'': $D_o^0 \bowtie_{dst_L = dst_L} D_o^1$.

\item 
According to the ``same-pointer condition'' $D_o^0 \bowtie_{dst_L = dst_L} D_o^1$,
we know that $D_o^0[dst_L]$ and $D_o^1[dst_L]$ point to the same tuple.
Considering that there already exists an edge $(D_o^1, dst_L, D_V^1)$,
an edge $(D_o^0, dst_L, D_V^1)$ will be constructed in the query graph.

\item 
All paths in the query graph that have the patterns:
\vspace{-0.8ex}{
\smaller
\begin{align*}
(D_V)-[out_L]\to(D_o)-[dst_L]\to(D_V)\\
(D_V)-[in_L]\to(D_i)-[src_L]\to(D_V)
\end{align*}
}
\vspace{-3.8ex}

will be recognized and duplicated oppositely.
The rest ``same pointer'' conditions will then be completed,
\ie $D_o^2[dst_L] = D_i^0[src_L]$ and 
$D_i^2[src_L] = D_i^1[src_L]$.
\end{enumerate}
}

\subsection{Plan Enumeration}

The ``exploration'' is represented as directed edge
with an additional function $F$ in the extended query graph,
for it to be still treated as an ordinary hyperedge in the plan enumeration process. 
Meanwhile, the ``same pointer condition'' will not be considered in the plan
enumeration for exploring neighbouring nodes.
That is, the edges with type ``\kw{duplicate}'' will be omitted when computing $N(v)$ and $N_E(v)$. 
They will only be used to resolve explorative condition.

However, as also mentioned in Section~\ref{subsec-query-optimization-with-exploration},
there are several additional restrictions on the enumerated CSG-CMP pair for them to be resolves as exploration operator, 
like ``duplicated nodes will not appear simultaneously''.
In this case, the CSG-CMP pairs exported by previous plan enumeration methods we directly apply here
can be illegal and break these restrictions.
Here, we simply add a filter to verify the CSG-CMP pairs
to get the legal ones without touching the details of the plan enumeration methods.

\begin{myfloat}[t]
	\hspace*{-1ex}
	\begin{minipage}{0.478\textwidth}
		\removelatexerror
		\IncMargin{0.2em}
		\LinesNotNumbered
		{\small
			\begin{algorithm}[H]
				\SetArgSty{textrm}
				\caption{Legal CSG-CMP pair}
				\label{legal-CSG-CMP}
				\Indentp{-2.5ex}
				\Indentp{1.4em}
				\textbf{Input:} Query graph $Q = (N, E, L, F)$, $CSG_0, CSG_1 \subseteq N$, and $D$.
				
				\textbf{Output:} Boolean value $legal$.
				
				// first check whether there are duplicated nodes
				
				\nextnr \For{$n_0 \in CSG_0$}{
					\nextnr \For{$n_1 \in CSG_1$}{
						\nextnr \If{$\langle n_0, n_1 \rangle \in D$ \Or $\langle  n_1, n_0 \rangle \in D$}{
							\nextnr \Return \False
						}
					}
				}
			
				// then check whether these two CSGs are connected by exploration edge
				
				\nextnr $e_e = null$  // to hold the exploration edge if it exists
				
				\nextnr \For{$v \in CSG_0 \cup CSG_1$}{
					\nextnr \For{$e \in N_E(v)$}{
						
						\nextnr $\langle l, A \rangle = L(e)$  // to find exploration edge connects them
						
						\nextnr \If{$l \neq explore$}{
							\nextnr \Continue
						}
					
						\nextnr $\langle n_s, n_d \rangle = F(e)$
					
						\nextnr \If{$n_s \in CSG_0$ \And $n_d \notin CSG_1$}{
							\nextnr \Continue // inside $CSG_0$ or toward an outside vertex
						}
						\nextnr \If{$n_s \in CSG_1$ \And $n_d \notin CSG_0$}{
							\nextnr \Continue // inside $CSG_1$ or toward an outside vertex
						}
					
						\nextnr \If{$e_e \neq null$}{
							\nextnr \Return \False ~ // more than one exploration edge
						}
						
						\nextnr $e_e = e$
					}
				}
				
				\nextnr \If{$e_e = null$} {
					// the CSG-CMP pair is not connected by exploration edge
					
					\nextnr \Return \True  
				}
			
				\nextnr $\langle n_s, n_d \rangle = F(e_e)$
				
				\nextnr \If{$n_d \in \{ CSG_0 \}$ \And $|\{ CSG_0 \}| > 1$} {
					\nextnr \Return \False ~ // toward the tuples in intermediate result
				}
			
				\nextnr \If{$n_d \in \{ CSG_1 \}$ \And $|\{ CSG_1 \}| > 1$} {
					\nextnr \Return \False ~ // toward the tuples in intermediate result
				}
			
				\nextnr \Return \True
				
			\end{algorithm}
		}
	\end{minipage}
	\vspace{-2.4ex}
\end{myfloat}

Algorithm~\ref{legal-CSG-CMP} shows such a filter.
For an extended query graph $Q = (N, E, L, F)$,
it accepts a CSG-CMP pair and a set denotes the duplicated nodes in $Q$: $D \subseteq N \times N$.
As an example, for the query graph in Figure~\ref{fig-hybrid-query},\eat{
$D = \{	\langle D_o^0, D_i^0 \rangle,  
	 	\langle D_i^0, D_o^0 \rangle,  
	 	\langle D_o^1, D_i^1 \rangle,  
	 	\langle D_i^1, D_o^1 \rangle,  
	 	\langle D_o^2, D_i^2 \rangle,  
	 	\langle D_i^2, D_o^2 \rangle \}$.}
$D = \bigcup_{k=0,1,2} \{	\langle D_o^k, D_i^k \rangle\}$.
It can then return a boolean value to select all legal CSG-CMP pairs for operator resolution.
\looseness=-1

The example below shows that enumerating query plan in 
such an enlarged unified space allows the 
access on graph and relational data to be considered in any order. 

\begin{example}
We next provide a detailed step-by-step explanation 
to show how the query plans $\text{Plan}_\text{RF}$,  $\text{Plan}_\text{GF}$ and $\text{Plan}_\text{H}$ in Figure~\ref{fig-hybrid-query}
can be enumerated.
\vspace{0.1ex}

\stitle{$\text{Plan}_\text{RF}$:}
\begin{enumerate}[leftmargin=4ex]
\item $CSG_{R0} \text{=} \{D_V^0\} \times CSG_{R1} \text{=} \{D\}$.

Resolve to operator: $D_V^0 \bowtie_{vid = uid} D$.

As a plain value join operator.

\item $CSG_{R2} \text{=} CSG_{R0} \cup CSG_{R1} \text{=} \{D_V^0, D\} \times CSG_{R3} \text{=} \{D_o^0\}$.

Resolve to operator: $R_{CSG_{R2}} \explore{D_V^0.out_L} D_o^0$.

As there is an edge: $(D_V^0, out_L, D_o^0)$ from $CSG_{R2}$ toward $CSG_{R3}$.

\item $CSG_{R4} \text{=} CSG_{R2} \cup CSG_{R3} \text{=} \{D_V^0, D, D_o^0\} \times CSG_{R5} \text{=} \{D_V^1\}$.

Resolve to operator: $R_{CSG_{R4}} \explore{D_o^0.dst_L} D_V^1$.

As there is an edge: $(D_o^0, dst_L, D_V^1)$ from $CSG_{R4}$ toward $CSG_{R5}$.

\item $CSG_{R6} \text{=} CSG_{R4} \cup CSG_{R5} \text{=} \{D_V^0, D, D_o^0, D_V^1\} \times CSG_{R7} \text{=} \{D_i^1\}$.

Resolve to operator: $R_{CSG_{R6}} \explore{D_V^1.in_L}_{ \psi (D_V^0.in_L)[src_L] \ni src_L} D_i^1$.

As there is an edge: $(D_V^1, in_L, D_i^1)$ from $CSG_{R6}$ toward $CSG_{R7}$,
and a condition $D_i^2[src_L] \text{=} D_i^1[src_L]$ exists
and an edge $(D_V^0,  in_L, D_i^2)$ explores from $CSG_{R6}$ to $D_i^2$,
an explorative condition, $\psi (D_V^0.in_L)[src_L] \ni src_L$, can then be incorporated into such 
an exploration.

\item $CSG_{R8} \text{=} CSG_{R6} \cup CSG_{R7} \text{=} \{D_V^0, D, D_o^0, D_V^1, D_i^1\} \times CSG_{R9} \text{=} \{D_V^2\}$.

Resolve to operator: $R_{CSG_{R8}} \explore{D_i^1.src_L} D_V^2$.

As there is an outward edge: $(D_i^1, src_L, D_V^2)$ from $CSG_{R8}$ toward $CSG_{R9}$.
\end{enumerate}

\stitle{$\text{Plan}_\text{H}$:}

\begin{enumerate}[leftmargin=4ex]
\item $CSG_{H0} \text{=} \{D_V^0\} \times CSG_{H1} \text{=} \{D_o^0\}$.

Resolve to operator: $D_V^0 \explore{out_L} D_o^0$.

As there is an edge: $(D_V^0, out_L, D_o^0)$ from $CSG_{H0}$ toward $CSG_{H1}$.

\item $CSG_{H2} \text{=} CSG_{H0} \cup CSG_{H1} \text{=} \{D_V^0, D_o^0\} \times CSG_{H3} \text{=} \{D_V^1\}$.

Resolve to operator: $R_{CSG_{H2}} \explore{D_o^0.dst_L} D_V^1$.

As there is an edge: $(D_o^0, dst_L, D_V^1)$ from $CSG_{H2}$ toward $CSG_{H3}$.

\item $CSG_{H4} \text{=} CSG_{H2} \cup CSG_{H3} \text{=} \{D_V^0, D_o^0, D_V^1\} \times CSG_{H5} \text{=} \{D\}$.

Resolve to operator: $R_{CSG_{H4}} \bowtie_{D_V^0.vid = uid} D$.

As a plain value join operator.

\item $CSG_{H6} \text{=} CSG_{H4} \cup CSG_{H5} \text{=} \{D_V^0, D_o^0, D_V^1, D\} \times CSG_{H7} \text{=} \{D_i^1\}$.

Resolve to operator: $R_{CSG_{H6}} \explore{D_V^1.in_L}_{ \psi (D_V^0.in_L)[src_L] \ni src_L} D_i^1$.

As there is an edge: $(D_V^1, in_L, D_i^1)$ from $CSG_{H6}$ toward $CSG_{H7}$,
and a condition $D_i^2[src_L] \text{=} D_i^1[src_L]$ exists
and an edge $(D_V^0,  in_L, D_i^2)$ explores from $CSG_{H6}$ to $D_i^2$,
an explorative condition, $\psi (D_V^0.in_L)[src_L] \ni src_L$, can then be incorporated into such 
an exploration.

\item $CSG_{H8} \text{=} CSG_{H6} \cup CSG_{H7} \text{=} \{D_V^0, D_o^0, D_V^1, D, D_i^1\} \times CSG_{H9} \text{=} \{D_V^2\}$\eat{.}
\looseness=-1

Resolve to operator: $R_{CSG_{H8}} \explore{D_i^1.src_L} D_V^2$.

As there is an edge: $(D_i^1, src_L, D_V^2)$ from $CSG_{H8}$ toward $CSG_{H9}$.
\end{enumerate}

\stitle{$\text{Plan}_\text{GF}$:}

\begin{enumerate}[leftmargin=4ex]
\item $CSG_{G0} \text{=} \{D_V^2\} \times CSG_{G1} \text{=} \{D_o^1\}$.

Resolve to operator: $D_V^2 \explore{out_L} D_o^1$.

As there is an edge: $(D_V^2, out_L, D_o^1)$ from $CSG_{G0}$ toward $CSG_{G1}$.

\item $CSG_{G2} \text{=} CSG_{G0} \cup CSG_{G1} \text{=} \{D_V^2, D_o^1\} \times CSG_{G3} \text{=} \{D_V^1\}$.

Resolve to operator: $R_{CSG_{G2}} \explore{D_o^1.dst_L} D_V^1$.

As there is an edge: $(D_o^1, dst_L, D_V^1)$ from $CSG_{G2}$ toward $CSG_{G3}$.

\item $CSG_{G4} \text{=} CSG_{G2} \cup CSG_{G3} \text{=} \{D_V^2, D_o^1, D_V^1\} \times CSG_{G5} \text{=} \{D_o^2\}$.

Resolve to operator: $R_{CSG_{G4}} \explore{D_V^2.out_L}_{ \psi (D_V^1.in_L)[src_L] \ni dst_L} D_o^2$.

As there is an edge: $(D_V^2, out_L, D_o^2)$ from $CSG_{G4}$ toward $CSG_{G5}$,
and a condition $D_i^0[src_L] \text{=} D_o^2[dst_L]$ exists
and an edge $(D_V^1,  in_L, D_i^0)$ explores from $CSG_{G4}$ to $D_i^0$,
an explorative condition, $\psi (D_V^1.in_L)[src_L] \ni dst_L$, can then be incorporated into such 
an exploration.

\item $CSG_{G6} \text{=} CSG_{G4} \cup CSG_{G5} \text{=} \{D_V^2, D_o^1, D_V^1, D_o^2\} \times CSG_{G7} \text{=} \{D_V^0\}$.

Resolve to operator: $R_{CSG_{G6}} \explore{D_o^2.dst_L} D_V^0$.

As there is an edge: $(D_o^2, dst_L, D_V^0)$ from $CSG_{G6}$ toward $CSG_{G7}$.

\item $CSG_{G8} \text{=} CSG_{G6} \cup CSG_{G7} \text{=} \{D_V^2, D_o^1, D_V^1, D_o^2, D_V^0\} \times CSG_{G9} \text{=} \{D\}$.

Resolve to operator: $R_{CSG_{G8}} \bowtie_{D_V^0.vid = uid} D$.

As a plain value join operator.
\end{enumerate}
\vspace{1.0ex}

This example is only to highlight the plan enumeration process, 
it will be shown below how the operators are resolved by accepting the corresponding CSG-CMP pair as input.
\end{example}

\eat{\stitle{Intersective exploration cost}.
Considering relations $R_A[a, b]$, $R_B[b, c]$ and $R_C[c, a]$, 
and join $R_A \bowtie R_B  \bowtie R_C$.
If there are corresponding $R_A^E$, $R_B^E$ and $R_C^E$,
where $\psi(R_B^E[a_L]) \subseteq R_A^E$ and $\psi(R_B^E[c_L]) \subseteq R_C^E$,
\st $R_B^E \explore{a_L}_{\psi(c_L)[a] = a} R_A^E$ = $R_A \bowtie R_B  \bowtie R_C$.
Then there are: 
\begin{align*}
C_{emm}(R_A^E \explore{b_L}_{\psi(c_L)[c] = c} R_B^E) \leq& C_{emm}((R_A \bowtie R_B)  \bowtie R_C) \\
C_{emm}(R_B^E \explore{c_L}_{\psi(a_L)[a] = a} R_C^E) \leq& C_{emm}((R_B \bowtie R_C)  \bowtie R_A) \\
C_{emm}(R_C^E \explore{a_L}_{\psi(b_L)[b] = b} R_A^E) \leq& C_{emm}((R_C \bowtie R_A)  \bowtie R_B)
\end{align*}
Proof:

Assume all considering hash join, then there would be:\eat{
\begin{align*}
	min(&C_{emm}(R_A \bowtie R_B  \bowtie R_C)) \\
   =min(&\tau |R_A| + \tau |R_B| + \tau |R_C| + |R_A \bowtie R_B| + |R_A \bowtie R_B  \bowtie R_C|,\\
        &\tau |R_A| + \tau |R_B| + \tau |R_C| + |R_B \bowtie R_C| + |R_A \bowtie R_B  \bowtie R_C|,\\
        &\tau |R_A| + \tau |R_B| + \tau |R_C| + |R_C \bowtie R_A| + |R_A \bowtie R_B  \bowtie R_C|)\\
   =min(&|R_A \bowtie R_B| + |R_B \bowtie R_C| + |R_C \bowtie R_A|) \\
   	 + & \tau |R_A| + \tau |R_B| + \tau |R_C| + |R_A \bowtie R_B  \bowtie R_C|
\end{align*}
\begin{align*}
& C_{emm}(R_B^E \explore{a_L}_{\psi(c_L)[a] = a} R_A^E) \\
=& \tau |R_B^E| + \tau min(|R_C^E|, |R_B^E[c_L]|) + \tau |R_B^E[a_L]| + |R_B^E \explore{a_L}_{\psi(c_L)[a] = a} R_A^E|\\
\leq& \tau |R_B| + \tau |R_C| + \tau |R_A \bowtie R_B| + |R_A \bowtie R_B  \bowtie R_C|
\end{align*}
Then, we have:
\begin{align*}
&C_{emm}(R_B^E \explore{a_L}_{\psi(c_L)[a] = a} R_A^E) \\
\leq & \tau |R_A| + \tau |R_B| + \tau |R_C| + |R_A \bowtie R_B  \bowtie R_C|\\
\leq & min(C_{emm}(R_A \bowtie R_B  \bowtie R_C))
\end{align*}

\stitle{Lemma.} Given a graph $G(V,E)$, the number of its triangles is $O(|E|^{1.5})$.

We then have:

$C_{emm}(R_B^E \explore{a_L}_{\psi(c_L)[a] = a} R_A^E) \in O(|E|^{1.5})$.
\\
\textbf{E.g. can achieve the worse-case-optimal complexity at triangle counting(?)}
}
\begin{align*}
  & C_{emm}((R_A \bowtie R_B) \bowtie R_C) \\
= & \tau |R_A| + \tau |R_B| + \tau |R_C| + |R_A \bowtie R_B| + |R_A \bowtie R_B  \bowtie R_C|
\end{align*}
and 
\begin{align*}
	 & C_{emm}(R_A^E \explore{b_L}_{\psi(c_L)[c] = c} R_B^E) \\
	=& \tau |R_A^E| + \tau min(|R_C^E|, |R_A^E[c_L]|) + \tau |R_A^E[b_L]| + |R_A^E \explore{b_L}_{\psi(c_L)[c] = c} R_B^E|\\
 \leq& \tau |R_A| + \tau |R_C| + \tau |R_A \bowtie R_B| + |R_A \bowtie R_B  \bowtie R_C|
\end{align*}

Then, we have:

$C_{emm}(R_A^E \explore{b_L}_{\psi(c_L)[c] = c} R_B^E) \leq C_{emm}((R_A \bowtie R_B) \bowtie R_C)$\\

Similarly, there are:
\begin{align*}
	C_{emm}(R_B^E \explore{c_L}_{\psi(a_L)[a] = a} R_C^E) \leq& C_{emm}((R_B \bowtie R_C)  \bowtie R_A) \\
	C_{emm}(R_C^E \explore{a_L}_{\psi(b_L)[b] = b} R_A^E) \leq& C_{emm}((R_C \bowtie R_A)  \bowtie R_B)
\end{align*}

Therefore, through the cost model, it can be seen that considering the pointer in the query plan
can always bring a cheaper plan.}

\eat{
\begin{example}
	\label{exp-cost-model}
	\warn{We estimate the cost of the three query plan, 
			$\text{Plan}_\text{RF}$, 
			$\text{Plan}_\text{GF}$ and 
			$\text{Plan}_\text{H}$, 
			listed in Figure~\ref{fig-hybrid-query}, assuming that
			they all utilizing the hash intersective exploration.
			We then have:\\
			$C_{emm}(\text{Plan}_\text{RF}) = $  \\
			$C_{emm}(\text{Plan}_\text{GF}) = $  \\
			$C_{emm}(\text{Plan}_\text{H})  = $  
			
			We then provides an set of assumption about the cardinality appeared in 
			above formulas to actually compute the cost of them: 
			\warn{$|...| = ...$ ...}.
			One can easily verify that these cardinalities do not conflict with 
			each other and there exists an actual data set that satisfy all them
			above.
			Under this setting, it can be calculated that $C_{emm}(P_{H}) < C_{emm}(P_{GF}) < C_{emm}(P_{RF})$
			which show that accessing the relational data in the middle 
			is more efficient than considering these two part of data separately.}
\end{example}

\begin{align*}
 &C_{emm}(\text{Plan}_\text{H}) \\
=& C_{emm}(CSG_{H8}) + \tau |CSG_{H8}| \\
=& C_{emm}(CSG_{H6}) + \tau | R_{CSG_6}[D_V^1.in_L] | + \tau | R_{CSG_6}[D_V^0.in_L] | + (1 + \tau) |CSG_{H8}| \\
=& C_{emm}(CSG_{H4}) + \tau |CSG_{H4}| + \tau|D| + |CSG_{H6}| + \tau | R_{CSG_6}[D_V^1.in_L] | + \tau | R_{CSG_6}[D_V^0.in_L] | + (1 + \tau) |CSG_{H8}| \\
\end{align*}}

\vspace{-1.0ex}
\subsection{Cost Estimation}

The example below provides an explanation of the
cost estimation.

\begin{example}
For the query plan $\text{Plan}_\text{RF}$,  $\text{Plan}_\text{GF}$ and $\text{Plan}_\text{H}$ in Figure~\ref{fig-hybrid-query},
we then provide a step-by-step explanation of how their cost can be estimated
with the cost model $C_{emm}$ in Section~\ref{subsec-query-optimization-with-exploration}.

\stitle{$\text{Plan}_\text{RF}$:}
\vspace{-1.0ex}
\begin{fleqn}
\begin{align*}
	C_{emm}(CSG_{R2})             	=&  \tau |D_V^0| + \tau |D| + |R_{CSG_{R2}}| \\
	C_{emm}(CSG_{R4})             	=&  C_{emm}(CSG_{R2}) + \tau |R_{CSG_{R2}}[D_V^0.out_L]| \\
	C_{emm}(CSG_{R6})             	=&  C_{emm}(CSG_{R4}) + \tau |R_{CSG_{R4}}[D_o^0.dst_L]| \\
									=&  C_{emm}(CSG_{R4}) + \tau |R_{CSG_{R4}}| \\
	C_{emm}(CSG_{R8})             	=&  C_{emm}(CSG_{R6}) + \tau | R_{CSG_{R6}}[D_V^1.in_L] | \\
									 &	+\tau \warn{| \Pi_{D_V^0.in_L} R_{CSG_{R6}}[D_V^0.in_L] |} + |R_{CSG_{R8}}| \\
	C_{emm}(\text{Plan}_\text{RF}) 	=&  C_{emm}(CSG_{R8}) + \tau | R_{CSG_{R8}}[D_i^1.src_L] | \\
									=&  C_{emm}(CSG_{R8}) + \tau | R_{CSG_{R8}}| \\
									=&  (1 + \tau) |R_{CSG_{R8}}| + \tau | R_{CSG_{R6}}[D_V^1.in_L] |\\ 
 								  	 & + \tau \warn{| \Pi_{D_V^0.in_L} R_{CSG_{R6}}[D_V^0.in_L] |} + \tau |R_{CSG_{R4}}| \\
 								  	 & + \tau |R_{CSG_{R2}}[D_V^0.out_L]| + |R_{CSG_{R2}}| + \tau |D_V^0| + \tau |D| 
\end{align*}
\end{fleqn}
\eat{
=	&  (1 + 2\tau) |R_{CSG_{R8}}| + \tau |R_{CSG_{R6}}[D_V^0.in_L] | + 2\tau |R_{CSG_{R4}}| \\
&	+ |R_{CSG_{R2}}| + \tau |D_V^0| + \tau |D| 

It can be seen that:

\begin{align*}
	 | R_{CSG_{R6}}[D_V^1.in_L]  | &= |R_{CSG_{R8}}| \\
	 | R_{CSG_{R2}}[D_V^0.out_L] | &= |R_{CSG_{R4}}|
\end{align*}
}

\stitle{$\text{Plan}_\text{H}$:}
\vspace{-1.0ex}
\begin{fleqn}
\begin{align*}		              
	C_{emm}(CSG_{H2})             =&  \tau |D_V^0| + \tau |D_V^0[out_L]| \\
	C_{emm}(CSG_{H4})             =&  C_{emm}(CSG_{H2}) + \tau |R_{CSG_{H2}}[D_o^0.dst_L]| \\
								  =&  C_{emm}(CSG_{H2}) + \tau |R_{CSG_{H2}}| \\
    C_{emm}(CSG_{H6})             =&  C_{emm}(CSG_{H4}) + \tau |D| + |R_{CSG_{H6}}| \\
    C_{emm}(CSG_{H8})             =&  C_{emm}(CSG_{H6}) + \tau | R_{CSG_{H6}}[D_V^1.in_L] | \\
	 							   &	+\tau \warn{| \Pi_{D_V^0.in_L} R_{CSG_{H6}}[D_V^0.in_L] |} + |R_{CSG_{H8}}| \\
	C_{emm}(\text{Plan}_\text{H}) =&  C_{emm}(CSG_{H8}) + \tau | R_{CSG_{H8}}[D_i^1.src_L] | \\
	 							  =&  C_{emm}(CSG_{H8}) + \tau | R_{CSG_{H8}}| \\
								  =&  (1 + \tau) |R_{CSG_{H8}}| + \tau | R_{CSG_{H6}}[D_V^1.in_L] |   \\
								   & + \tau \warn{| \Pi_{D_V^0.in_L} R_{CSG_{H6}}[D_V^0.in_L] |} + |R_{CSG_{H6}}| \\
								   & + \tau |R_{CSG_{H2}}| + \tau |D_V^0[out_L]| + \tau |D_V^0|  + \tau |D|
\end{align*}
\end{fleqn}
\eat{ \\
	=	&  (1 + 2\tau) |R_{CSG_{H8}}| + \tau |R_{CSG_{H6}}[D_V^0.in_L]| + |R_{CSG_{H6}}| \\							
	&	+ 2\tau | R_{CSG_{H2}} | + \tau |D_V^0|  + \tau |D|
	
It can be seen that:
	\begin{align*}
		| R_{CSG_{R6}}[D_V^1.in_L]  | &= |R_{CSG_{H8}}| \\
		| D_V^0[out_L] |              &= |R_{CSG_{H2}}|
\end{align*}}

\stitle{$\text{Plan}_\text{GF}$:}
\vspace{-1.0ex}
\begin{fleqn}
\begin{align*}									
	C_{emm}(CSG_{G2})             	=& \tau |D_V^2| + \tau | D_V^2[out_L] | \\
	C_{emm}(CSG_{G4})             	=& C_{emm}(CSG_{G2}) + \tau | R_{CSG_{G2}}[D_o^1.dst_L] | \\
									=& C_{emm}(CSG_{G2}) + \tau | R_{CSG_{G2}} | \\
	C_{emm}(CSG_{G6})             	=& C_{emm}(CSG_{G4}) + \tau | R_{CSG_{G4}}[D_V^2.out_L] | \\
									 &	+\tau \warn{| \Pi_{D_V^1.in_L} R_{CSG_{G4}}[D_V^1.in_L] |} + |R_{CSG_{G6}}| \\
	C_{emm}(CSG_{G8})             	=& C_{emm}(CSG_{G6}) + \tau | R_{CSG_{G6}}[D_o^2.dst_L] | \\
									=& C_{emm}(CSG_{G6}) + \tau | R_{CSG_{G6}}| \\
	C_{emm}(\text{Plan}_\text{GF})  =& C_{emm}(CSG_{G8}) + \tau |D| + | R_{CSG_{G8}}|  \\
								  	=& | R_{CSG_{G8}}| + (1 + \tau) |R_{CSG_{G6}}|  +\tau | R_{CSG_{G4}}[D_V^2.out_L] |\\
									  & + \tau \warn{| \Pi_{D_V^1.in_L} R_{CSG_{G4}}[D_V^1.in_L] |} \\
									  & + \tau | R_{CSG_{G2}} | + \tau |D| + \tau |D_V^2| + \tau | D_V^2[out_L] |
\end{align*}
\end{fleqn}

\warn{As a simplification, for all relation $T$ and their field $A$, 
	we assume $|\Pi_{A}(T)| \approx |T|$.
	Meanwhile, for arbitrary relations $T$ and $T'$, and pointer field $p$ in $T'$,
	we assume $| T[T'.p] | \approx |T| \frac{|T'[p]|}{|T'|}$,
	which also implies the simple histogram method we utilized to estimate 
	the {\em ``selectivity''} of exploration.}

We then have:
\begin{fleqn}
\begin{align*}
	C_{emm}(\text{Plan}_\text{RF}) \approx	
	&  (1 + \tau) |R_{CSG_{R8}}| \\
	& + \tau | R_{CSG_{R6}}|  (\frac{|D_V^1[in_L]|}{|D_V^1|} + \frac{|D_V^0[in_L]|}{|D_V^0|}) + \tau |R_{CSG_{R4}}|  \\
	& + |R_{CSG_{R2}}|(1 + \tau \frac{|D_V^0[out_L]|}{|D_V^0|}) + \tau |D_V^0| + \tau |D| \\
	C_{emm}(\text{Plan}_\text{H})  \approx	
	&  (1 + \tau) |R_{CSG_{H8}}| \\
	& + |R_{CSG_{H6}}|(\tau\frac{|D_V^1[in_L]|}{|D_V^1|} + \tau\frac{|D_V^0[in_L]|}{|D_V^0|} + 1) \\
	&	+ \tau |R_{CSG_{H2}}| + \tau |D_V^0| (1 + \frac{|D_V^0[out_L]|}{|D_V^0|}) + \tau |D| \\
	C_{emm}(\text{Plan}_\text{GF}) \approx  
	& | R_{CSG_{G8}}| + (1 + \tau) |R_{CSG_{G6}}| \\
	&+ \tau | R_{CSG_{G4}}| (\frac{|D_V^1[in_L]|}{|D_V^1|} + \frac{|D_V^2[out_L]|}{|D_V^2|})  \\
	& + \tau | R_{CSG_{G2}} | + \tau |D_V^2| (1 + \frac{|D_V^2[out_L]|}{|D_V^2|}) + \tau |D|
\end{align*}
\end{fleqn}

To actually compute such a cost, we then specify a set of statistics,
one can verify that there exists a data set that satisfies them 
especially when there are pulled down filter conditions for label checking contained in these nodes in the query graph:
\begin{align*}
&|R_{CSG_{R8}}| =                 |R_{CSG_{H8}}| =                   |R_{CSG_{G8}}| =          10^4\\
&|R_{CSG_{R6}}| =                 |R_{CSG_{H6}}| =                   |R_{CSG_{G6}}| = 2 \times 10^3\\
&|R_{CSG_{R4}}| = 2\times10^5,                                    \; |R_{CSG_{G4}}| = 10^4 \\
&|R_{CSG_{R2}}| =                 |R_{CSG_{H2}}| = 10^4,          \; |R_{CSG_{G2}}| = 10^3 \\
& \frac{|D_V^0[in_L]|}{|D_V^0|} = \frac{|D_V^0[out_L]|}{|D_V^0|} = \frac{|D_V^1[in_L]|}{|D_V^1|} = \frac{|D_V^2[out_L]|}{|D_V^2|} = 100 \\
& |D_V^0| = |D_V^2| = |D| = 10^3
\end{align*}

Then, there are:

\vspace{0.4ex}
$C_{emm}(\text{Plan}_\text{H}) \approx 0.34 \; C_{emm}(\text{Plan}_\text{RF}) \approx 0.27 \; C_{emm}(\text{Plan}_\text{GF})$
\vspace{0.4ex}

That is, on this set of data with this designated query, the plan $\text{Plan}_\text{H}$  has the lowest cost
as it pulls the join on relation down to the pattern matching process.
\end{example}

\eat{\stitle{Exploration cardinality estimation}.
There is actually a more lightweight
method to estimate the cost of a plan contain exploration.
That is:

$|R_{CSG}[R.A]| \approx |R_{CSG}| \frac{|R[A]|}{|R|}$

The average $\frac{|R[A]|}{|R|}$ can be precomputed for 
each pointer field and be utilized for the optimizer.}

\eat{
There are:
\begin{align*}
	C_{emm}(\text{Plan}_\text{RF}) = 10^4 * 1.2 
						       + 2 * 10^3 * 0.2 * (100 + 100) 
						       + 2 * 10^5 * 0.2 
						       +     10^4 * (1 + 0.2 * 100)
						       +     10^3 * 0.2
						       +     10^3 * 0.2
						       
    C_{emm}(\text{Plan}_\text{H}) = 10^4 * 1.2 
						      + 2 * 10^3 * (0.2 * (100 + 100) + 1)
						      +     10^4 * 0.2
						      +     10^3 * 0.2 * (1 + 100)
						      +     10^3 * 0.2
						      
	C_{emm}(\text{Plan}_\text{GF}) = 10^4 * 1
						       + 2 * 10^3 * 1.2
						       +     10^4 * 0.2 * (100 + 100)
						       +     10^3 * 0.2 
						       +     10^3 * 0.2 * (1 + 100)
						       +     10^3 * 0.2
\end{align*}}

\eat{\subsection{Cost Model}
\label{subsec-cost-model}
To take the introduced operator $\psi_L$ into consideration for query optimization, a cost model, $C_{emm}$, is proposed here which is extended from $C_{mm}$.
$C_{mm}$ is a very simple cost model proposed in~\cite{leis2015good} that considers only the operation and omits the memory access as in~\cite{manegold2002generic}, which can achieve a comparable performance as the most advanced ones when considering main memory only.
Instead of defining recursively, $C_{emm}$ would evaluate the cost straightforwardly in an equivalent way, defined as:
{
	\small
	\begin{equation*}
		C_{emm}(Plan) = \Sigma_{op \in Plan}  E_{times}(op) * E_{cost}(op) 
	\end{equation*}
}
where $E_{times}(op)$ represents the cardinality estimated for each operator while $E_{cost}(op)$ denotes the cost estimated for each single execution.

Cardinality estimation is supposed to be the base of the cost model.
But with an observation, we can find that all strategises proposed before~\cite{park2020g, kim2021combining,yang2021huge,harmouch2017cardinality} are allowed to be directly utilized here. 
The cardinality relates only to the query pattern, which can be seen that $c_0$ to $c_4$, across Figure~\ref{fig-physical-query}(a) to Figure~\ref{fig-physical-query}(d), all represents the same value.
It allows the cardinality to be estimated with any existed method
to support our cost model and therefore the estimation strategy itself is not discussed here.

The essential of our model is estimating the cost of each operation, as:
{\small
	\begin{align*}
		C_{emm}(T) = 
		\begin{cases}
			C_{emm}(T_1) + \tau |T_1[A]| 	& \text{(if $T = T_1 \cexplore{A} S$, nested loop)}  \\
			C_{emm}(T_1) + \tau |S| + |T| 	& \text{(if $T = T_1 \cexplore{A} S$, hash)}  \\
			\intertext{$C_{emm}(T_1) + \tau |T_1[A]| \times \frac{1}{2} |T_1[A']| / |T_1| $}
			\intertext{\qquad \qquad \ \  \text{(if $T = T_1 \explore{A}_{\psi(T_{1}[A'])[a] \ni b} S$, nested loop)}} 
			\intertext{$C_{emm}(T_1) + \tau |T_1[A]| + \tau |T_1[A']| +  |T|$}
			\intertext{\qquad \qquad \ \  \text{(if $T = T_1 \explore{A}_{\psi(T_{1}[A'])[a] \ni b} S$, hash)}} 
		\end{cases}
	\end{align*}
}
Same as in~\cite{leis2015good}, $\tau \leq 1$ represents the cost of table scan compared to the lookup in hash table,
while $\lambda$ denotes the relative cost of lookup in index.

In addition to query optimization, evaluating physical plans under these two different cost models, $P_0$, $P_1$ and $P_2$ in $C_{emm}$, physical plans for $P_3$ in $C_{mm}$, \textcolor{red}{can provide a quantitative answer to one of the most fundamental questions: \emph{why processing data with graph structure instead of directly within the relational form?}} especially considering that lots of graphs are directly constructed from the relational data.



\begin{example}
	\label{example-cost-model}
	It should be remarked in ahead that the cost model here is not fair to our model since the lake of memory access consideration completely discards the effort in Section.\ref{sec-memory}.
	Even though, it is still enough to show the proposed model outperforms the conventional relational model. The cost of the plan in the proposed model can be evaluated as:
	{\small
		\begin{align*}
			C_{emm}(P_0) = |V| & + c_0 \cdot \tau E(v.{out}) \\
			& + c_1 \cdot \tau E(v.{out}) \\
			& + c_2 \cdot \tau E(v.{ in}) \\
			& + c_3 \cdot \lambda \\
			C_{emm}(P_1) = |V| & + c_0 \cdot E(v.{out}) \\
			& + c_0 \cdot \tau E(v.{out}) \\
			& + c_1 \cdot \tau E(v.{out}) \\
			& + c_2 \cdot (\tau E(v.{ in}) + c_4 / c_2 ) \\
			C_{emm}(P_2) = & 2 (|V| + c_0 \cdot \tau E(v.{out}) + c_1 \cdot \tau E(v.{out}))  \\
			& + c_4
	\end{align*}}
	The plan in Figure~\ref{fig-physical-query}(d) is a logical one, by implement the join operators as \emph{hash join} and \emph{index join}, we can get physical plan $P_{HJ}$ and $P_{INL}$ respectively, whose cost can be evaluated as:
	{\small
		\begin{align*}
			C_{mm}(P_{HJ})  = &\tau \cdot |E| + \tau \cdot |E| + c_2 \\
			+ &\tau \cdot |E| + c_3 \\
			+ &\tau \cdot |E| + c_4 \\
			C_{mm}(P_{INL}) = &\tau \cdot |E| + \lambda \cdot |E| \cdot max(\frac{c_2}{|E|}, 1) \\
			+ &\lambda \cdot c_2 \cdot max(\frac{c_3}{c_2}, 1) \\
			+ &\lambda \cdot c_3 \cdot max(\frac{c_4}{c_3}, 1) 
	\end{align*}}
	with the cardinality simply estimated as:
	{\small
		\begin{align*}
			c_0 &= |V| &\\
			c_1 &= c_0 * E(v.{out})& \\
			c_2 &= c_1 * E(v.{out})& \\
			c_3 &= c_2 * E(v.{ in})& \\
			c_4 &= c_3 * \sigma    & , 0  \leq \sigma \leq 1 \\
			E(v.{out}) &= E(v.{in}) = \frac{|E|}{|V|} &
	\end{align*}}
	By setting $d = E(v.{out}) = E(v.{in})$, we have:
	{\small
		\begin{align*}
			C_{emm}(P_0)    &= |V| ( (\tau + \lambda) d^3 +  \tau    d^2 +  \tau d + 1) \\
			C_{emm}(P_1)    &= |V| ( (\tau +  \sigma) d^3 +  \tau    d^2 + 2\tau d + 1) \\
			C_{emm}(P_2)    &= |V| (          \sigma  d^3 + 2\tau    d^2 + 2\tau d + 2) \\
			C_{mm}(P_{HJ})  &= |V| ( (  1  + \sigma ) d^3 +  \tau    d^2 + 4\tau d) \\
			C_{mm}(P_{INJ}) &= |V| (       2 \lambda  d^3 +  \lambda d^2 +  \tau d)
	\end{align*}}
	
	
	We set $\lambda = 2$ and $\tau = 0.2$ as in~\cite{leis2015good}, it should be noticed that such a configuration is also not fair to our model since both the large hash table and index can be costly when for intermediate result.
	Meanwhile by assuming the data graph is generated randomly which has no locality at all, there are $\sigma = \frac{|E|}{|V|^2}$. Together, we have:
	{\small
		\begin{align*}
			C_{emm}(P_0)     &= |V| ( 2.2 d^3 + 0.2 d^2 + 0.2 d + 1) \\
			C_{emm}(P_1)     &= |V| ( (  0.2  + \frac{d}{|V|} ) 
			d^3 + 0.2 d^2 + 0.4 d + 1) \\
			C_{emm}(P_2)     &= |V| ( \frac{d}{|V|}
			d^3 + 0.4 d^2 + 0.4 d + 2) \\
			C_{ mm}(P_{HJ})  &= |V| ( (  1  + \frac{d}{|V|} ) 
			d^3 + 0.2 d^2 + 0.8 d) \\
			C_{ mm}(P_{INJ}) &= |V| (  4  d^3 +  2  d^2 + 0.2 d)
		\end{align*}
	}
	
	It can be seen that:
	$C_{emm}(P_2) < C_{emm}(P_1) < C_{ mm}(P_{HJ}) < C_{emm}(P_0) < C_{ mm}(P_{INJ})$. More quantitatively, with $d$ set to $8$, which is in range of a typical data graph, on an large enough graph such that $d/|V| \approx 0$, we have:
	{\small
		\textcolor{red}{
			\begin{align*}
				C_{emm}(P_1) & \approx \frac{1}{ 5  } C_{ mm}(P_{HJ}) \\
				C_{emm}(P_2) & \approx \frac{1}{20.5} C_{ mm}(P_{HJ})
		\end{align*}}
	}
	which shows that even without considering the optimization on memory access, even the graph itself does not have any kind of locality at all, the proposed model can still have a lower cost compared to the conventional query process. 
\end{example}






\textcolor{red}{
	Such an example roughly shows how a quantitative analysis is possible and proves that processing data with graph structure is meaningful.}
With such an analysis method, one can quantitatively know how much benefits can be brought by organizing data as graph structure, and can decompose a query into two parts, \textcolor{red}{on graph and on relational database respectively, as a hybrid query for better overall performance}.

However, it should be noticed that the cost model here is unfair for our proposal while the memory access cost omitted here can be much more important compared to general SQL query. Since the partial result can be exponential and is typically very large in practice, the cost of \emph{hash join} is therefore been largely under-evaluated.
By considering the memory layout in Section.\ref{sec-memory}, an improved memory model with  the consideration of memory access pattern as in~\cite{manegold2002generic} would be proposed in following work.}

\subsection{Operator Resolution}

The above discussions are all about CSG and will generate a tree of them that
representing the optimized join order.
According to the query optimization workflow,
a query plan will then be resolved corresponding to it.
The crucial points for the operator resolution process here
to departure from its counterpart for conventional relational operator
are highlighted in Section~\ref{subsec-query-optimization-with-exploration}.
Below is an overall description of the entire resolution process.

\begin{myfloat}[H]
	\vspace{-1.2ex}
	\hspace*{-1ex}
	\begin{minipage}{0.478\textwidth}
		\removelatexerror
		\IncMargin{0.2em}
		\LinesNotNumbered
		{\small
			\begin{algorithm}[H]
				\SetArgSty{textrm}
				\caption{\ResolveOp}
				\label{operator-resolve}
				\Indentp{-2.5ex}
				\Indentp{1.4em}
				\textbf{Input:} Query graph $Q = (N, E, L, F)$, $CSG_0, CSG_1 \subseteq N$.
				
				\textbf{Output:} Operator $op$.
				
				// resolve explorative join condition, as \ResolveExp can only consider exploration condition in only direction
				
				\nextnr $\theta = \text{\ResolveExp}(CSG_0, CSG_1) \cup \text{\ResolveExp}(CSG_1, CSG_0)$
				
				// resolve plain value join condition, \ResolveCond can consider the condition from two directions
				
				// will not consider the edges with type \kw{duplicate}
				
				\nextnr $\theta = \theta \cup \text{\ResolveCond}(CSG_0, CSG_1)$
				
				// resolve operator, find whether these two CSGs are connected by an exploration edge
				
				\nextnr $e_e = null$  // to hold the exploration edge if it exists
				
				\nextnr \For{$v \in CSG_0 \cup CSG_1$}{
					\nextnr \For{$e \in N_E(v)$}{
						
						\nextnr $\langle l, A \rangle = L(e)$
						
						\nextnr \If{$l \neq explore$}{
							\nextnr \Continue
						}
						
						\nextnr $\langle n_s, n_d \rangle = F(e)$
						
						\nextnr \If{$n_s \in CSG_0$ \And $n_d \notin CSG_1$}{
							\nextnr \Continue // inside $CSG_0$ or toward an outside vertex
						}
						\nextnr \If{$n_s \in CSG_1$ \And $n_d \notin CSG_0$}{
							\nextnr \Continue // inside $CSG_1$ or toward an outside vertex
						}
						
						\nextnr $e_e = e$
					}
				}
				
				\nextnr \If{$e_e = null$} {
					
					\nextnr \Return $R_{CSG_0} \bowtie_\theta R_{CSG_1}$ // is a plain value join 
				}
				
				\nextnr $\langle l, A \rangle = L(e_e)$, $\langle n_s, n_d \rangle = F(e_e)$ // is an exploration
				
				\nextnr \If{$n_s \in CSG_0$} {
					\nextnr \Return $R_{CSG_0} \explore{A}_\theta R_{CSG_1}$
				}
				\nextnr \Return $R_{CSG_1} \explore{A}_\theta R_{CSG_0}$
				
			\end{algorithm}
		}
	\end{minipage}
	\vspace{-2.4ex}
\end{myfloat}

\ResolveOp in Algorithm~\ref{operator-resolve} describes the overall operator resolution process,
where the aggression is omitted.
By accepting a CSG-CMP pair as input, explorative conditions and plain value join conditions
will be first resolved, by calling \ResolveExp and \ResolveCond respectively.
After that, an operator will be generated according to the edges connect the two CSGs
and be integrated with the resolved conditions.
That is, if the two input CSGs are connected by an exploration edge,
an exploration operator will then be resolved.
Otherwise, a plain-value join operator will be resolved.
Meanwhile, it should be noticed that according to the query graph construction process
and the CSG-CMP pairs filter atop of the plan enumeration method,
any two CSGs that are considered for operator resolution here can only be connected 
by a single exploration edge or several edges for plain join conditions.

\eat{As we directly utilized the conventional there can be ,}

We then show how the explorative conditions can be resolved
according to the input CSG-CMP pair through \ResolveExp in Algorithm~\ref{explorative-condition-resolve}.
Corresponding to the description in Section~\ref{subsec-query-optimization-with-exploration},
outward edges from one CSG will be first explored
to find the nodes not contained in both CSGs.
For these explored nodes, 
it will be checked whether there are edges connecting them to the other CSG.
If so, those edges will be considered as explorative join conditions.
The \ResolveCond resolves plain join conditions 
along the same line as its relational counterpart and will not be further explained here.

\eat{
	The algorithm given above provides a solution to 
	generate a single operator.
	
	As described in Section~\ref{subsec-query-optimization-with-exploration},
	the general optimization workflow would export a tree of CSG for the optimal query plan,
	it would be further rewritten into a logical plan.
	
	\begin{myfloat}[H]
		\hspace*{-1ex}
		\begin{minipage}{0.478\textwidth}
			\removelatexerror
			\IncMargin{0.2em}
			\LinesNotNumbered
			{\small
				\begin{algorithm}[H]
					\SetArgSty{textrm}
					\caption{Resolve join condition}
					\label{condition-resolve}
					\Indentp{-2.5ex}
					\Indentp{1.4em}
					\textbf{Input:} CSG-CMP: $CSG_0$ and $CSG_1$.
					
					\textbf{Output:} Join condition set $\theta$.
					
					\nextnr $\theta = \{ \}$
					
					\nextnr \For{$n_0 \in \{V_0\}$}{
						\nextnr \For{$e \in n_0.out$}{
							\nextnr \If{$e.dst \in \{V_1\}$}{
								\nextnr $E = E \cup e$
							}
						}
					}
					
					\nextnr \For{$n_1 \in \{V_1\}$}{
						\nextnr \For{$e \in n_1.out$}{
							\nextnr \If{$e.dst \in \{V_0\}$}{
								\nextnr $E = E \cup e$
							}
						}
					}
					
					\nextnr \Return $\theta$
					
				\end{algorithm}
			}
		\end{minipage}
		\vspace{-3.4ex}
\end{myfloat}}

With the above operator resolution process, 
a logical query plan can then be rewritten from the optimized CSG tree accordingly.

\begin{myfloat}[H]
	\vspace{-1.2ex}
	\hspace*{-1ex}
	\begin{minipage}{0.478\textwidth}
		\removelatexerror
		\IncMargin{0.2em}
		\LinesNotNumbered
		{\small
			\begin{algorithm}[H]
				\SetArgSty{textrm}
				\caption{\text{\ResolveExp}}
				\label{explorative-condition-resolve}
				\Indentp{-2.5ex}
				\Indentp{1.4em}
				\textbf{Input:} Query graph $Q = (N, E, L, F)$, $CSG_0, CSG_1 \subseteq N$.
				
				\textbf{Output:} Explorative condition set $\theta^I$.
				
				\nextnr $\theta^I = \{\}$
				
				\nextnr \For{$n_0 \in CSG_0$}{
					\nextnr \For{$e \in N_E(n_0)$}{
						
						\nextnr $\langle l, A \rangle = L(e)$
						
						\nextnr \If{$l \neq \kw{explore}$} {
							\nextnr \Continue // is not an exploration edge
						}
					
						\nextnr $\langle n_s, n_d \rangle = F(e)$
						
						\nextnr \If{ $n_d \in CSG_0 \cup CSG_1$}{
							\nextnr \Continue // has already been considered
						}
						// to find the edges connect $n_d$ and nodes in $CSG_1$
						
						\nextnr \For{$e_c \in N_E(n_d)$}{
							
							\nextnr $\langle l_c, A_c \rangle = L(e_c)$
							
							\nextnr \If{$l_c = \kw{explore}$} {
								\nextnr \Continue
							}
							
							\nextnr \If{$N(e_c) \subseteq CSG_1 \cup \{ n_d \}$}{
								// $e_c$ can be resolved as explorative join condition
								
								\nextnr $\theta^I = \theta^I \cup \{ \psi(A)[A_c] \ni A_c \}$
							}
						}
					}
				}
				
				\nextnr \Return $\theta^I$
				
			\end{algorithm}
		}
	\end{minipage}
	\vspace{-1.4ex}
\end{myfloat}

\subsection{Query Execution}

Introducing the exploration operator into the query plan 
allows the access on graph and relational data to be executed
seamlessly without switching to another execution engine
to avoid performance degradation, 
{\em instead of merely satisfying a taste of elegance}.

A typical example is demonstrated in 
Figure~\ref{fig-hybrid-query} 
but was not clearly explained in Section~\ref{subsec-query-optimization-with-exploration}
due to the limitation of space,
which will be explained with a step-by-step example here in detail.

\begin{example}
Considering the three stages of the pipelined execution for $\text{Plan}_\text{H}$ in Figure~\ref{fig-hybrid-query},
that access across the relational and graph data
without breaking the pipelined execution process.

\begin{enumerate}[leftmargin=4ex]
\item 
In $\text{stage}_\text{1}$, an exploration operator $R_{CSG_{H2}} \explore{D_o^0.dst_L} D_V^1$ is executed.
Field $D_o^0.dst_L$ is held in the pipeline intermediate result $R_{CSG_{H2}}$
which points to relation $D_V^1$.
During the exploration process, 
all tuples $t_{H2} \in R_{CSG_{H2}}$ will be explored to iterate over all tuples 
$t_V^1 \in \psi(t_{H2}[D_o^0.dst_L]) \subseteq D_V^1$ 
to grab field $t_V^1[in_L]$ into the next stage 
of the pipelined intermediate result till it is filled.

\item 
In $\text{stage}_\text{2}$, an ordinary join operator $R_{CSG_{H4}} \bowtie_{D_V^0.vid = D.uid} D$ is executed.
Field $D_V^0.vid$ is held in the pipeline intermediate result $R_{CSG_{H4}}$.
All tuples $t_{H4} \in R_{CSG_{H4}}$ will then be iterated,
to join on all tuples $t_R \in D$ \st $t_{H4}[D_V^0.vid] = t_R[uid]$.

\item 
In $\text{stage}_\text{3}$, an intersective exploration operator is executed,
$R_{CSG_{H6}} \explore{D_V^1.in_L}_{ \psi (D_V^0.in_L)[src_L] \ni src_L} D_i^1$,
to continue the pattern matching process.
Field $D_V^0.in_L$ and $D_V^1.in_L$ are held in the pipeline intermediate result $R_{CSG_{H6}}$.
All tuples $t_{H6} \in R_{CSG_{H6}}$ will be iterated,
to intersect all tuples 
$t_i^2 \in \psi (t_{H6}[D_V^0.in_L]) \subseteq D_i^2$ and 
$t_i^1 \in \psi (t_{H6}[D_V^1.in_L]) \subseteq D_i^1$,
\st $t_i^1[src_L] = t_i^2[src_L]$.
\end{enumerate}
\vspace{0.4ex}

In the external level,
it can be seen that $\text{stage}_\text{1}$ to $\text{stage}_\text{3}$ explore 
the graph data, 
relational data and back to the graph data subsequently without breaking the pipelined execution.
\end{example}
\vspace{-4.4ex}

\eat{the generated hybrid query plan can contain both the .
	introducing the 
	
	The unified representation
	allows the hybrid query plan to be executed without switching across different execution engines.
	engines
	engine
	
	with the \ModelName-based query evaluation workflow.
	A step-by-step explanation of such an execution process
	is explained below.
	
	The generated query plan can contain both the 
	hybrid query plan and the .}

\eat{
The example above shows that
eliminating the cross model access as in Figure~\ref{fig-RX-model}
can bring actual benefit to hybrid query evaluation,
instead of merely satisfying a taste of elegance.

by representing both the graph and the relational data uniformly
with the \ModelName model, 
the cross model access is thus eliminated
which allows the hybrid query plan to be executed without switching across different execution engines.

It can be seen that the entire hybrid query plan, 
consist of access on both relational and graph data,
can be executed without interrupting the pipelined execution process
or switching to another relational/native graph-based engine.
	
\stitle{$\text{Stage}_\text{1}$}.
The pipeline intermediate result holds the field $e_0.dst_L$, 
and can be .

Holds $e_0.dst$
The field $D_V^1[in_L]$ would be explored
through the pointer in $e_0.dst$ and be grabbed into the next stage 
of the pipelined intermediate result till it is filled.

\stitle{$\text{Stage}_\text{2}$}.}

\begin{figure}[H]
	\centerline{\includegraphics[scale=1.0]{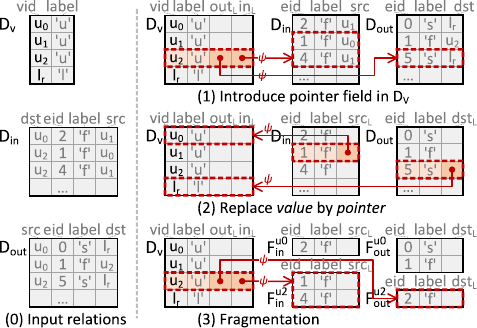}}
	\centering
	\vspace{-2.4ex}
	\caption{A step-by-step demonstration of the conversion process from relations to \ModelName model}
	\label{fig-RG-convert}
	\vspace{-1.4ex}
\end{figure}

\subsection{Graph Representation in RG Model}

The example below provides a detailed explanation 
to illustrate how graph-structured data can be represented in the \ModelName model.
Different from the description in Section~\ref{subsec-define-model}
for encoding graph-structured data,
the example here offers a (conceptional) conversion process
from a relational representation of the graph data, 
to emphasize the extensions that the \ModelName model introduces into the relational model.

\eat{And to emphasize the difference it introduces to the relational model.
by converting from a conventional relational representation,
with vertex and edge relations, of graph.
It also provides a clearer explanation of the fragmentation 
process that was only roughly mentioned in Section~\ref{subsec-define-model}.
}

\begin{example}
Taking $G$ in Figure~\ref{fig-example} as an example, 
Figure~\ref{fig-RG-convert} provides a step-by-step explanation
to store the graph $G$ in the \ModelName model
from a relational representation of it.
As a simplification, we do not consider the property on 
both vertices and edges here.

\begin{enumerate}[leftmargin=4ex]
\setcounter{enumi}{-1}
\item 	The relational representation of $G$ consists of three relations: 
		$\at{D_{V}(vid, label)}$ to hold all vertices in $G$;
		$\at{D_{in}(dst, eid, label, src)}$ to hold all edges in $G$;
		$\at{D_{out}(src, eid, label, dst)}$, as a duplication of $\at{D_{in}}$, 
		or a virtual view of it.
\item	Add pointer field $out_L$ (resp. $in_L$) into $D_V$ 
		to connect to all outward (resp. input) edges for each vertex in it,
		which makes it become $\at{D_V(vid, label, out_L, in_L)}$.
		Field $src$ in $D_{in}$ 
		and $dst$ in $D_{out}$ can then be removed, 
		for them to become 
		$\at{D_{in}(eid, label, dst)}$ and 
		$\at{D_{out}(src, eid, label)}$ respectively.
		It thence reduces the duplication between $D_{in}$ and $D_{out}$.
\item	Replace the field $src$ in $D_{in}$ (resp. $dst$ in $D_{out}$) as $src_L$ (resp. $dst_L$),
		that is, for each tuple $t \in D_{in}$ (resp. $t \in D_{out}$),
		replace the {\em value} of its field  $t[src] \in \mathbb{V}$ (resp. $t[dst] \in \mathbb{V}$)
		with pointer $t[src_L] \in \mathbb{P}$ (resp. $t[dst_L] \in \mathbb{P}$)
		towards the same vertex tuple.
		We can then get $\at{D_{in}(eid, label, dst_L)}$ and 
		$\at{D_{out}(src_L, eid, label)}$.
		
\item	Fragment the extended relation $D_{in}$ and $D_{out}$ according to the pointers in 
		$out_L$ and $in_L$ in $D_V$, as mentioned in Section~\ref{subsec-define-model}.
\end{enumerate}
\vspace{0.8ex}

One can verify that the \ModelName representation obtained in 
the above process is equivalent to the one resulted in Example~\ref{exa-conversion}.
\end{example}

It should be noticed that 
the extended relations $D_\at{V}$, $D_\at{out}$, $D_\at{in}$
and the {\em dereference function $\psi$} in 
$\D_G^E$ have actually already been obtained at $\text{stage}_\text{2}$.
The following $\text{stage}_\text{3}$ is only to complete the fragmentation
to support an efficient exploration through the pointer.
It is also noteworthy that the fragmentation performed in $\text{stage}_\text{3}$
is actually at the logical level, without specifying the memory layout or the physical designs.

\begin{figure}[H]
	\begin{center}
		\vspace{-4.3ex}
		\begin{minipage}[t]{\textwidth / 2}
			\subfigure[\small \dblp]{\label{fig-dblp-storage}
				{\includegraphics[width=2.8cm]{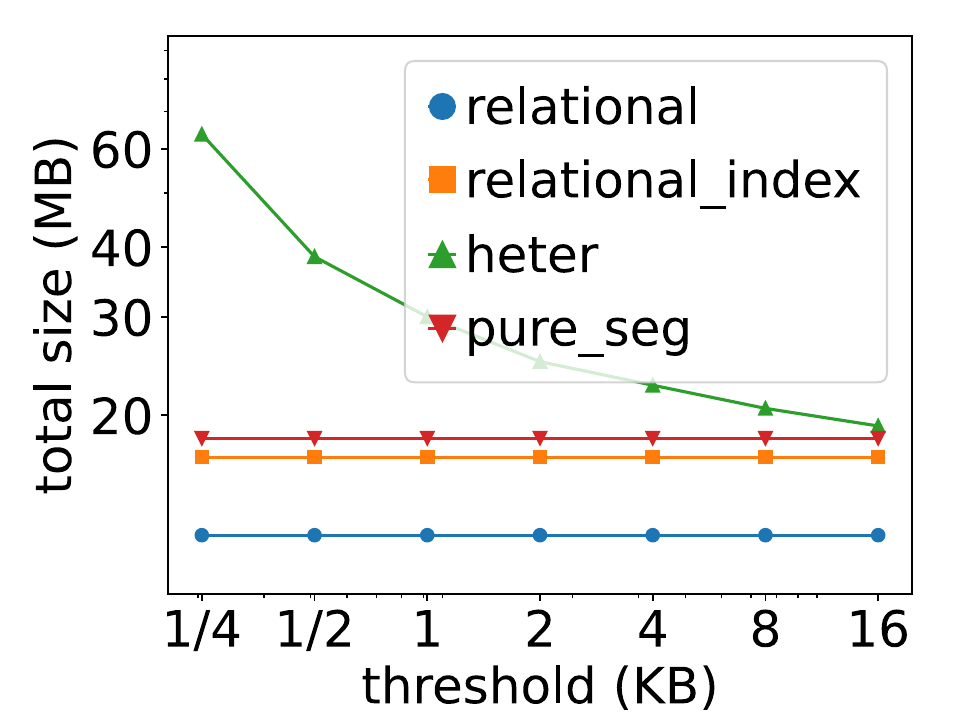}}}
			\hspace{-0.2cm} 
			\subfigure[\small \imdb]{\label{fig-imdb-storage}
				{\includegraphics[width=2.8cm]{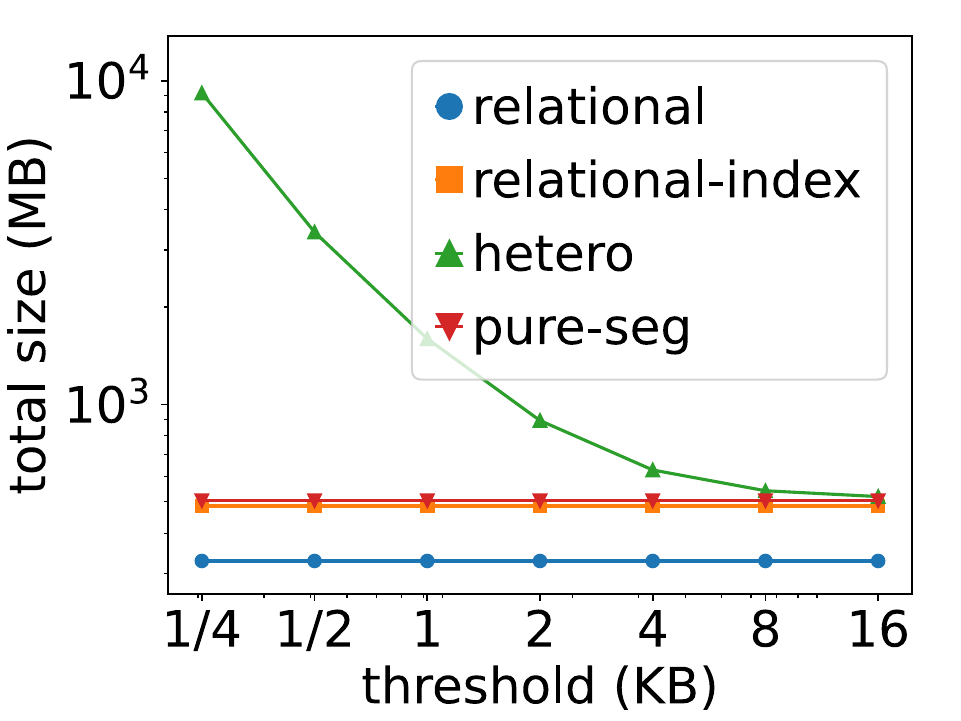}}}
			\hspace{-0.2cm} 
			\subfigure[\small \yago]{\label{fig-yago-storage}
				{\includegraphics[width=2.8cm]{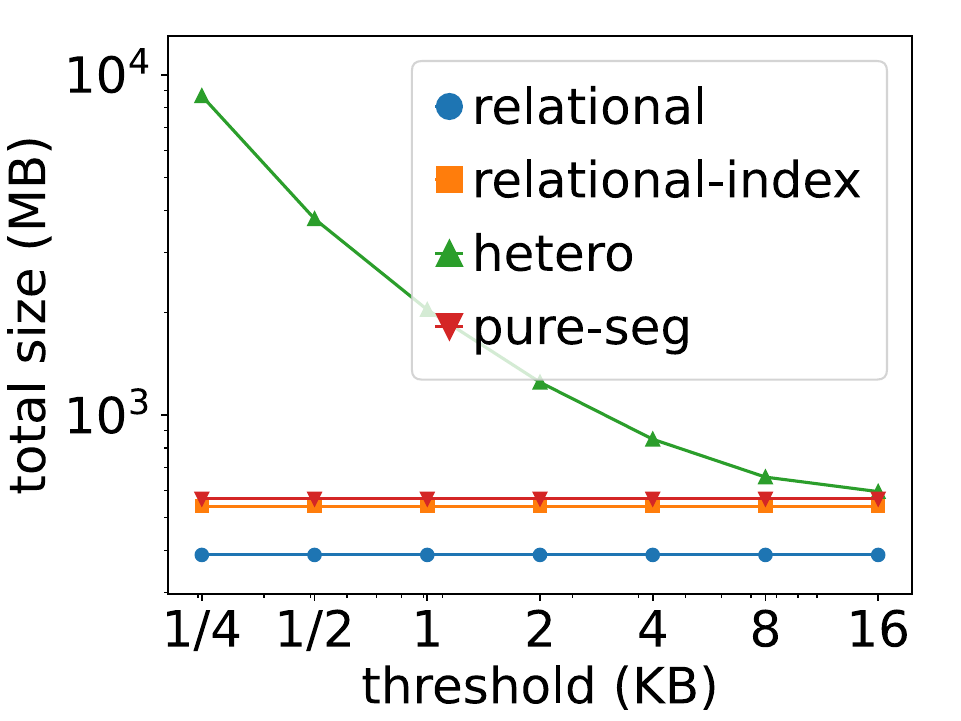}}}
			\vspace{-2.4ex}
		\end{minipage}
	\end{center}
	\vspace{-2.4ex}
	\caption{Heterogeneous fragment storage estimation}
	\label{fig-storage-exp}
	\vspace{-2.3ex}
\end{figure}

\begin{figure}[H]
	\vspace{-2.0ex}
	\begin{center}
		\vspace{-4.8ex}
		\begin{minipage}[t]{\textwidth / 2}
			\subfigure[\small \dblp]{\label{fig-dblp-incremental}
				{\includegraphics[width=2.8cm]{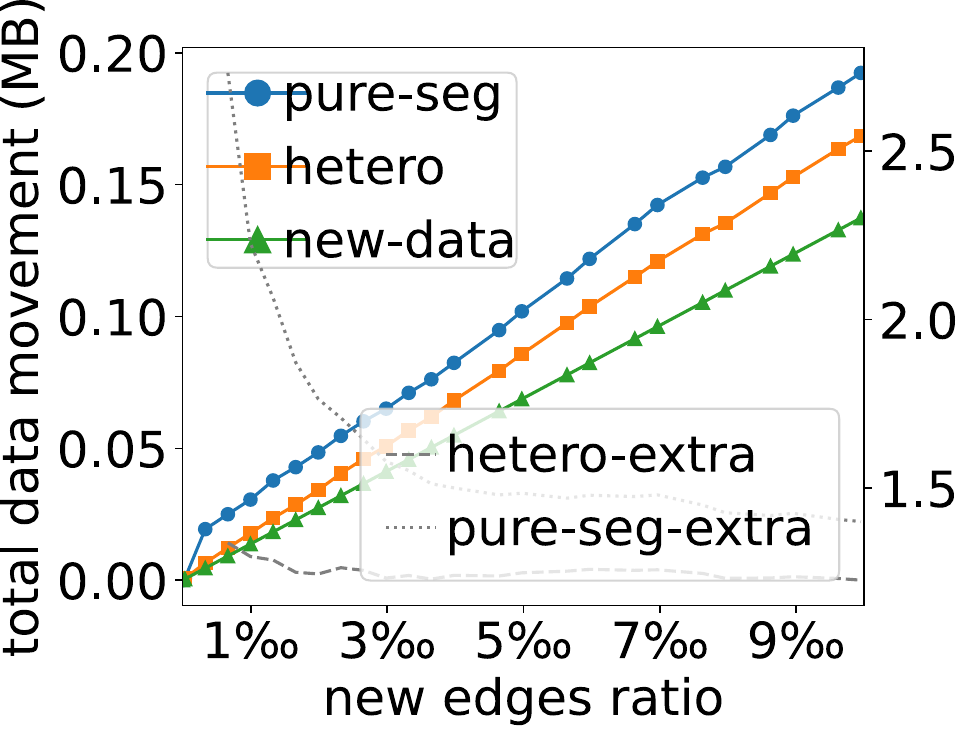}}}
			\hspace{-0.2cm} 
			\subfigure[\small \imdb]{\label{fig-imdb-incremental}
				{\includegraphics[width=2.8cm]{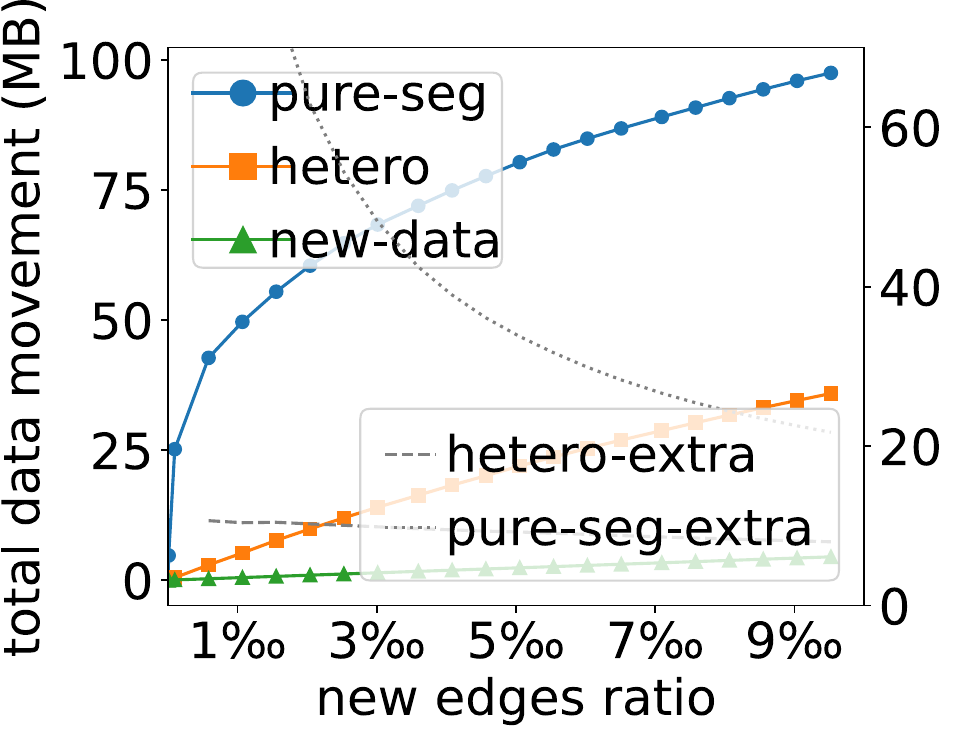}}}
			\hspace{-0.2cm} 
			\subfigure[\small \yago]{\label{fig-yago-incremental}
				{\includegraphics[width=2.8cm]{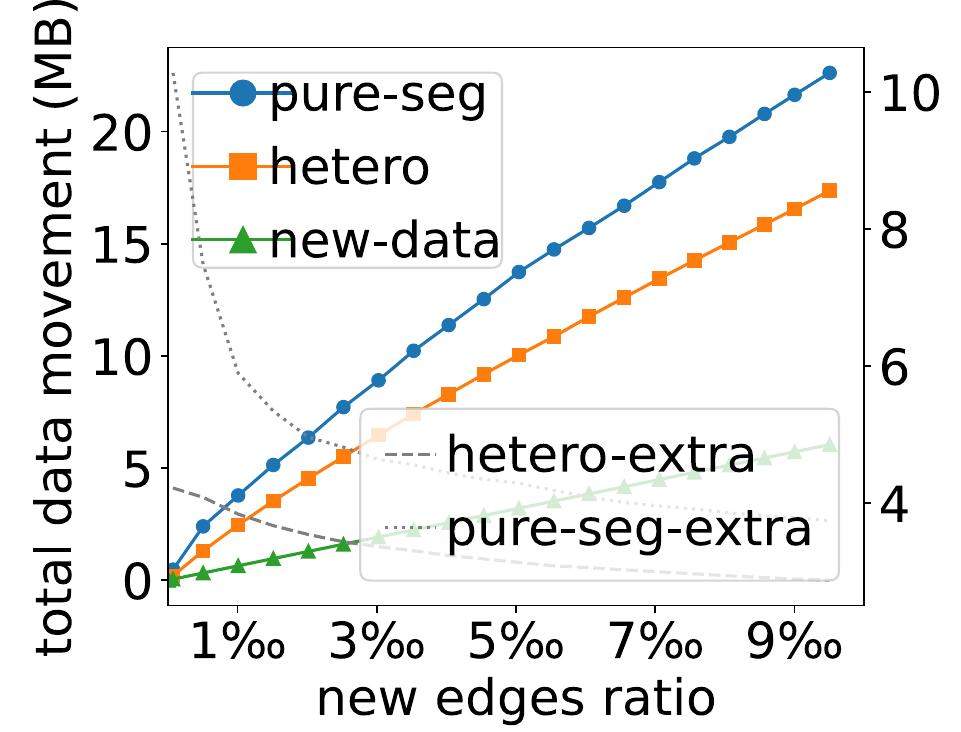}}}
			\vspace{-2ex}
		\end{minipage}
	\end{center}
	\vspace{-2.4ex}
	\caption{Incremental data movement estimation}
	\label{fig-incremental-exp}
	\vspace{-2.3ex}
\end{figure}

Meanwhile, it can also be clearly verified here that the $\D_G^E$ is in the regular form.
That is, all pointers in field $out_L$ (resp. $in_L$) of $D_V$
reference to the tuples in $D_{out}$ (resp. $D_{in}$);
the pointers in both fields $src_L$ (resp. $dst_L$) of $D_{in}$ (resp. $D_{out}$)
all point to tuple in $D_V$.
Therefore, given a pointer $p \in \mathbb{P}$,
the relation that holds $\psi(p)$ can be looked-up according to the underlying field of $p$.
A node can then be introduced into the query graph as an alias of that relation
to correspond to the explorative condition as in Appendix~\ref{appendix-query-graph-construct}.

\vspace{-0.4ex}
\subsection{Evaluation on Memory Management}

We evaluated the proposed heterogeneous fragment management strategy, 
on its storage cost and maintenance cost.

\stitle{Exp-6: Storage cost}. 
As described in Section~\ref{sec-model},
the extended relations need to be fragmented
to efficiently support the exploration of a set of tuples.
However, such a process will inevitably introduce
additional costs, 
including the storage for the additional pointer fields and the internal fragmentation inside a block.
We evaluated the cost introduced by the heterogeneous fragment management strategy, denoted as \kw{hetero}, 
with the following four kinds of baseline:
(1) \kw{relational}, stores graph data in 
	vertex relation $D_{V}(vid, label)$ and 
	edge relation $D_{E}(eid, src, label, dst)$;
(2) \kw{relational\text{-}index}, further considers two indexes on $src$ and $dst$ field on the edge relation $D_{E}$ respectively;
(3) \kw{pure\text{-}seg}, stores graph data in the \ModelName model and treats all fragments as segments, even the very large ones, and will not divide them into blocks;
(4) \kw{pure\text{-}block}, stores graph data in the \ModelName model and allocates at least one block for each fragment.
We set the block size as \kw{64KB} and varied the predefined {\em block-segment threshold} 
for heterogeneous fragment management mentioned in Section~\ref{subsec-physical-implementation}, 
to evaluate the cost introduced in such a process.
\looseness=-1

Evaluated with real-life data, 
we demonstrated in Figure~\ref{fig-storage-exp} that storing graph data in the \ModelName model
with the \kw{pure\text{-}seg} setting
took
51.5\%, 
49.9\% and 
46.1\% 
additional space on \dblp, \imdb and \yago respectively
compared to the \kw{relational} baseline.
However, when taking the storage for the indexes data into account,
\ie comparing \kw{pure\text{-}seg} to \kw{relational\text{-}index},
the additional space required on them would be reduced to
9.08\%, 
5.01\% and 
6.52\%  
respectively.
It should be noticed that the indexes on edge relation
took a large portion of the storage since we did not considered heavy property on both 
vertex and edge in the graph data.
Considering the two additional indexes is equivalent to consider the columns $src$ and $dst$
in the edge relation twice which can largely increase the required space.
Therefore, it can be seen that the space utilized for storing graph data in the \ModelName model 
is basically equivalent to storing the same set of data in relations and then add 
indexes on them.

We then vary the predefined {\em block-segment threshold} to evaluate the extra storage cost  
that the internal fragmentation 
in the heterogeneous memory management strategy can bring.
As shown in Figures~\ref{fig-dblp-storage} to~\ref{fig-yago-storage} , 
when the threshold is set to {\kw 8KB},
\ie the fragments smaller than {\kw 8KB} will all be managed as segments,
\kw{hetero} took
24.5\%, 
11.5\% and 
21.7\% additional space compared to \kw{pure\text{-}seg} on these three data graphs respectively.
Meanwhile, the heterogeneous management also significantly reduced (< 1/100) 
the space required for the \kw{pure\text{-}block} strategy
(not drawn).

\vspace{-0.6ex}
\stitle{Exp-7: Maintenance cost}. 
We then evaluated the maintenance cost and the reduction of data movement amount
achieved by the heterogeneous fragment management strategy.
For each data graph $G$, we generated a $\Delta G$ on it 
and compared the data movement amount required to accept such an update
by the heterogeneous fragment management strategy, denoted as \kw{hetero},
with two kinds of baselines:
(1) in \kw{new\text{-}data}, we simply calculated the size of new inserted edges;
(2) in \kw{pure\text{-}seg}, we calculated the data movement amount required when all fragments are managed as segments, 
	where all introduced edges would uniformly require their underlying segments to be reallocated.
We also estimated the ratio of extra data movement required by these two strategies,
\ie compared \kw{hetero} and \kw{pure\text{-}seg} to \kw{new\text{-}data},
denoted as \kw{hetero\text{-}extra} and \kw{pure\text{-}seg\text{-}extra} respectively.
\looseness=-1
	
Figure~\ref*{fig-incremental-exp} 
shows the data movement amount required for segment reallocation when different amounts of new edges were
inserted into the data graph that had been stored in the \ModelName model.
Owing to their topological characteristics,
it can be seen that
the data movement amount in \kw{pure\text{-}seg} baseline
were not extraordinarily larger than the \kw{new\text{-}data} on both \dblp and \yago.
That is, on these two data graphs,
the introduced edges were all added into the relatively small fragments and thus does not introduce
much extra data movement.
However, the situation is completely different on \imdb,
as the introduced edges can easily connect to vertices with a large degree,
the required data movement to reallocate the underlying segment thence 
becomes unacceptable even with only a very few edges inserted.
Meanwhile, on all data graphs, the ratio of extra data movement 
required in both fragment management strategises were reduced
as the number of inserted edges increases.
This is due to the size of the tuples inserted into the segments 
is becoming comparable to the data that these segments originally held.

The experimental result shows that the heterogeneous
memory management strategy can significantly reduce the extra data movement enforced by
introducing a small amount of edges.
With the block size set to \kw{64KB} and the predefined {\em block-segment} threshold set to \kw{8KB},
such a strategy can reduce the extra data movement in \kw{pure\text{-}seg\text{-}extra} by
1.71x, 
9.49x and 
1.54x on these three data graphs respectively
when 1\textperthousand~new edges are inserted into them.
Combining with the storage cost estimated in Exp-6, it can be seen that 
such a heterogeneous strategy fairly achieved a trade-off between these two strategies. 
We defer an in-depth study to apply more advanced dynamic memory management 
strategies~\cite{wilson1995dynamic} to future work
to further reduce the extra data movement here.

\eat{
It should be emphasized that the experiment here is to evaluate the 
advantages that can be brought by the heterogeneous fragment management strategy 
proposed in Section~\ref{subsec-physical-implementation},
evidenced by the difference between 
\kw{hetero\text{-}extra} and \kw{pure\text{-}seg\text{-}extra}.
Only a most native strategy to manage the allocation/deallocation of segment itself is adopted
here, which simply reallocates each resized segment without 
verifying whether there are sufficient neighbouring space for them to be enlarged in-place.
We defer an in-depth study to apply more advanced dynamic memory management 
strategies~\cite{wilson1995dynamic} to future work
to further reduce the extra data movement here.}

\eat{Meanwhile, considering the undirected pointer can avoid the 
and pointer assignment and bring \tbf x, \tbf x and \tbf x 
efficiently improvement on above maintenance process
on \dblp, \imdb and \yago respectively when accepting 30000 new edges.
In particular, if for each vertex, only one of its inward (resp. outward) edges is inserted/deleted,
such a speedup would reach \tbf x, \tbf x and \tbf x 
on these three data graphs respectively,
as the pointer re-assignment contributes a large portion here.
}

\subsection{Notation Collection}

We collect the frequent used notations in 
Table~\ref{tab-notation-collection},
ordered by the location of their first appearance.

\begin{table}
	\centering
	\begin{smaller}
		\centering
		\renewcommand{\arraystretch}{1.25}
		\begin{tabular}{|c|c|c|} 
			\hline
			{\bf  }  
			& {\bf Notation}
			& {\bf Conception}  \\ 
			\hline
			\multirow{5}{*}{\rotatebox[origin=c]{90}{Section~\ref{sec-SQL}}}  &   
			\makecell{$\mathbb{V}$} & 
			\makecell{domain of plain values} \\ \cline{2-3}
			&   
			\makecell{$P[\bar A](G)$} & 
			\makecell{graph pattern query} \\ \cline{2-3}
			&   
			\makecell{$\join_\delta$} & 
			\makecell{join on semantic connection} \\ \cline{2-3}
			&   
			\makecell{\RAd} & 
			\makecell{proposed relational algebra dialect} \\ \cline{2-3}
			&   
			\makecell{\SQLd} & 
			\makecell{proposed \SQL dialect} \\
			\hline 
			\multirow{12}{*}{\rotatebox[origin=c]{90}{Section~\ref{subsec-define-model}}}  &   
			\makecell{$\mathbb{P}$} & 
			\makecell{domain of pointers} \\ \cline{2-3}
			&   
			\makecell{$\R^E$} & 
			\makecell{\ModelName schema} \\ \cline{2-3}
			&   
			\makecell{$\D^E$} & 
			\makecell{extended database} \\ \cline{2-3}
			&   
			\makecell{$\phi$} & 
			\makecell{injective partial reference function} \\ \cline{2-3}
			&  
			\makecell{$\psi$} &  
			\makecell{dereference function} \\ \cline{2-3}
			&   
			\makecell{$\R^E_G$} & 
			\makecell{\ModelName schema of graph $G$} \\ \cline{2-3}
			&   
			\makecell{$\D_G^E$} & 
			\makecell{extended relation of graph $G$} \\ \cline{2-3}
			&   
			\makecell{$D_\at{V}$} & 
			\makecell{extended relation of vertices} \\ \cline{2-3}
			&   
			\makecell{$D_\at{out}$} & 
			\makecell{extended relation of outgoing edges} \\ \cline{2-3}
			&   
			\makecell{$D_\at{in}$} & 
			\makecell{extended relation of incoming edges} \\ \cline{2-3}
			&   
			\makecell{$F^v_\at{out}$} & 
			\makecell{fragment of $D_\at{out}$ of all outgoing edges from $v$} \\ \cline{2-3}
			&   
			\makecell{$F^v_\at{in}$} & 
			\makecell{fragment of $D_\at{in}$ of all incoming edges toward $v$} \\ 
			\hline 
			\multirow{5}{*}{\rotatebox[origin=c]{90}{Section~\ref{subsec-RX-model-operator}}}  &   
			\makecell{$\theta^I$} & 
			\makecell{notation of explorative condition} \\ \cline{2-3}
			&   
			\makecell{$\psi(A)[B] \ni C$} & 
			\makecell{form of an explorative condition} \\ \cline{2-3}
			&   
			\makecell{$S_1 \cexplore{A} S_2$} & 
			\makecell{the (logical) exploration operator} \\ \cline{2-3}
			&   
			\makecell{$S_1 \ijoin S_2$} & 
			\makecell{the (logical) join with explorative condition} \\ \cline{2-3}
			&   
			\makecell{$S_1 \iexplore S_2$} & 
			\makecell{the (logical) intersective exploration} \\ 
			\hline 
			\multirow{6}{*}{\rotatebox[origin=c]{90}{Section~\ref{subsec-query-optimization-with-exploration}}}  &   
			\makecell{$(S_1, A, S_2)$} & 
			\makecell{form of an exploration edge} \\ \cline{2-3}
			&   
			\makecell{$|S|$} & 
			\makecell{cardinality estimated for relation $S$} \\ \cline{2-3}
			&   
			\multirow{2}{*}{\makecell{$|S[A]|$}} & 
			\multirow{2}{*}{\makecell{cardinality estimated for data \\ linked by field $A$ in $S$}} \\
			&& \\ \cline{2-3}
			&   
			\makecell{$C(v_i)$} & 
			\makecell{candidate set generated for pattern vertex $v_i$} \\ \cline{2-3}
			&   
			\makecell{$D^C_{v_i}$} & 
			\makecell{separated relation to store $C(v_i)$} \\ 
			\hline 
		\end{tabular}
	\end{smaller}
	\caption{Summary of notations}\label{tab-notation-collection}
	\vspace{-4.7ex}
\end{table}

\begin{table*}
	\centering
	\begin{smaller}
		\centering
		\renewcommand{\arraystretch}{1.25}
		\begin{tabular}{|c|c|c|c|c|c|c|} 
			\hline
			\multirow{2}{*}{\bf \rotatebox[origin=c]{90}{Level} } &  
			\multicolumn{2}{c|}{\bf Relational Model} &  
			\multicolumn{2}{c|}{\bf Relational Genetic Model} &  
			\multicolumn{2}{c|}{\bf Graph Model} \\ \cline{2-7}
			& {structural} & { operational} 
			& {structural} & { operational} 
			& {structural} & { operational} \\ 
			\hline
			\rotatebox[origin=c]{90}{~External~}	&	\makecell{relational \\ views}  &  \makecell{\SQL etc.}  
			& 	\makecell{relational views \\ + graph views}  &  \makecell{\SQLd} 
			& 	\makecell{graph views}  &  \makecell{graph query \\ languages}	\\ 
			\hline
			\rotatebox[origin=c]{90}{~Logical~}	    &	\makecell{base \\ relations}  &  \makecell{logical \\ operators}  
			& 	\makecell{extended \\ relations}  &  \makecell{logical operators \\ + exploration}  
			&	\makecell{base graphs}  &  \makecell{logical graph \\ operators}	\\ 
			\hline
			\rotatebox[origin=c]{90}{~Physical~}	&	\makecell{physical \\ designs}  &  \makecell{physical \\ operators} 
			& 	\makecell{physical designs \\ + pointers}  &  \makecell{physical operators \\ + exploration}  
			&	\makecell{physical designs \\ for graph}  &  \makecell{physical graph \\ operators}	\\ 
			\hline
		\end{tabular}
	\end{smaller}
	\caption{A comparison of different data models under the perspective of the three level architecture}\label{tab-model-comparison}
	\vspace{-3.7ex}
\end{table*}

\vspace{-0.6ex}
\subsection{Query Language Comparison}

\eat{
Generally speaking, the \ModelName model 
enhances the relational model on its 
{\em ``structural part''}~\cite{codd1980data, codd1981relational}
to consider logical-level 
pointers and the induced exploration operator (Section~\ref{subsec-RX-model-operator}).
It should be highlighted that those pointers are at the logical level
as what they reference to are tuples, the logical level objects,
which thus 
allow them to appear in logical query plans to facilitate the evaluation of
graph pattern queries; \eg 
we will introduce ``exploration'' operators in query plans
based on such pointers.
Without affecting the foundations of relational model,
such a mechanisms can also be directly 
embedded into the logical query plan (see Section~\ref{subsec-query-optimization-with-exploration}) 
to evaluate pattern queries with a same efficiency as in the native graph store.

We then carry out a brief argumentation of the proposed \ModelName model
to show why it is optimal for our purpose.
After that, we discuss the expressiveness and the usability of such a query system
and compare it with other data models.

\subsubsection*{Argumentation}

To argue that the \ModelName model is the optimal for our purpose,
we first collect a set of goals for a model to satisfy for unifying the graph and relational model,
and further analysis the demands they pose to the required model.
Of course, since the argument itself is not a clear and strict theory,
such an argumentation cannot be the same sound as a theoretical proof.

\stitle{Goals}. We collect a set of requirements for a data model to satisfy
to uniformly manage the graph and relational data,
which are:
\begin{enumerate}
	\item	\label{req-uniform-access}
	Provide a uniform access and query interface for both relational and graph data,
	without exposing the complexity internal structures to the external view. 
	\item	\label{req-uniform-evaluate}
	Provide a unified seamless evaluation workflow for the hybrid declarative queries across graph and relational data.
	\item	\label{req-comparable-performance}
	Achieve a comparable performance w.r.t. native graph-based methods (resp. relational methods) on graph queries (resp. relational queries).
\end{enumerate}

\stitle{Demands}.
By further analysing the goals listed above, 
we can then collect some demands of the required model:

\mbi
\item
To achieve (\ref{req-uniform-evaluate}) and (\ref{req-comparable-performance}) at the same time, 
the relation-side workloads demand the query evaluation workflow provided by the required moded
to be a {\em ``sublation''} of the counterpart for relational query,
\ie to support all the existing technologies for the relational query evaluation.

\item
To achieve (\ref{req-uniform-access}) and (\ref{req-comparable-performance}) at the same time,
the graph-side workloads demand the required model to support an exploration operation at the logical level,
while those connections among the logical level components that support those explorations 
need to be hidden in the external view.

\item 
According to the reflections in Section~\ref{subsec-reflection} and Section~\ref{sec-practice},
the required model needs to be a {``sublation''} of the relational model;
otherwise, several solutions need to be reinvented upon such a model for the challenges that have already been
addressed by the relational model.
\mei

Conclusively, 
it can be seen that the proposed \ModelName model achieves all these goals with the minimal modifications
to the relational model
without introducing various messy conceptions for graph~\cite{angles2008survey, stonebraker2005goes},
like {\em ``vertex/entity''}, {\em ``edge/relationship''}, {\em ``label/attribute''} etc.,
which allows our model to be considered as the {\em optimal} under such a criteria.}

\eat{
	That is, it needs to support all the methods for relational query evaluation methods for the query
	
	To achieve (1) and (2), it should be a {\em ``sublation''} of the relational model; 
	to achieve (3), it should support the exploration operator.
	The proposed \ModelName model achieves all these three requirements with minimal modification to 
	the relational model, which allows it to be stated as the optimal one although there cannot 
	be a theoretical proof of such a statement.
	
	As for its potential drawbacks, the only thing we can think about is that 
	it might still be challenged in the same way as the relational model.
	The logic/physic separation provided by the relational model was actually
	used to attack it in 1970s, to state that it is impossible to be implemented 
	efficiently until System R and Ingres [125] were developed.
	To this end, we use enormous effort to bring our own ``Ingres'' along with the
	proposed model instead of trying to publish it alone, as a ``Codd-style'' paper 
	full of philosophical discussion is no way to be accepted today.

\subsubsection*{Query Language}}

Again, instead of providing a fully-featured and well-standardised query language 
for the community to follow,
\SQLd is proposed here mainly as a demo language to verify the concept of hybrid queries 
across graph and relational data,
and also to make our discussions on query evaluation more concentrated.
It should also be noted again that such a language is defined on the external-level of the \ModelName model
as shown in Figure~\ref{fig-overview} and Table~\ref{tab-model-comparison},
where the introduced logical-level pointers are hidden
to avoid introducing unacceptable complexity.

However, by saying that \SQLd is a demo language,
we do not mean that it is overly simplified and 
overlooks the key features of various graph query languages,
or is in an unusable shape.
In fact, a large portion of those features provided by various graph query languages 
are equivalently achieved in \SQLd by combing few key features from both the graph-side 
and relational-side query languages.
The most significant limitation on the expressiveness of \SQLd appears in the lack of support for
the recursive feature, which is far beyond the scope that can be discussed here
but can have some alternative temporary solution and will be systemically studied in future work.

\eat{
Therefore, it is possible for \SQLd to be regarded as a core to be extended
into a fully-featured and usable query language.

While the remaining ones,
which are far beyond the scope can be discussed here, 
can also be considered in future works.
Therefore, it is possible for \SQLd to be regarded as a core to be extended
into a fully-featured and usable query language.}

Below, we compare the expressiveness and the usability of \SQLd
with some graph query languages:

\stitle{Expressiveness}.
Following the same categorisation of query features and semantics in~\cite{angles2017foundations},
we first compare the expressiveness of \SQLd with other graph query languages.
We listed the comparison on pattern matching queries and navigational queries
in Table~\ref{tab-pattern-matching-compare} and Table~\ref{tab-navigational-compare} respectively.
Meanwhile, although they are not collected in the same table,
considering the connection between the RDF model and the relational model,
it can be seen that the other graph algebras built atop of the relational model~\cite{Jeffrey1}
can, at most, achieve an equivalent expressiveness as SPARQL.

Comparing to the various graph query languages,
the most significant limitations of \SQLd appear on the navigational queries,
since it can only describe the path with fixed-length patterns which might even hard to be called
a real {``navigational''} query.
The navigational queries in those graph query languages
are mainly described by path with different kinds of regular expressions,
which can be, and can only be, equivalently achieved in \SQL with a recursive extension~\cite{sakr2021future}.
Although a consideration of the recursive features is far beyond the scope that can be discussed here,
previous works~\cite{Verticagraph} also provide an alternative way to support 
recursive exploration,
\ie repeatedly running a script until it reaches a fixed point.
However, such an approach is only a temporary solution and is far from considering all the 
optimization methods available for recursive relational queries~\cite{bancilhon1985magic},
we thus defer a systemic study on this to further work.

In conclusion, as a demo language to prove the concept of hybrid query across graph and relational data,
\SQLd can fairly express the core concepts of the graph query languages especially on pattern queries.
Meanwhile, the limitations on navigational queries in \SQLd
can have some temporary solutions as in~\cite{Verticagraph}, 
and a systematic study on this can be achieved by incorporating more classical works on recursive relational queries~\cite{bancilhon1985magic}, 
which is deferred to future work.

\eat{which are mainly described in those various graph query languages with recursive features.
	
	as a dialect of \SQL, \SQLd is most similar to SPARQL
	
	Moreover, previous works~\cite{Verticagraph} provides a continent way to support 
	the recursive exploration,
	\ie repeatedly running a script till it reach the fixed point.}

\stitle{Usability}.
We further analysis the usability of such a demo language, more specifically,
on its composability, ease of pattern specification, and simplicity of its query interface.

\etitle{Composability}.
The composability is to say that the output of a query can be used as the input of another query,
which is overlooked by the graph query languages in the early stage~\cite{sparql, PGQL}.
The recent query languages~\cite{angles2018GCORE}, and recent versions of some evolving query languages~\cite{francis2018cypher},
achieve such a property in different ways
by exporting the result of a query as a graph~\cite{angles2018GCORE}
or passing {``table-graphs''}~\cite{francis2018cypher}.
As a dialect of \SQL, \SQLd then naturally retains its capability of query composition,
that is, it can naturally export the result of the query on 
the graph-side data as relations for it to be regarded as the input of another query.

\etitle{Pattern specification}.
There are majorly three styles to specify the pattern to be matched on data graph 
in previous graph query languages,
namely, the ``SQL-like'' style~\cite{sparql};
the ``functional'' style~\cite{TinkerPop};
and the ``ASCII-art'' style~\cite{GQL-standard,SQL-standard,  angles2018GCORE, alin2022Graph, francis2023researcher, francis2018cypher, PGQL}.
Among them, the ``ASCII-art'' style is undoubtedly the easiest-to-use and the most widely-recognized one
in the community of graph query language,
as it is utilized in a major portion of the query languages from both academia and industry,
which is thus followed in \SQLd.

\etitle{Query interface}.
An important lesson learnt from the {``great battle''} between the relational and the CODASYL model~\cite{Bachman1969CODASYL} is that 
the complicated structures is almost impossible for the practitioners to understand~\cite{stonebraker2005goes}.
Therefore, in \ModelName model, 
as shown in Figure~\ref{fig-overview},
the similar pointer structures are introduced, 
but they only appear on the logical level and are hidden from the external view.
That is, for the practitioners who are trying to query on the data by writing declarative queries, 
they can only view on the certain relations and certain graphs,
without seeing the extended relations with complicated pointer structures behind them on logical level.

\begin{table}
	\centering
	\begin{smaller}
		\centering
		\renewcommand{\arraystretch}{1.5}
		\begin{tabular}{|c|c|c|} 
			\hline
			{\bf Language} &  
			{\bf Supported patterns} &  
			{\bf Semantics} \\ 
			\hline
			{SPARQL~\cite{sparql}}	&	{all complex graph patterns}  &  {homomorphism-based, bags}	\\ 	\hline
			{Cypher~\cite{francis2018cypher}}	&	{all complex graph patterns}  &  {no-repeated-edges,  bags}	\\	\hline
			{Gremlin~\cite{TinkerPop}}	&	\makecell{all complex graph patterns 
				\\ without explicit optional}  &  {homomorphism-based, bags}	\\	\hline
			{\SQLd}		&	{all complex graph patterns}  &  {homomorphism-based, bags}	\\	\hline
		\end{tabular}
	\end{smaller}
	\caption{A comparison of semantics for pattern matching}\label{tab-pattern-matching-compare}
	\vspace{-3.3ex}
\end{table}

\begin{table}
	\centering
	\begin{smaller}
		\centering
		\renewcommand{\arraystretch}{1.75}
		\begin{tabular}{|c|c|c|c|} 
			\hline
			{\bf Language} &  
			{\bf Path expressions} &  
			{\bf Semantics} &  
			{\bf Choice of output} \\ 
			\hline
			{SPARQL}	&	\makecell{more than \\ RPQs}			&  \makecell{arbitrary paths,\\ sets}		&	{boolean / nodes} \\ 	\hline
			\makecell{Cypher}	&	\makecell{fragment of \\ RPQs}			&  \makecell{no repeated\\edges, bags}	&	\makecell{boolean / nodes \\ / paths / graphs} \\	\hline
			{Gremlin}	&	\makecell{more than \\ RPQs}			&  \makecell{arbitrary paths,\\ bags}		&	{nodes / paths} \\	\hline
			{\SQLd}		&	\makecell{fixed-length 
				\\ path queries}	&  \makecell{fixed-length\\ paths, bags}	&	{boolean / nodes} \\	\hline
		\end{tabular}
	\end{smaller}
	\caption{A comparison of semantics for navigational queries}\label{tab-navigational-compare}
	\vspace{-3.7ex}
\end{table}

By doing so, those logical level pointers can be utilized
to support a more efficient exploration without exposing themselves
to the external view and can thus keep the query interface simple.
An illustration of the \ModelName model
according to the three level architecture of data model~\cite{date1999introduction, steel1975ansi, codd1990relational}
is demonstrated in Table~\ref{tab-model-comparison}.
It can be seen that at the external level, \ModelName is a combination of the 
graph and relational data,
while \SQLd is a demonstration of a unified query language to access across them.

However, such a distinction between the graph and relational data
would be eliminated at the logical level, where both sides of data are represented uniformly 
with extended relations as shown in Section~\ref{subsec-define-model},
which thus provides a unified plan space for the access on graph and relational data
to be considered in any order without involving cross-model access~\cite{gimadiev2021combined, hassan2018grfusion, vyawahare2018hybrid}.
The practitioners then only need to concentrate on 
how to express their query without considering 
it can be evaluated efficiently, as the optimizer behind provides a unified plan space to select the optimal one from them
(see Section~\ref{subsec-query-optimization-with-exploration}).
\vspace{1ex}

\eat{
	Meanwhile, 
	as a demo language,
	such a query language also successfully demonstrates how a unified description 
	of the hybrid queries across graph and relational data can be provided.
	Such a plan space also further enables the access on graph and relational data
	to be uniformly considered in any order,
	for an optimal query plan to be generated and be further evaluated seamlessly without involving cross-model access~\cite{gimadiev2021combined, hassan2018grfusion, vyawahare2018hybrid} (see Section~\ref{subsec-query-optimization-with-exploration}).
	
	the relational model can be generated from the .
	
	Meanwhile, as a 
	
	This makes it possible to seamlessly manage relational
	data and graph data together without cross-model access
	\cite{gimadiev2021combined, hassan2018grfusion, vyawahare2018hybrid}. Moreover,
	it provides  a larger space for optimizing hybrid queries, \eg
	pushing relational join conditions down into 
	the pattern matching.
	while the extensions it introduces are also all easy for the practitioners to accept:
	The ``ASCII-art'' style description of pattern query is ~\cite{alin2022Graph};
	The ``map'' clause for ER can be simply regarded as a special kind of {\em join} without specifying conditions.
	

	It can then be seen that the proposed \ModelName model achieves all these three requirements with minimal modifications
	to the relational model without .
	Meanwhile, the \ModelName model can also achieve an equivalent expressiveness 
	compared to previous graph algebras that are built atop of the relational model~\cite{zhao2017all}.
	The recursion feature~\cite{sakr2021future} is far beyond the scope of this paper but can be somewhat alternatively achieved as in~\cite{Verticagraph}.
	We defer a systemic study of this to future work.
	
	We will see in Section~\ref{subsec-logical-exploration} that the pointers can
	
	In fact, by combing few key features from both the graph and relational query languages,
	a large portion of the features they provide can expressed.
	While the rest of those features can be considered with a consideration of 
	the recursive which is much beyond the scope that can be discussed here.

	
	
}

Conclusively, as a demo language, \SQLd shows that it is possible to provide a query 
across graph and relational data, which:
\mbi

\item 	has the composability to uniformly allow both the query 
		result from both graph and relational data
		as the input of the other query;

\item	allows the pattern to be match on the data graph 
		to be specified with a widely accepted ``ASCII-art'' style,
		while supporting most of the key features
		in the various graph query languages;

\item	has a unified query interface without exposing the 
		complicated pointer structures at the external level, 
		with a capability for automatic optimization
		to liberate the programmers from considering the detailed execution procedure;
\mei

By showing those core features, 
\SQLd proves its potential to be considered as a core language,
and to be further extended into a fully-featured, well-standardized practical query
language.

\end{document}